\documentclass[12pt,a4paper]{article}

\usepackage[english]{babel}
\usepackage[dvips]{graphicx}

\usepackage{yhmath}
\usepackage{amsfonts}
\usepackage{amsmath}
\usepackage[single]{accents}

\usepackage{array}
\usepackage{verbatim}
\usepackage{graphics}
\usepackage{subcaption} 
\usepackage{epsfig}
 
\usepackage{tabularx}
\usepackage{mathcomp}
\usepackage{multirow}
\usepackage{url}

\usepackage{placeins} 
\usepackage{relsize}

\usepackage{natbib}
\usepackage{abeCommands}

\newcommand{\mm}{\widetilde{m}}
\usepackage{hyperref}
\hypersetup{
    colorlinks=true,
	linktocpage=true,
	pdfstartpage=1, pdfstartview=FitV,
    breaklinks=true, pdfpagemode=UseNone, pageanchor=true, pdfpagemode=UseOutlines,
    plainpages=false, bookmarksnumbered, bookmarksopen=true, bookmarksopenlevel=1,
    hypertexnames=true,
    pdfhighlight=/I,
    linkcolor=RoyalBlue, 
	citecolor=webgreen, 
    urlcolor=DarkBlue, 
    pdftitle={High order time integrators for the simulation of charged particle motion in magnetic quadrupoles},%
    pdfauthor={},
    pdfsubject={},
    pdfkeywords={}
}

\begin{document}

\title{High order time integrators for the simulation of charged particle motion in magnetic quadrupoles}

\author{Abele Simona $^{(1)}$, Luca Bonaventura$^ {(1)}$\\
Thomas Pugnat  $^{(2)}$, Barbara Dalena $^{(2)}$}

\maketitle

 \begin{center}
{\small
 
(1) MOX -- Modelling and Scientific Computing, \\
Dipartimento di Matematica, Politecnico di Milano \\
Via Bonardi 9, 20133 Milano, Italy\\
{\tt abele.simona@polimi.it, luca.bonaventura@polimi.it }\\
{$ \ \ $ }\\
 (2)  Commissariat \`a l'\'Energie Atomique et aux \'Energies Alternatives \\
 Saclay \\                        
{\tt  thomas.pugnat@gmail.com, barbara.dalena@cea.fr}\\
 }
\end{center}

\date{}


\vspace*{0.5cm}


\vspace*{0.5cm}

\pagebreak

\abstract{Magnetic quadrupoles are essential components of particle accelerators like the Large Hadron Collider.
In order to study numerically the stability of the particle beam crossing a quadrupole, a large number of particle revolutions in the accelerator must be simulated, thus leading to the necessity to preserve numerically invariants of motion over a long time interval and to a substantial computational cost, mostly related to the repeated  evaluation of the magnetic vector potential. In this paper, in order to reduce this cost, we first consider a specific gauge transformation that allows to reduce significantly the number of vector potential evaluations. We then analyze the sensitivity of the numerical solution  to the interpolation procedure required to compute magnetic vector potential data from gridded precomputed values at the locations required by high order time integration methods. Finally, we compare several high order integration techniques, in order to assess their accuracy and efficiency for these long term simulations. Explicit high order Lie methods are considered, along with implicit high order symplectic integrators  and conventional explicit Runge Kutta methods. Among symplectic methods,
high order Lie integrators  yield optimal results in terms of cost/accuracy ratios, but non symplectic
Runge Kutta methods perform remarkably well even in very long term simulations. Furthermore, the accuracy
of the field reconstruction and interpolation techniques are shown to be limiting factors for the accuracy of the
particle tracking procedures.}


\pagebreak

\section{Introduction}
\label{intro} \indent
 
Magnetic quadrupoles are key components  of  particle accelerators that are used to focus particle beams.
 In high energy circular accelerators, the quality of their magnetic fields
 can influence the overall beam stability.  
    The beam stability is measured in terms of the dynamic aperture, 
    defined as the region in the phase space outside which a particle is considered as lost from the beam. The dynamic aperture is estimated by 
    solving numerically the Hamilton equations describing the
      particle trajectories   for a large number 
    of accelerator revolutions  (typically, more than $10^5$). Therefore, 
    it is important to use very efficient time integration methods which also guarantee good long-term preservation
    of dynamical invariants of motion.
    	The action of a quadrupole can be approximately described using a linear combination of the position and momenta of the particles at the inlet that yields the position and momenta at the outlet \cite{carey:1988}.
	 This approach yields a good approximation if the particles travel near the quadrupole center (small apertures) and it is exact in regions where the magnetic field is constant along the longitudinal axis $z$ and only 
	 the main quadrupole field is present. In realistic cases, however,  the magnetic field has a more complex structure which involves non-uniformity along $z$ and harmonics of higher order, see e.g. \cite{forest:1998}.
	 These inhomogeneities of the field at the quadrupole ends, known as fringe field, can lead to a non-linear dependency  of the position and momenta of the particles at the outlet from the position and momenta at the inlet. In this paper, we focus on the numerical problems encountered when modelling these non linear dependencies in an accurate and efficient way, as necessary for the design of the large aperture quadrupoles foreseen for the HL-LHC project \cite{rossi:2011}.  	The accurate numerical
	solution of the complete Hamilton equations is mandatory in this case. A preliminary study on 
	the applicability of second-order methods based on the  Lie algebra integrators proposed in \cite{wu:2003} has been carried out in \cite{dalena:2014} for the case of a realistic quadrupole.
	 	 
		  In this work, we first consider a specific gauge transformation that allows to reduce by 
		  approximately 50\%  the computational cost of each vector potential evaluation, thus significantly enhancing the efficiency of any numerical approximation method employed for the particle trajectory simulation.
We then compare several high order integration techniques, which   allow to maintain sufficiently high accuracy
even with relatively large integration step values, 
in order to assess their accuracy and efficiency for these long-term simulations. 
Explicit high order Lie methods \cite{wu:2003} are considered along with implicit high order symplectic integrators \cite{hairer:2006}  and more conventional, non symplectic explicit Runge-Kutta methods.  
  
 In the case of  realistic  vector potentials, the errors induced by the vector potential reconstruction and interpolation become significant and reduce the highest possible accuracy that can be attained. Furthermore, since in realistic cases the magnetic vector potential evaluation is more costly, numerical methods which require less evaluations, such as the second-order Lie method, appear to be more competitive. On the other hand, experiments
 with idealized fields show that,
if these errors could be reduced,
  higher order methods could be advantageous  and the speed gain obtained with the horizontal-free Coulomb gauge would enhance their efficiency. In particular, the explicit fourth-order Runge-Kutta appears to be the most efficient method
  and the fourth-order Lie the most efficient among  symplectic methods.  
   A particularly interesting aspect of the results obtained is the fact that non symplectic methods
appear to be competitive with symplectic ones even on relatively long integrations, when stability of the computed trajectories and energy conservation are considered. Indeed,  the spurious energy
losses appear to be more closely related to the errors in the representation of the magnetic vector potential
than to those introduced by the time integrators.
This unexpected result warrants more detailed investigation.

	  A detailed outline of the paper is as follows.
	In section \ref{quadrups}, we introduce the Hamiltonian that describes the motion of a charged particle inside a magnetic quadrupole and the corresponding Hamilton equations. 
	In section \ref{ode_rev}, the numerical methods used in this work are briefly reviewed,
	along with the specific issues that arise when the vector potential values are only available at
	given sampling intervals, so that appropriate
	interpolation procedures must be employed if high order methods are to be applied.	
	In section \ref{vecpot}, we review the approach employed to represent the magnetic vector potential.
	We also show how the non-uniqueness of the vector potential can be exploited to identify
	 a  gauge that allows   to reduce the number of vector potential evaluations.	
	Numerical results for a simple vector potential that can be expressed analytically are presented
	in section \ref{analyticCase}. An assessment of  time integration methods
	 on a realistic case is presented in section
	 \ref{realCase}.
	 Finally, some conclusions are drawn in section
	\ref{conclu}, where we also discuss possible developments of this work.   	
	 
\section{Charged particle motion in magnetic quadrupoles}
\label{quadrups} \indent

Magnetic quadrupoles are devices  in which  a stationary magnetic field with cylindrical symmetry is generated.
Near the quadrupole center, where the particles travel, the field is a solution of the Maxwell equations in the vacuum and in absence of charges and currents
	 
\begin{equation}
\label{maxwell}
\begin{aligned}
    \Nabla \cdotp \vec{E} & = 0 & \qquad  \Nabla \cdotp \vec{B} &= 0 \\
    \Nabla \times \vec{E} & = 0 & \qquad  \Nabla \times \vec{B} &= 0. 
\end{aligned}
\end{equation}
		Since the magnetic field $\vec{B}$ is irrotational   and therefore conservative 
		on simply connected domains,  there exist a magnetic scalar potential $\psi$ and a  magnetic vector potential $\vec{A}$ such that
    $ \vec{B} = \Nabla \times \vec{A} = \Nabla \psi. $
The magnetic scalar potential is defined up to constants and the magnetic vector potential is defined up to   gradients of scalar functions, so that   given a scalar function $\lambda$ and defining $\vec{A'} = \vec{A} + \Nabla \lambda$, one has $\Nabla \times \vec{A'} =  \Nabla \times (\vec{A} + \Nabla \lambda) = \Nabla \times \vec{A} = \vec{B}. $ 
        A gauge transformation $\vec{A'} = \vec{A} + \Nabla \lambda $
        does not change the magnetic field but can yield a more convenient representation of the vector potential
        for specific purposes.    
    The second      Maxwell equation can be rewritten as
    \begin{equation}
    \label{eq:LaplPotEq}
      0 = \Nabla \cdotp \vec{B} = \evA{\Nabla \cdotp ( \Nabla} \psi \evA{)} = \evA{\Delta} \psi ~,
    \end{equation}
    which means that the scalar potential satisfies the Laplace equation. Many high accuracy
    numerical methods are available to solve this equation starting from appropriate boundary conditions
    and taking into account also the real geometry of the accelerator, see e.g. \cite{corno:2016}, \cite{lipnikov:2011}.
    However, in the approach most commonly used in accelerator physics, see   e.g.  \cite{dragt:1997},   \cite{venturini:1999}, the  magnetic vector potential $\vec{A}$ is represented  by a power series
 in the transversal coordinates, whose coefficients are functions
       of the longitudinal coordinate. A detailed presentation of this approach is given in section \ref{vecpot}.

		The motion of a charged particle   in a magnetic quadrupole is described in terms of its position $\qq(t) = \parT{x, \, y, \, z}^T$ and its canonical momentum $\pp(t) = \parT{p_x, \, p_y, \, p_z}^T$. In the case of a high energy particle accelerator, the particles speed is very close to the speed of light. The relativistic Hamiltonian is given by
	\begin{equation}
		\label{eq:REMHamil}
		\HH\parT{ \qq, \, \pp}
		= \sqrt{m^2 c^4 + c^2 \parT{\pp - Q \, \vec{A}(\qq)}^2 }, 
	\end{equation}
	where  $m$ denotes the rest mass of the particle,  $c$ is the speed of light and $Q$ the particle charge \cite{panofsky:1962}. 
	The mechanical momenta are denoted by
	$\pp^M = m \gamma \vec{v}, $
	where $\vec{v}$ is the particle speed and 
	$\gamma = 1/\sqrt{1 -  \|\vec{v}\|^2/c^2} $ is the  
	Lorentz factor.
The canonical momenta $\pp$ are related to the mechanical momenta through:
\[
    \pp = \pp^M + Q \, \vec{A}(\qq).
\]
Introducing the state vector $\ww = \parT{\qq^T, \, \pp^T}^T$,  the relativistic Hamilton equations 
can be written as 
	\begin{equation}
	\label{eq:REMJgH}
	\begin{aligned}
		\dt{\ww} & = \mat{J} \, \Nabla \HH \parT{\ww} = \\
		& = 
		\begin{bmatrix}
		\dfrac{c^2\parT{\pp - Q\,  \vec{A}\parT{\vec{\qq}}}}{\sqrt{m^2 c^4 + c^2 \parT{\pp - Q \, \vec{A}(\qq)}^2}} \\[1.5em]
		\dfrac{Q \, c^2}{\sqrt{m^2 c^4 + c^2 \parT{\pp - Q \, \vec{A}(\qq)}^2}}\, \Jac{\vec{A}}{\qq}{\qq}^T  \parT{\pp - Q \, \vec{A}\parT{\qq}} 
	\end{bmatrix}~.
	\end{aligned}
	\end{equation}
	Here we denote  
	$\Jac{\vec{A}}{\qq}{\qq}= \partial \vec{A}/\partial \vec{\qq} $  and we set
	\begin{equation}
    \label{eq:matj}
\mat{J} =
    \begin{bmatrix}
        \mathbf{0}& \mathbf{I} \\
        - \mathbf{I}  & \mathbf{0} \\
    \end{bmatrix},
\end{equation}
where  $\mathbf{0},\mathbf{I}$ are the zero and identity matrix, respectively.
For the specific problem at hand,  it is   convenient to assume that the  $z$ axis is the symmetry axis of the
quadrupole and to
 use the longitudinal coordinate $z$ as independent variable instead of time. This change of independent variable leads to the Hamiltonian
	\begin{flalign*}
	&   {\cal F} \deff - \sqrt{\dfrac{p_t^2}{c^2} - m^2 c^2 - \parT{p_x - Q \, A_x}^2
	 - \parT{p_y - Q \, A_y}^2} -  Q \, A_z ~,
	\end{flalign*}
	where now $p_t \deff -\HH$ is the conjugate momentum of $t$.  The dynamical variables are
	 $\ww = \parT{x, \, y, \, t, \, p_x, \, p_y, \, p_t}^T$ and $z$ is now the independent variable. It can be noticed that this new Hamiltonian does not describe an autonomous system, due to the fact that the magnetic vector potential depends on the independent variable $z$.
	
	The magnetic field along the center of a quadrupole is null. Therefore, a particle traveling along the $z$ axis with speed $v_z^0$ is not influenced by the quadrupole. In the following we will indicate the quantities related to the reference particle using the superscript $0$. The trajectory described by this particle is called the reference orbit and it is identified by $x^0 = y^0 = p_x^0 = p_y^0 = 0$, $t^0 = \dfrac{z}{v_z^0} = \dfrac{z}{c \, \beta^0}$ and $p_t^0  = - \gamma^0 m c^2$. It is convenient to describe the motion of a general particle   in terms of deviations with respect of the reference orbit
$	\ww^d = \ww - \ww^0$.
	This change of variables is a canonical transformation which leads to the new 
	Hamiltonian  
	\begin{align*}
	 {\cal F}^{d} & =  - \sqrt{\dfrac{\parT{p_t^0 + p_t^d}^2}{c^2} - m^2 c^2 - \parT{p_x - Q \, A_x}^2 - \parT{p_y - Q \, A_y}^2} -  \\
	& \phantom{ = } \,\,  - Q \, A_z - \dfrac{p_t^d}{ c\,\beta^0}~.
	\end{align*}
	
	Moreover, it is convenient to introduce a specific scaling of the deviation variables.
	In particular, all the position variables will be scaled by a fixed length $L,$ usually denoted as the bunch length, while the momentum variables are scaled by  the module of the reference mechanical momentum 
	$$p^0 = \sqrt{\parT{\dfrac{p_t^0}{c}}^2 - m^2 c^2}.$$
	 In this work, we will use a reference length $L=1\,\text{m}$ and we will use as a reference the momentum of a proton with rest mass $m_0 \sim 9.38\,\text{MeV}/\text{c}^2 \sim 1.67\cdot10^{-27} \,\text{kg}$ and total energy $E = 7\,\text{TeV} \sim 1.12 \cdot 10^{-6} \,\text{J}$. Therefore the reference momentum is given by:
    \begin{equation}
        p^0 = 7\, \dfrac{\text{TeV}}{\text{c}} \sim 3.74 \cdot 10^{-15} \, \text{kg} \cdot \dfrac{\text{m}}{\text{s}}~.
    \end{equation}
	The scaled, non dimensional variables are denoted by $X,Y,\tau $ and $P_x,P_y,P_{\tau}, $ respectively,
	where $\tau    = c t^d/L, $ $P_\tau   =  p_t^d/(p^0 \,c).$
	  	 Another important quantity that will be used in this work is the mechanical momentum deviation
	\[
	\delta = \dfrac{\abs{\pp} - p^0}{p^0} = \dfrac{\abs{\pp}}{p^0} - 1~,
	\]
	which is related to $P_\tau$ by the relation
$$		\delta = \sqrt{1 - \dfrac{2 P_\tau}{\beta^0} + P_\tau^2} - 1~.$$
	Replacing the canonical pair $\parT{\tau, \, P_\tau}$ by $\parT{\ell, \, \delta},$ 
	one obtains the Hamiltonian
	\begin{equation}
	\label{eq:HRBis}
    	\widetilde{ {\cal F}}   = -\sqrt{\parT{1 + \delta}^2 - \parT{P_x - \widetilde{A}_x}^2 - \parT{P_y - \widetilde{A}_y}^2} - \widetilde{A}_z - \delta,
    		\end{equation}
	where $\vec{\widetilde{A}}\parT{X, Y, Z} = \dfrac{Q}{p^0}\vec{A}\parT{L \,X, L \, Y, L \, Z}$.
	For relativistic particles, the momenta in the transversal plane are much smaller than the total momentum module, i.e. 
	\[\parT{1 + \delta}^2 \gg \parT{P_x - \widetilde{A}_x}^2 + \parT{P_y - \widetilde{A}_y}^2 ~. \]
 Therefore, the so called paraxial approximation can be introduced, that amounts to substituting the square root in equation \eqref{eq:HRBis} with its Taylor expansion truncated at the first order:
	\begin{align}
	\widetilde{ \widetilde {\cal F}}& = -\sqrt{\parT{1 + \delta}^2 - \parT{P_x - \widetilde{A}_x}^2 - \parT{P_y - \widetilde{A}_y}^2} - \widetilde{A}_z - \delta = \nonumber\\ 
	& = - \parT{1 + \delta} \sqrt{1 - \dfrac{\parT{P_x - \widetilde{A}_x}^2}{\parT{1 + \delta}^2} - \dfrac{\parT{P_y - \widetilde{A}_y}^2}{\parT{1 + \delta}^2}} - {\widetilde{A}_z} - {\delta} \approx \nonumber \\ 
	& \approx - {\parT{1 + \delta}} \parT{1 - \dfrac{\parT{P_x - \widetilde{A}_x}^2}{2 \parT{1 + \delta}^2} - \dfrac{\parT{P_y - \widetilde{A}_y}^2}{2 \parT{1 + \delta}^2}} - {\widetilde{A}_z} - {\delta} = \nonumber \\ 
	& = {-{1}} - {2\delta} + \dfrac{\parT{P_x - \widetilde{A}_x}^2}{2 \parT{1 + \delta}} + \dfrac{\parT{P_y - \widetilde{A}_y}^2}{ 2 \parT{1 + \delta}} - {\widetilde{A}_z}~,  \label{eq:parApp6D}
	\end{align}
	where the constant terms can be neglected because they do not influence the Hamilton equations. 
	In general, due to the fact that the vector potential depends on the independent variable $Z$, this Hamiltonian describes a $6 $ dimensional non-autonomous system. To obtain an autonomous system, it is possible to introduce a new canonical pair $\parT{Z, \, P_z}$ and a new independent variable $\sigma$:
	\begin{equation}
		\label{eq:parApp8D}
		\KK \parT{\qq, \, \pp; \sigma} =  \dfrac{\parT{P_x - \widetilde{A}_x}^2}{2 \parT{1 + \delta}} +  \dfrac{\parT{P_y - \widetilde{A}_y}^2}{ 2 \parT{1 + \delta}}   - {\widetilde{A}_z} - {2\delta}  + {P_z},
	\end{equation}
	where $\qq = \parT{X, \, Y, \ell, \, Z}$ and $\pp = \parT{P_x, \, P_y, \, \delta, \, P_z}$.
	In this case, the resulting Hamilton equations are:
	\begin{equation}
	\label{eq:ODE8Dparax}
	\begin{aligned}
		\dt{\ww}  = \mat{J} \, \grad \KK =  
		\begin{bmatrix}
		\dfrac{P_x - \widetilde{A}_x}{\delta + 1} \\[1.1em]
		\dfrac{P_y - \widetilde{A}_y}{\delta + 1} \\[1.1em]
		- \dfrac{\parT{P_x - \widetilde{A}_x}^2}{2\parT{\delta + 1}^2} - \dfrac{\parT{P_y - \widetilde{A}_y}^2}{2\parT{\delta + 1}^2} - 2\\[1.1em]
		1 \\[1.1em]
		 \pder{\widetilde{A}_x}{X} \, \dfrac{\parT{P_x - \widetilde{A}_x}}{\delta + 1} + \pder{\widetilde{A}_y}{X} \, \dfrac{\parT{P_y - \widetilde{A}_y}}{\delta + 1} + \pder{\widetilde{A}_z}{X} \\[1.1em]
		 \pder{\widetilde{A}_x}{Y} \, \dfrac{\parT{P_x - \widetilde{A}_x}}{\delta + 1} + \pder{\widetilde{A}_y}{Y} \, \dfrac{\parT{P_y - \widetilde{A}_y}}{\delta + 1} + \pder{\widetilde{A}_z}{Y} \\[1.1em]
		0 \\[1.1em]
		 \pder{\widetilde{A}_x}{Z} \, \dfrac{\parT{P_x - \widetilde{A}_x}}{\delta + 1} + \pder{\widetilde{A}_y}{Z} \, \dfrac{\parT{P_y - \widetilde{A}_y}}{\delta + 1} + \pder{\widetilde{A}_z}{Z} \\[1.1em]
		\end{bmatrix}~.
	\end{aligned}
	\end{equation}
	Moreover, it can be noticed that the Hamiltonian does not depend on $\ell,$ so that  the partial derivative of $\KK$ with respect to $\ell$ is zero. As a consequence, $\delta$ is a constant of motion, equal to the initial value, denoted by the subscript $0$,  
	$\delta_0.$ If the evolution of the variable $\ell$ is not needed, the canonical pair $\parT{\ell, \, \delta}$ can be neglected, considering $\delta_0$ as a parameter and reducing again the size of the phase space. In this case, the Hamiltonian is still given by \eqref{eq:parApp8D} but, since the dynamical variables are now $\ww = \parT{X, \, Y, Z, \, P_x, \, P_y, \, P_z}$, the Hamilton equations become:
	\begin{equation}
	\label{eq:ODE6Dparax}
	\begin{aligned}
		\dt{\ww}  = \mat{J} \, \grad \KK  
		= 
		\begin{bmatrix}
		\dfrac{P_x - \widetilde{A}_x}{\delta_0 + 1} \\[1.1em]
		\dfrac{P_y - \widetilde{A}_y}{\delta_0 + 1} \\[1.1em]
		1 \\[1.1em]
		 \pder{\widetilde{A}_x}{X} \, \dfrac{\parT{P_x - \widetilde{A}_x}}{\delta_0 + 1} + \pder{\widetilde{A}_y}{X} \, \dfrac{\parT{P_y - \widetilde{A}_y}}{\delta_0 + 1} + \pder{\widetilde{A}_z}{X} \\[1.1em]
		 \pder{\widetilde{A}_x}{Y} \, \dfrac{\parT{P_x - \widetilde{A}_x}}{\delta_0 + 1} + \pder{\widetilde{A}_y}{Y} \, \dfrac{\parT{P_y - \widetilde{A}_y}}{\delta_0 + 1} + \pder{\widetilde{A}_z}{Y} \\[1.1em]
		 \pder{\widetilde{A}_x}{Z} \, \dfrac{\parT{P_x - \widetilde{A}_x}}{\delta_0 + 1} + \pder{\widetilde{A}_y}{Z} \, \dfrac{\parT{P_y - \widetilde{A}_y}}{\delta_0 + 1} + \pder{\widetilde{A}_z}{Z} \\[1.1em]
		\end{bmatrix}~.
	\end{aligned}
	\end{equation}
	A further simplification can be achieved noticing that $P_z$  is decoupled from the other dynamical variables,
	so that its computation   can be neglected if we are only interested in the dynamics of the transversal variables, reducing the number of equations \eqref{eq:ODE6Dparax} to the four ones associated to $X$, $Y$, $P_x$ and $P_y$.

 \section{Review of high order numerical methods for ODE problems}
\label{ode_rev} \indent
    In this section, some high order numerical methods for the  
    solution of a first order system $\dt{\vec{y}}(t)    = \vec{f}(t, \vec{y}) $
    will be reviewed, in view of their application to the solution of the
      Hamilton equations. A  more detailed presentation
      of the relevant numerical methods  can be found for example in \cite{hairer:2006}.
      
      An important feature of
      Hamiltonian flows is their symplectic property, which can be defined more precisely as follows.
      \begin{defi}
          A differentiable map $\vec{g}:U \rightarrow \R^{2d}$, where $U \subset \R^{2d}$ is an open set,
           is called symplectic if:
          \begin{equation*}
              \mathcal{J}_{\vec{g}}\parT{\ww}^T \mat{J} \mathcal{J}_{\vec{g}}\parT{\ww}
               = \mat{J} \qquad \forall \ww \in U~,
          \end{equation*}
          where $\mathcal{J}_{\vec{g}}$ is the Jacobian matrix of $\vec{g}$ and $\mat{J}$ is the matrix \eqref{eq:matj}.
      \end{defi}
      When this  property is  preserved by the numerical method, i.e., if the one-step map $\Phi_{\Delta t} : y_0 \mapsto \Phi_{\Delta t}(y_0)=y_1$ is symplectic, quadratic invariants of motion are preserved, thus
      ensuring in principle a good behaviour for long-term simulations. 
      The first symplectic techniques that will be considered are Runge-Kutta methods, which  can be written in general  as  
	\begin{defi}
		Let $b_i, \, a_{ij} \, (i,\,j = 1, \, \ldots, \, s)$ be real numbers and let $c_i = \sum_{j = 1}^s a_{ij}$.
		A s-stage Runge-Kutta method is given by:
		\begin{eqnarray}
		\vec{y}_{n + 1} & =& \vec{y}_n + \Delta t \sum_{i = 1}^s b_i    \, \vec{f}(t_n + \Delta t \, c_i, \vec{u}_i)   \label{eq:rkStep} \nonumber \\
		\vec{u}_i       & = &\vec{y}_n + \Delta t \sum_{i = 1}^s a_{ij} \, \vec{f}(t_n + \Delta t \, c_i, \, \vec{u}_j), \quad i = 1, \, \ldots, \, s. \label{eq:uiRK}
		\end{eqnarray}
	\end{defi}
	Runge-Kutta methods are often summarized via 
	 the so called  Butcher tableau, in which all the coefficients are arranged as:
	\begin{equation*}
		\begin{array}{c | c  c  c}
		c_1    & a_{1, \, 1} & \cdots  & a_{1, \, s} \\
		\vdots & \vdots      & \ddots  & \vdots      \\[-0.5em]
		c_s    & a_{s, \, 1} & \cdots  & a_{s, \, s} \\
		\hline                                       \\[-2.2em]
		& b_1         & \cdots  & b_s
		\end{array}~.
	\end{equation*}
	A Runge-Kutta method is explicit if $ a_{i j} = 0 $ for $j \geq i.$
    The following theorem gives a sufficient condition for a Runge-Kutta method to be symplectic \cite{hairer:2006}.
	\begin{theorem}
		\label{th:sympRK}
		If the coefficients of a Runge-Kutta method satisfy:
		\begin{equation}
		b_i \, a_{ij} + b_j \, a_{ji} = b_i \, b_j   \quad \forall i, \, j = 1, \, \ldots, \, s
		\end{equation}
		then the method is symplectic.
	\end{theorem}
	Gauss methods are particular implicit Runge-Kutta methods, some of which satisfy the condition of theorem \eqref{th:sympRK} and are thus symplectic. The midpoint method can be interpreted as the second-order Gauss method, characterized by the   Butcher tableau
	\begin{equation*}
	\renewcommand*{\arraystretch}{2.4}
	\begin{array}{c | c}
	\dfrac{1}{2} & \dfrac{1}{2}  \\
	\hline                                                                                                       \\[-3.0em]
	& \dfrac{1}{2}               \\
	\end{array}~.
	\end{equation*}
	The fourth-order Gauss method, considered in this paper,
	 is characterized by the Butcher tableau
	\begin{equation*}
		\renewcommand*{\arraystretch}{2.4}
		\begin{array}{c | c  c  c}
		\dfrac{1}{2} - \dfrac{\sqrt{3}}{6} & \dfrac{1}{4}                       & \dfrac{1}{4} - \dfrac{\sqrt{3}}{6} \\
		\dfrac{1}{2} + \dfrac{\sqrt{3}}{6} & \dfrac{1}{4} + \dfrac{\sqrt{3}}{6} & \dfrac{1}{4}                       \\[0.5em]
		\hline                                                                                                       \\[-3.0em]
		& \dfrac{1}{2}                       & \dfrac{1}{2}                       \\
		\end{array}~.
	\end{equation*}
	The sixth-order Gauss method is instead characterized by the   Butcher tableau
	\begin{equation*}
		\renewcommand*{\arraystretch}{2.4}
		\begin{array}{c | c  c  c}
		\dfrac{1}{2} - \dfrac{\sqrt{15}}{10} & \dfrac{5}{36} & \dfrac{2}{9} - \dfrac{\sqrt{15}}{15} & \dfrac{5}{36} - \dfrac{\sqrt{15}}{30} \\
		\dfrac{1}{2}                         & \dfrac{5}{36} + \dfrac{\sqrt{15}}{24} & \dfrac{2}{9} & \dfrac{5}{36} - \dfrac{\sqrt{15}}{24} \\
		\dfrac{1}{2} + \dfrac{\sqrt{15}}{10} & \dfrac{5}{36} + \dfrac{\sqrt{15}}{30} & \dfrac{2}{9} - \dfrac{\sqrt{15}}{15} & \dfrac{5}{36} \\[0.5em]
		\hline                                                                                                       \\[-3.0em]
		& \dfrac{5}{18}                       & \dfrac{4}{9}      & \dfrac{5}{18}                 \\
		\end{array}~.
	\end{equation*}	
	Implicit methods require the solution of a nonlinear system of equations at each time step. This can be done using either the Newton or  the fixed-point method. Even though Newton's method is usually superior, numerical results show that the latter is faster. This behaviour can be justified by the fact that, for the problems at hand, both methods require a small number of iterations to achieve convergence, 
	but the Newton method implies   higher initial costs related to the evaluation of Jacobian matrix,
	see e.g. the discussion in \cite{hairer:2006}.
	Also the
	 best known fourth-order explicit Runge-Kutta method has been considered in this work.
	 This method is not symplectic and it is characterized by the   Butcher tableau
	\begin{equation*}
		\begin{array}{c | c  c  c  c}
		0            &              &              &              &              \\[0.2em]
		\dfrac{1}{2} & \dfrac{1}{2} &              &              &              \\[.5em]
		\dfrac{1}{2} & 0            & \dfrac{1}{2} &              &              \\[0.1em]
		1            & 0            & 0            & 1            &              \\
		\hline                                                                   \\[-1.5em]
		& \dfrac{1}{6} & \dfrac{1}{3} & \dfrac{1}{3} & \dfrac{1}{6} \\
		\end{array}~.
	\end{equation*}

	If $\HH$ is the Hamiltonian ruling the evolution of an autonomous system, then the exact solution of the Hamilton equations can be formally represented as
	\begin{equation}
		\vec{w}(t) = \expp{t\, \lie{- \HH}}\vec{w}_0.
	\end{equation}
Here, the notation of  equation \eqref{eq:REMJgH} is used, $\lie{}$ denotes the Lie operator and the exponentiation of a Lie operator is called Lie transformation \cite{dragt:1997}. 
	The  methods based on Lie algebra techniques most widely applied in accelerator physics
employ a second-order approximation of the Lie transformation. Higher order Lie methods are then built using the procedure introduced by Yoshida in    \cite{yoshida:1990} and further discussed in \cite{wu:2003}.
	The first step is to split the Hamiltonian $\HH$ in $s$ solvable parts
	\begin{equation*}
		\HH = \sum_{i = 1}^s \HH_i
	\end{equation*}
	such that $\expp{\lie{\HH_i}} $ can be computed exactly for  $ i=1, \, \ldots, \, s$. This is true if $\lie{\HH_i}$ is nilpotent of order two (i.e. $\lie{\HH_i}^{\!k}~ \ww = \vec{0} $ for $k \geq 2$), because in this case the exponential series reduces to a finite sum. A second order approximation is  then given by
	\begin{equation}
		\label{eq:LieApp2}
	\begin{aligned}
		& \expp{\Delta t\, \lie{- \HH}} = \\
		& \hspace{1cm} = \expp{\dfrac{\Delta t}{2} \, \lie{- \HH_1}}\expp{\dfrac{\Delta t}{2} \, \lie{- \HH_2}} \ldots \\
		& \hspace{1cm} \ldots\expp{\Delta t \, \lie{- \HH_s}} \expp{\dfrac{\Delta t}{2} \, \lie{- \HH_{s - 1}}} \ldots \\
		& \hspace{1cm} \ldots\expp{\dfrac{\Delta t}{2} \, \lie{- \HH_1}} + o\parT{\Delta t^2}.
	\end{aligned}
	\end{equation}
	Denoting by  $\mathcal{M}_2 (\Delta t)$ the approximation \eqref{eq:LieApp2} and by 
	  $\mathcal{M}_{2n} (\Delta t)$  an approximation of order $2n$, an approximation of order $2n + 2$ can be built as follows
	\begin{equation}
		\mathcal{M}_{2n + 2} (\Delta t) = \mathcal{M}_{2n} (\alpha_1 \, \Delta t) \mathcal{M}_{2n} (\alpha_0 \, \Delta t) \mathcal{M}_{2n} (\alpha_1 \, \Delta t)~,
	\end{equation}
	where $\alpha_0 = - \dfrac{2^{{1}/{2n + 1}}}{2 - 2^{{1}/{(2n + 1)}}}$ and $\alpha_1 = \dfrac{1}{2 - 2^{{1}/{(2n + 1)}}}$. In this work, methods of order $4$ and $6$ have been considered, with $(\alpha_0, \alpha_1)$ pairs given by
	\begin{equation*}
		\parT{- \dfrac{2^{1/3}}{2 - 2^{1/3}}, \, \dfrac{1}{2 - 2^{1/3}}} \quad \text{and} \quad \parT{- \dfrac{2^{1/5}}{2 - 2^{1/5}}, \, \dfrac{1}{2 - 2^{1/5}}}~,
	\end{equation*}
	respectively.
	
	Taking into account the discussion in section \ref{quadrups}, the map $\mathcal{M}_2$, applied to $\ww_n$ yields the following algorithm
	\begin{equation}
    \label{eq:lieAlg4D}
	\resizebox{0.90\linewidth}{!}{%
		$\setcounter{MaxMatrixCols}{18}\begin{matrix}
		\ww_{n +  1/11} & = & \ww_{n} ~ + & \dfrac{\sigma}{2}	& \bigg( & 0, & 0, & \pder{\widetilde{A}_z}{X}, & \pder{\widetilde{A}_z}{Y} & \bigg)^T ~;\\
		\ww_{n +  2/11} & = & \ww_{n +  1/11} ~ + &  									& \bigg( & 0, & 0, & \widetilde{A}_x, &  \mathlarger{\int} \pder{\widetilde{A}_x}{Y} \, \dd X & \bigg)^T ~;\\
		\ww_{n +  3/11} & = & \ww_{n +  2/11} ~ + &  \dfrac{\sigma}{2} 	& \bigg( & \dfrac{P_x}{1 + \delta_0}, &  0, & 0, & 0 & \bigg)^T ~;\\ 
		\ww_{n +  4/11} & = & \ww_{n +  3/11} ~ + &  										& \bigg( & 0, & 0, & -\widetilde{A}_x, & - \mathlarger{\int} \pder{\widetilde{A}_x}{Y} \, \dd X & \bigg)^T ~;\\ 
		\ww_{n +  5/11} & = & \ww_{n +  4/11} ~ + &  									& \bigg( & 0, & 0, &  + \mathlarger{\int} \pder{\widetilde{A}_y}{X} \, \dd Y, & +\widetilde{A}_y & \bigg)^T ~;\\ 
		\ww_{n +  6/11} & = & \ww_{n +  5/11} ~ + &  {\sigma} 					& \bigg( & 0, & \dfrac{P_y}{1 + \delta_0}, & 0, & 0 & \bigg)^T ~;\\ 
		\ww_{n +  7/11} & = & \ww_{n +  6/11} ~ + &  										& \bigg( & 0, & 0, &  - \mathlarger{\int} \pder{\widetilde{A}_y}{X} \, \dd Y, & -\widetilde{A}_y & \bigg)^T ~;\\ 
		\ww_{n +  8/11} & = & \ww_{n +  7/11} ~ + &  									& \bigg( & 0, & 0, & \widetilde{A}_x, &  \mathlarger{\int} \pder{\widetilde{A}_x}{Y} \, \dd X & \bigg)^T ~;\\
		\ww_{n +  9/11} & = & \ww_{n +  8/11} ~ + &  \dfrac{\sigma}{2} 	& \bigg( & \dfrac{P_x}{1 + \delta_0}, & 0, & 0, & 0 & \bigg)^T ~;\\ 
		\ww_{n + 10/11} & = & \ww_{n +  9/11} ~ + &  										& \bigg( & 0, & 0, & -\widetilde{A}_x, & - \mathlarger{\int} \pder{\widetilde{A}_x}{Y} \, \dd X & \bigg)^T ~;\\ 
		\ww_{n + 1} & = & \ww_{n +  10/11} ~ + & \dfrac{\sigma}{2} & \bigg( & 0, & 0, & \pder{\widetilde{A}_z}{X}, & \pder{\widetilde{A}_z}{Y} & \bigg)^T ~.
		\end{matrix}$%
	}
	\end{equation}
 	
	\medskip
	In the case of particle motion inside a magnetic quadrupole,
	the ODE system is given by \eqref{eq:ODE6Dparax}. The magnetic vector potential is written in the form \eqref{eq:genVecPot} and, in many practical applications, only its sampled values at equally spaced locations in
	  $Z$ are available.
	On the other hand, all the  methods introduced require the magnetic vector potential evaluation at $Z$ values   different from the sampled ones.  
For some methods, like the midpoint  and explicit Runge Kutta method, only   the
evaluation at $Z = Z_{n} + \Delta Z / 2 $ is required, so that interpolation of the sampled data
can be avoided if a $ \Delta Z $ is employed for computation that is twice that of the data.
	However,  in general an interpolation is needed   in order to provide the magnetic vector potential $A$ evaluated at the points needed by each specific ODE solver. 
	This introduces a further source of error whose quantification is not a straightforward task. 
	Some proposals to compute intermediate values will be compared in  section \ref{analyticCase},
	extending the preliminary results   in  \cite{pugnat:2015}.

\section{Representation of the magnetic vector potential}
\label{vecpot}
In this section,  the approach used in this work for the reconstruction of
 the magnetic vector potential  will be introduced. 
The reader is referred to \cite{dragt:1997} for a complete presentation of this technique.
Due to the geometry of the quadrupole, it is natural to describe its magnetic field using cylindrical coordinates $\parT{\rho, \, \phii, \, z}$. Due to the periodicity of the field in the angular variable $\phii$, it is then possible to expand the angular dependence using Fourier series
\begin{equation}
\label{eq:bfield}
    \vec{B}\parT{\rho, \, \phii, \, z} =  \sum_{m=1}^{\infty} \vec{B}_m \parT{\rho, \, z} \, \sin{m\, \phii} + \vec{A}_m \parT{\rho, \, z} \, \cos{m\, \phii}~.
\end{equation}

The field harmonics $\vec{A}_m$ and $\vec{B}_m$   are the basis of the vector potential approximation.
 Exploiting the quadrupole symmetries, it is possible to show that only the harmonics associated to certain values of $m$ are different from zero, in particular those with $m = 2 \, (2j + 1)$ for $ j \geq 0.$    
 The magnetic scalar potential $\psi$ satisfies the Laplace equation \eqref{eq:LaplPotEq}. This
 allows to derive from \eqref{eq:bfield} the representation
\begin{equation}
\label{eq:PsiMcossin}
\resizebox{0.75\textwidth}{!}{$
    \begin{aligned}
    \psi(\rho, \, \phii, \, z) 	& = \sum_{m = 1}^{\infty} \sin{m \, \phii} \sum_{\ell = 0}^{\infty} {(-1)^\ell} {\dfrac{m!}{2^{2 \ell} \ell ! \parT{\ell + m}!}\rho^{2 \ell + m}} C_{m,\, s}^{[2\ell]}(z) + \\ 
    & \hspace{0.5mm}+ \sum_{m = 1}^{\infty} \cos{m \, \phii} \sum_{\ell = 0}^{\infty} {(-1)^\ell} {\dfrac{m!}{2^{2 \ell} \ell ! \parT{\ell + m}!}\rho^{2 \ell + m}} C_{m,\, c}^{[2\ell]}(z)    = \\
    & = \sum_{m = 1}^{\infty} \psi_{m, \,s} + \psi_{m, \,c} = \psi_{s} + \psi_{c}~,
    \end{aligned}
    $}
\end{equation}
where $C_{m,\, s}^{[2\ell]}(z)$ (and $C_{m,\, c}^{[2\ell]}(z)$) are known functions called  normal (skew) generalized gradients. The radial component of the harmonics $A_{\rho, \, m}$ and $B_{\rho, \, m}$ (see equation \eqref{eq:bfield}), measured at a certain radius $R_{an}, $ also known as radius of analysis, are denoted from now on simply by  $A_{ m}$ and $B_{ m}.$ They can be used to compute the generalized gradients using the following formula
\begin{equation}
\label{eq:grad}
\begin{aligned}
    C^{[n]}_{m, \, c} (z) & = \dfrac{i^{n}}{2^m m!}  \dfrac{1}{\sqrt{2 \pi}} \int\limits_{-\infty}^{+\infty} 
    \dfrac{k^{m+n-1}}{I_m'(R_{an} \,k)} \,{\ft{A}_m(R_{an},\, k)}\, e^{ikz} \,\dd k~, \\
    C^{[n]}_{m, \, s} (z) & = \dfrac{i^{n}}{2^m m!}  \dfrac{1}{\sqrt{2 \pi}} \int\limits_{-\infty}^{+\infty} 
    \dfrac{k^{m+n-1}}{I_m'(R_{an}\, k)}\, {\ft{B}_m(R_{an},\, k)}\, e^{ikz}\, \dd k~.
\end{aligned}
\end{equation}
Here, $I_m'$ denotes the modified Bessel function of the first kind, while ${\ft{A}_m(R_{an},\, k)}$ and ${\ft{B}_m(R_{an}, \,k)}$ denote the Fourier transforms of ${{A}_m(R_{an}, \, z)}$ and ${{B}_m(R_{an}, \, z)}$, respectively. It is possible to express different quantities, such as the harmonics or the magnetic potentials, using either the normal or the skew generalized gradients. Depending on which generalized gradient is used, these quantities
are labelled as normal ($s$) or skew ($c$). 
Also the magnetic vector potential   be defined using normal and skew terms:
\begin{equation}
    \vec{A} = \sum_{m=1}^{\infty} (\vec{A}^{m, \, s} + \vec{A}^{m, \, c})~.
\end{equation}
Using the generalized gradients, it is possible to derive the expression for a first  vector potential gauge, called azimuthal-free gauge, for which $A_\varphi \equiv 0$:
\begin{equation}
\label{eq:vecPotAF}
\resizebox{0.75\textwidth}{!}{$
    \begin{aligned}
    A_x^{m, \, s}  & = \phantom{-}  \cos{\phii}  \dfrac{\cos{m\, \phii}}{m} \sum_{\ell = 0}^{\infty}\dfrac{\parT{-1}^\ell m!}{2^{2\ell}\ell!\parT{\ell + m}!}C_{m, \, s}^{[2\ell + 1]}(z) \rho^{2\ell + m + 1} ~;\\
    A_x^{m, \, c}  & = - \cos{\phii}  \dfrac{\sin{m\, \phii}}{m} \sum_{\ell = 0}^{\infty}\dfrac{\parT{-1}^\ell m!}{2^{2\ell}\ell!\parT{\ell + m}!}C_{m, \, c}^{[2\ell + 1]}(z) \rho^{2\ell + m + 1} ~;\\
    A_y^{m, \, s}  & =  \phantom{-} \sin{\phii}  \dfrac{\cos{m\, \phii}}{m} \sum_{\ell = 0}^{\infty}\dfrac{\parT{-1}^\ell m!}{2^{2\ell}\ell!\parT{\ell + m}!}C_{m, \, s}^{[2\ell + 1]}(z) \rho^{2\ell + m + 1} ~;\\
    A_y^{m, \, c}  & = - \sin{\phii}  \dfrac{\sin{m\, \phii}}{m} \sum_{\ell = 0}^{\infty}\dfrac{\parT{-1}^\ell m!}{2^{2\ell}\ell!\parT{\ell + m}!}C_{m, \, c}^{[2\ell + 1]}(z) \rho^{2\ell + m + 1} ~;\\
    A_z^{m, \, s}     & = -\dfrac{\cos{m\, \phii}}{m} \sum_{\ell = 0}^{\infty}\parT{2\ell + m}\dfrac{\parT{-1}^\ell m!}{2^{2\ell}\ell!\parT{\ell + m}!}C_{m, \, s}^{[2\ell ]}(z) \rho^{2\ell + m} ~;\\
    A_z^{m, \, c}     & = \phantom{-}\dfrac{\sin{m\, \phii}}{m} \sum_{\ell = 0}^{\infty}\parT{2\ell + m}\dfrac{\parT{-1}^\ell m!}{2^{2\ell}\ell!\parT{\ell + m}!}C_{m, \, c}^{[2\ell ]}(z) \rho^{2\ell + m} ~.
    \end{aligned}
    $}
\end{equation}
 If we require that a vector potential is divergence-free   $\grad \cdot \widehat{\vec{A}} = 0,$ we obtain the so called Coulomb gauge.   The symmetric Coulomb gauge $\widehat{\vec{A}}$ belongs to this category and can be expressed as
\begin{equation}
\label{eq:vecPotCouGauge}
\resizebox{0.75\textwidth}{!}{$
    \begin{aligned}
    \widehat{A}_x^{m, \, s}  & = \phantom{-} \dfrac{\cos{\parT{m + 1}\phii}}{2} \sum_{\ell = 0}^{\infty}\dfrac{\parT{-1}^\ell m!}{2^{2\ell}\ell!\parT{\ell + m + 1}!}C_{m, \, s}^{[2\ell + 1]}(z) \rho^{2\ell + m + 1} ~;\\
    \widehat{A}_x^{m, \, c}  & =          -  \dfrac{\sin{\parT{m + 1}\phii}}{2} \sum_{\ell = 0}^{\infty}\dfrac{\parT{-1}^\ell m!}{2^{2\ell}\ell!\parT{\ell + m + 1}!}C_{m, \, c}^{[2\ell + 1]}(z) \rho^{2\ell + m + 1} ~;\\
    \widehat{A}_y^{m, \, s}  & = \phantom{-} \dfrac{\sin{\parT{m + 1}\phii}}{2} \sum_{\ell = 0}^{\infty}\dfrac{\parT{-1}^\ell m!}{2^{2\ell}\ell!\parT{\ell + m + 1}!}C_{m, \, s}^{[2\ell + 1]}(z) \rho^{2\ell + m + 1} ~;\\
    \widehat{A}_y^{m, \, c}  & = \phantom{-} \dfrac{\cos{\parT{m + 1}\phii}}{2} \sum_{\ell = 0}^{\infty}\dfrac{\parT{-1}^\ell m!}{2^{2\ell}\ell!\parT{\ell + m + 1}!}C_{m, \, c}^{[2\ell + 1]}(z) \rho^{2\ell + m + 1} ~;\\
    \widehat{A}_z^{m, \, s}  & =          -  \cos{m\, \phii} \sum_{\ell = 0}^{\infty}\dfrac{\parT{-1}^\ell m!}{2^{2\ell}\ell!\parT{\ell + m}!}C_{m, \, s}^{[2\ell ]}(z) \rho^{2\ell + m} ~;\\
    \widehat{A}_z^{m, \, c}  & = \phantom{-} \sin{m\, \phii} \sum_{\ell = 0}^{\infty}\dfrac{\parT{-1}^\ell m!}{2^{2\ell}\ell!\parT{\ell + m}!}C_{m, \, c}^{[2\ell ]}(z) \rho^{2\ell + m} ~.
    \end{aligned}
    $}
\end{equation}
The $x$ and the $y$ components can be written as follows
\begin{equation}
\begin{aligned}
\widehat{A}_x^{\mm, \, s}  & = \phantom{-} \cos{\mm \, \phii} \sum_{\ell = 0}^{\infty}\dfrac{\parT{-1}^\ell \mm!}{2^{2\ell}\ell!\parT{\ell + \mm}!}B_{\mm, \, s}^{[2\ell]}(z) \rho^{2\ell + \mm} ~;\\
\widehat{A}_x^{\mm, \, c}  & = \phantom{-}  \sin{\mm \, \phii} \sum_{\ell = 0}^{\infty}\dfrac{\parT{-1}^\ell \mm!}{2^{2\ell}\ell!\parT{\ell + \mm}!}B_{\mm, \, c}^{[2\ell]}(z) \rho^{2\ell + \mm} ~;\\
\widehat{A}_y^{\mm, \, s}  & = \phantom{-} \sin{\mm \, \phii} \sum_{\ell = 0}^{\infty}\dfrac{\parT{-1}^\ell \mm!}{2^{2\ell}\ell!\parT{\ell + \mm}!}B_{\mm, \, s}^{[2\ell]}(z) \rho^{2\ell + \mm} ~;\\
\widehat{A}_y^{\mm, \, c}  & =          - \cos{\mm \, \phii} \sum_{\ell = 0}^{\infty}\dfrac{\parT{-1}^\ell \mm!}{2^{2\ell}\ell!\parT{\ell + \mm}!}B_{\mm, \, c}^{[2\ell]}(z) \rho^{2\ell + \mm} ~,
\end{aligned}
\label{eq:vecPotSCCarCompBis}
\end{equation}
where $\mm = m + 1$ and
\begin{align*}
B_{\mm, \, s}^{[2\ell]}(z) & = \phantom{-} \dfrac{1}{2 \mm} \, C_{\mm - 1, \, s}^{[2\ell + 1]}(z) ~;\\
B_{\mm, \, c}^{[2\ell]}(z) & =          -  \dfrac{1}{2 \mm} \, C_{\mm - 1, \, c}^{[2\ell + 1]}(z) ~.
\end{align*}
Finally, via a gauge transformation $\bar{\vec{A}} = \widehat{\vec{A}} + \grad \lambda,$
a new form of the vector potential can be derived, such that   $\bar{A}_x \equiv 0.$
The derivation of this so called horizontal-free Coulomb gauge 
is described in detail \cite{dragt:1997} and summarized in the following.
As it will be shown later in this section, the property $\bar{A}_x \equiv 0 $
implies that using this representation for the vector potential leads to
a significant reduction in the computational cost of each right hand side evaluation
in the numerical solution of system \eqref{eq:ODE6Dparax}.

Notice  that a gauge transformation is equivalent to a canonical transformation, see e.g. \cite{kotkin:2013}. 
In particular, if we consider a Hamiltonian $\KK$, similar to \eqref{eq:parApp8D}, with vector potential $\vec{A}$ and dynamical variables $\qq$ and $\pp$ then, using a new vector potential $\vec{A}' = \vec{A} + \grad \lambda$ we obtain a Hamiltonian $\KK'$ in the same form of $\KK$ with dynamical variables:
\begin{equation}
\label{eq:vpCanTrans}
\begin{aligned}
    \vec{Q} & = \qq~;	\\
    \vec{P} & = \pp + Q \grad \lambda~,
\end{aligned}
\end{equation}
where $Q$ denotes again the particle charge.
To derive the horizontal free Coulomb gauge transformation, we build a harmonic function $\lambda$ as
\begin{equation}
\label{eq:lmb}
\begin{aligned}
\lambda & =       \sum_{m=0}^{\infty} \left[\sin{m\, \phii} \lambda_{m, \, c} +
    \cos{m\, \phii}\lambda_{m, \, s}\right],
\end{aligned}
\end{equation}
where
$$
\lambda_{m, \, c}=\sum_{\ell = 0}^{\infty}\dfrac{\parT{-1}^\ell m!}{2^{2\ell}\ell!\parT{\ell + m}!}L_{m, \, c}^{[2\ell]}(z) \rho^{2\ell + m},
$$
$$
\lambda_{m, \, s}=\sum_{\ell = 0}^{\infty}\dfrac{\parT{-1}^\ell m!}{2^{2\ell}\ell!\parT{\ell + m}!}L_{m, \, s}^{[2\ell]}(z) \rho^{2\ell + m},
$$
and  the coefficients $L_{m,\, s/c}^{[2\ell]}(z)$ are related to the coefficients $C_{m,\, s/c}^{[2\ell]}(z)$ by the following relations
\begin{equation}
\begin{aligned}
L_{m + 1, \, s}^{[2\ell]} (z) & = \dfrac{1}{m + 1} \parQ{\dfrac{1}{4m} L_{m - 1, \, s}^{[2\ell + 2]}(z) - B_{m, \, s}^{[2\ell]}(z)}\\
L_{m + 1, \, c}^{[2\ell]} (z) & = \dfrac{1}{m + 1} \parQ{\dfrac{1}{4m} L_{m - 1, \, c}^{[2\ell + 2]}(z) - B_{m, \, c}^{[2\ell]}(z)}
\end{aligned}
\end{equation} 
and $L_{m, \, s/c}^{[0]}(z) \equiv 0$ for $m \leq 2$.
It is possible to show that $\partial_x \lambda_{s/c} = -\widehat{A}_x^{s/c}$ leading to the desired horizontal-free Coulomb gauge:
\begin{equation}
\label{eq:vecPotHFC}
\begin{aligned}
\bar{A}_x^{s} & = 0~ \ \ \  \bar{A}_x^{c}  = 0~;\\
\bar{A}_y^{s} & = \widehat{A}_y^{ s} + \partial_y \lambda_s~;\\
\bar{A}_y^{c} & = \widehat{A}_y^{ c} + \partial_y \lambda_c~;\\
\bar{A}_z^{s} & = \widehat{A}_z^{ s} + \partial_z \lambda_s~;\\
\bar{A}_z^{c} & = \widehat{A}_z^{ c} + \partial_z \lambda_c~,
\end{aligned}
\end{equation}
where 
\begin{equation}
    \resizebox{0.75\textwidth}{!}{$
    \begin{aligned}
    \partial_y \lambda_{m, \, c} & = - \sum_{\ell = 0}^{\infty} \dfrac{\parT{-1}^\ell}{2^{2\ell} \ell! \parT{\ell + m}!} \parG{\parT{m + 1}L_{m + 1, \, c}^{[2\ell]} + \dfrac{1}{4m} L_{m - 1, \, c}^{[2\ell + 2]} } \\ 
    & \phantom{=} \times \rho^{2\ell + m} \cos{m \, \phii} ~;\\
    \partial_y \lambda_{m, \, s} & = - \sum_{\ell = 0}^{\infty} \dfrac{\parT{-1}^\ell}{2^{2\ell} \ell! \parT{\ell + m}!} \parG{\parT{m + 1}L_{m + 1, \, s}^{[2\ell]} + \dfrac{1}{4m} L_{m - 1, \, s}^{[2\ell + 2]} } \\ 
    &  \phantom{=} \times  \rho^{2\ell + m} \sin{m \, \phii}~,
    \end{aligned}
    $}
\end{equation}
and $\partial_z \lambda$ is obtained from \eqref{eq:lmb}:
\begin{equation}
\begin{aligned}
\partial_z \lambda & = \sum_{m=0}^{\infty} \sin{m\, \phii} \sum_{\ell = 0}^{\infty}\dfrac{\parT{-1}^\ell m!}{2^{2\ell}\ell!\parT{\ell + m}!}L_{m, \, c}^{[2\ell + 1]}(z) \rho^{2\ell + m} + \\
& \hspace{0.45mm} + \sum_{m=0}^{\infty} \cos{m\, \phii} \sum_{\ell = 0}^{\infty}\dfrac{\parT{-1}^\ell m!}{2^{2\ell}\ell!\parT{\ell + m}!}L_{m, \, s}^{[2\ell + 1]}(z) \rho^{2\ell + m}~.
\end{aligned}
\end{equation}
 All the previous vector potential descriptions, when expressed in Cartesian coordinates, take the  form
\begin{equation}
\label{eq:genVecPot}
    \vec{A} \parT{x, \, y, \, z} =  \sum_{i, \, j} \vec{a}_{i, \, j}(z) \, x^i \, y^j~,
\end{equation}
where the coefficients $\vec{a}_{i, \, j}(z)$ depend on the longitudinal coordinate $z$. 
In practical cases, the  series expansions in the previous formulae  are truncated to a finite number of terms,
which depends on the number of harmonics used to describe the field, while the maximum number of generalized gradients derivatives $ND$  determines the range of the indices $i$ and $j$ in \eqref{eq:genVecPot}.
Therefore, for a given $z=\widehat{z}$ value, the evaluation time of  \eqref{eq:genVecPot}
is proportional to the number of coefficients $\vec{a}_{i, \, j}(\widehat{z}) $ retained,
which can therefore be used to estimate the computational cost entailed by each representation.

Using the horizontal-free gauge, all the terms in the previous series expansions corresponding to the $x$ component of the vector potential are null, while the number of terms corresponding to the 
 other two components is similar in the Coulomb and horizontal free gauges. 
The number of coefficients required by the magnetic vector potential at a specific $z$ location, in different gauges, using the harmonics $m\in\{2, \, 6, \, 10, \, 14\}$ and different generalized gradient derivatives are shown
in table \ref{tab:vpCoeffHarmo2-6-10-14}.

    \begin{table}[!htb]
        \centering
        \begin{tabular}{|c|c|c|c|c|c|}
         \cline{3-6}
            \multicolumn{2}{c|}{~}          & \multicolumn{2}{c|}{ND = 2}& \multicolumn{2}{c|}{ND = 16} \\
            \cline{3-6}
            \multicolumn{2}{c|}{~}          & Normal  & Skew   & Normal & Skew \\
            \hline                          
            \multirow{2}{*}{$A_x$} & AF     & $20$    & $16$   & $112$  & $105$\\   
            & HFC    & $0$     & $0$    & $0$    & $0$\\
            \hline                             
            \multirow{2}{*}{$A_y$} & AF     & $20$    & $16$   & $112$  & $105$\\   
            & HFC    & $20$    & $20$   & $119$  & $112$\\
            \hline                             
            \multirow{2}{*}{$A_z$} & AF     & $40$    & $36$   & $128$  & $120$\\   
            & HFC    & $44$    & $32$   & $135$  & $113$\\  
            \hline                             
            \multirow{3}{*}{$TOT$} & AF     & $80$    & $68$   & $352$  & $330$\\   
            & HFC    & $64$    & $52$   & $254$  & $225$ \\
            \cline{2-6}                               
            & HFC/AF & $0.80$  & $0.76$ & $0.72$ & $0.68$    \\  
             \cline{1-6}               
        \end{tabular}
        \vskip 0.3cm
        \caption{Number of vector potential coefficients using the harmonics $m\in\{2, \, 6, \, 10, \, 14\}$, for different gauges and number of generalized gradients derivatives.}
        \label{tab:vpCoeffHarmo2-6-10-14}
    \end{table}
    It can be noticed that the horizontal-free Coulomb gauge requires in general
    between 20\% and 25\% less coefficients
   with respect to the azimuthal-free gauge.

    The effective reduction in computational cost achievable by this reformulation
    is however dependent on the specific features of each numerical method.  Indeed,
    Runge-Kutta methods require the evaluation of the right hand side of Hamilton equations $\mat{J} \, \grad \KK$, while Lie methods evaluate the map $\mathcal{M}_2$. Therefore, the speed-up of Runge-Kutta and Lie methods, achieved with the different gauges, is related to the number of vector potential coefficients required by $\mat{J} \, \grad \KK$ and $\mathcal{M}_2,$ respectively. In table \ref{tab:vpCoeffRatio}, the number of evaluations of each vector potential component required by $\mat{J} \, \grad \KK$ and by $\mathcal{M}_2$ are shown. Moreover, in the last columns, the ratios between the number of vector potential coefficients using the horizontal-free Coulomb gauge and the azimuthal-free gauge are shown, when considering harmonics $m \in\{ 2, \, 6, \, 10, \, 14\}$ and $ND = 16$. It can be noticed that the map $\mathcal{M}_2$, and therefore Lie methods, benefit more from the change of gauge, because they require more evaluations of the $A_x$ component.
    \begin{table}[!htb]
        \newcolumntype{L}[1]{>{\raggedright\let\newline\\\arraybackslash\hspace{0pt}}m{#1}}
        \newcolumntype{C}[1]{>{\centering\let\newline\\\arraybackslash\hspace{0pt}}m{#1}}
        \newcolumntype{R}[1]{>{\raggedleft\let\newline\\\arraybackslash\hspace{0pt}}m{#1}}
        \centering
        \begin{tabular}{|c|C{1cm}C{1cm}C{1cm}|C{1.2cm}C{1.2cm}|}
         \cline{2-6}
            \multicolumn{1}{c|}{~} & \multicolumn{3}{c|}{Number of evaluations} & \multicolumn{2}{c|}{Ratio HFC/AF}\\
            \cline{2-6}
            \multicolumn{1}{c|}{~} & $A_x$ & $A_y$ & $A_z$ & Norm & Skew \\
            \hline
            $\mat{J} \, \grad \KK$ & $3$ & $3$ & $2$ & $0.666$ & $0.646$ \\
            $\mathcal{M}_2$        & $8$ & $4$ & $4$ & $0.539$ & $0.517$   \\
            \hline                
        \end{tabular}
          \vskip 0.3cm
        \caption{Number of function evaluation for each vector potential component (left) and ratio of function evaluations between horizontal free Coulomb gauge (HFC) and azimuthal free gauge (AF), for standard symplectic methods (first row) and Lie algebra based methods (second row). Numbers refer to the case of
         harmonics $m\in\{2, \, 6, \, 10, \, 14 \}$ and $ND=16$ generalized gradient derivatives.}
        \label{tab:vpCoeffRatio}
    \end{table}
The overall efficiency of each method also depends on the total 
 number of  evaluations of $\mat{J}\grad\KK$ or $\mathcal{M}_2$ required at each step. These
 are reported in table \ref{tab:methEval}.
Moreover, it is important to notice that implicit methods require a certain number of fixed point iterations, between $5$ and $8$, that increase significantly their computational cost per time step. 
For a comprehensive efficiency comparison, the total computational cost of each method
must be compared to the accuracy level it allows to achieve and to the accuracy level that is actually
required for accelerator design. This comparison will be attempted in sections \ref{analyticCase} and \ref{realCase}. 

 \begin{table}
    \centering
    \newcommand{\aA}{$\vec{A}_\Box$}
    \newcommand{\aadx}{$\partial_X\vec{A}_\Box$}
    \newcommand{\aady}{$\partial_Y\vec{A}_\Box$}
    \newcommand{\aadxx}{$\partial_{XX}\vec{A}_\Box$}
    \newcommand{\aadxy}{$\partial_{XY}\vec{A}_\Box$}
    \newcommand{\aadyy}{$\partial_{YY}\vec{A}_\Box$}
    \newcommand{\ff}{$\mat{J} \grad \KK$}
    \newcommand{\dff}{$\mathcal{J}_{\mat{J} \grad \KK}$}
    \newcommand{\nfp}{N_{fp}}
    \newcommand{\mII}{$\mathcal{M}_2$}
    \begin{tabular}{|l|cc|}
    \hline
        &	\ff					&	\mII	\\
        \hline
        midPoint	&	$1 \cdot \nfp$	&			\\
        rk4			&	$4$				&			\\
        gauss4		&	$2 \cdot \nfp$	&			\\
        gauss6		&	$3 \cdot \nfp$	&			\\
        lie2		&					&	$1$		\\
        lie4		&					&   $3$		\\
        lie6		&					&	$9$		\\	
        \hline
    \end{tabular}
    \vskip 0.3cm
    \caption{Number of evaluations of $\mat{J}\nabla\KK$ or $\mathcal{M}_2$ required by each numerical method.
 $\nfp $ denotes the number of fixed point iterations required by implicit methods. }
    \label{tab:methEval}
\end{table}

In the following sections, only normal quadrupoles  will be considered 
(i.e. with null skew generalized gradients), so that the notation will be simplified
 using $C_m^{[n]}$ instead of $C_{m, \, s}^{[n]}$.

\begin{figure}[!htb]
     \centering
     \begin{tabular}{cc}
         \includegraphics[width=0.45\textwidth]{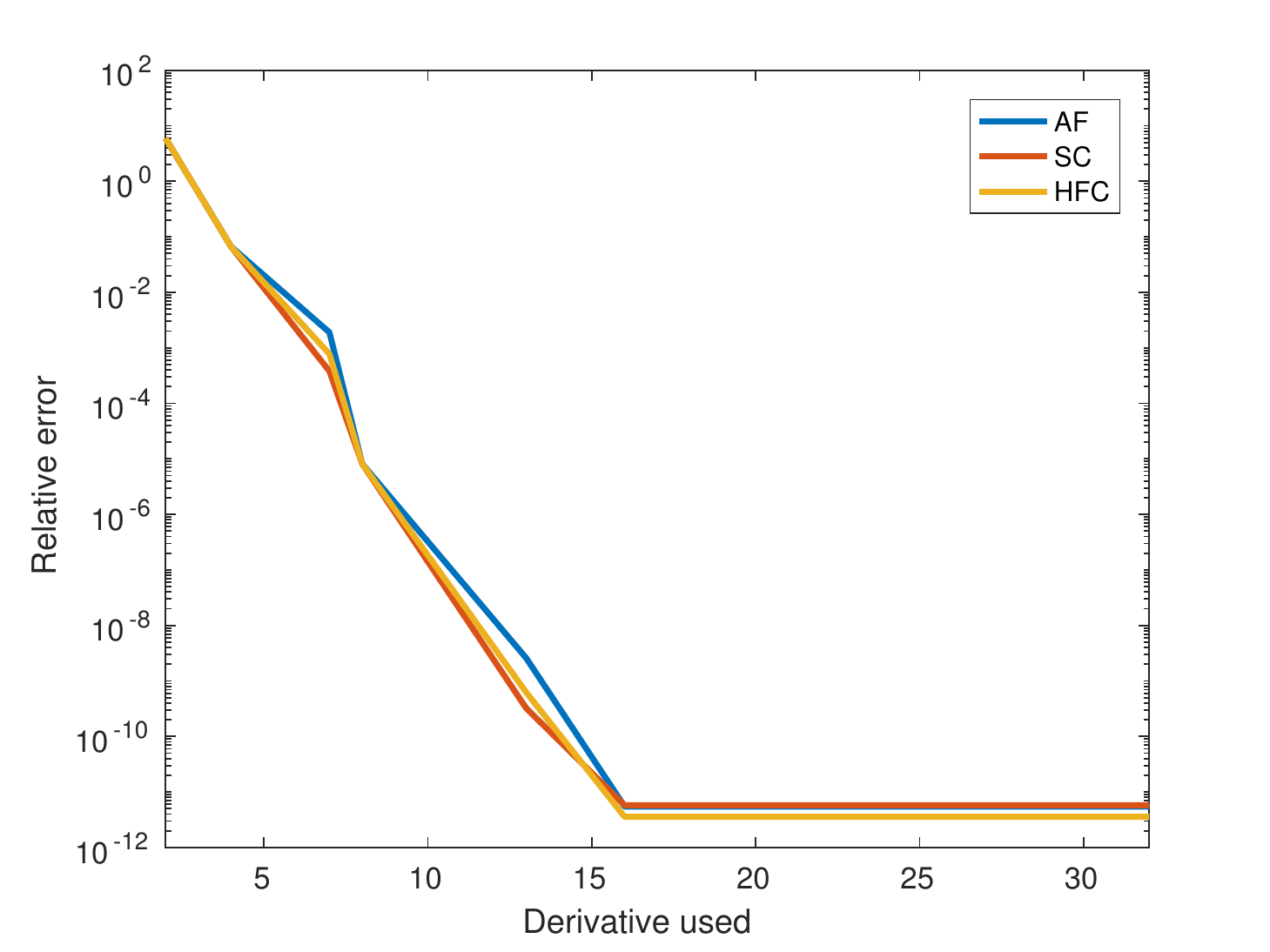}
         &
         \includegraphics[width=0.45\textwidth]{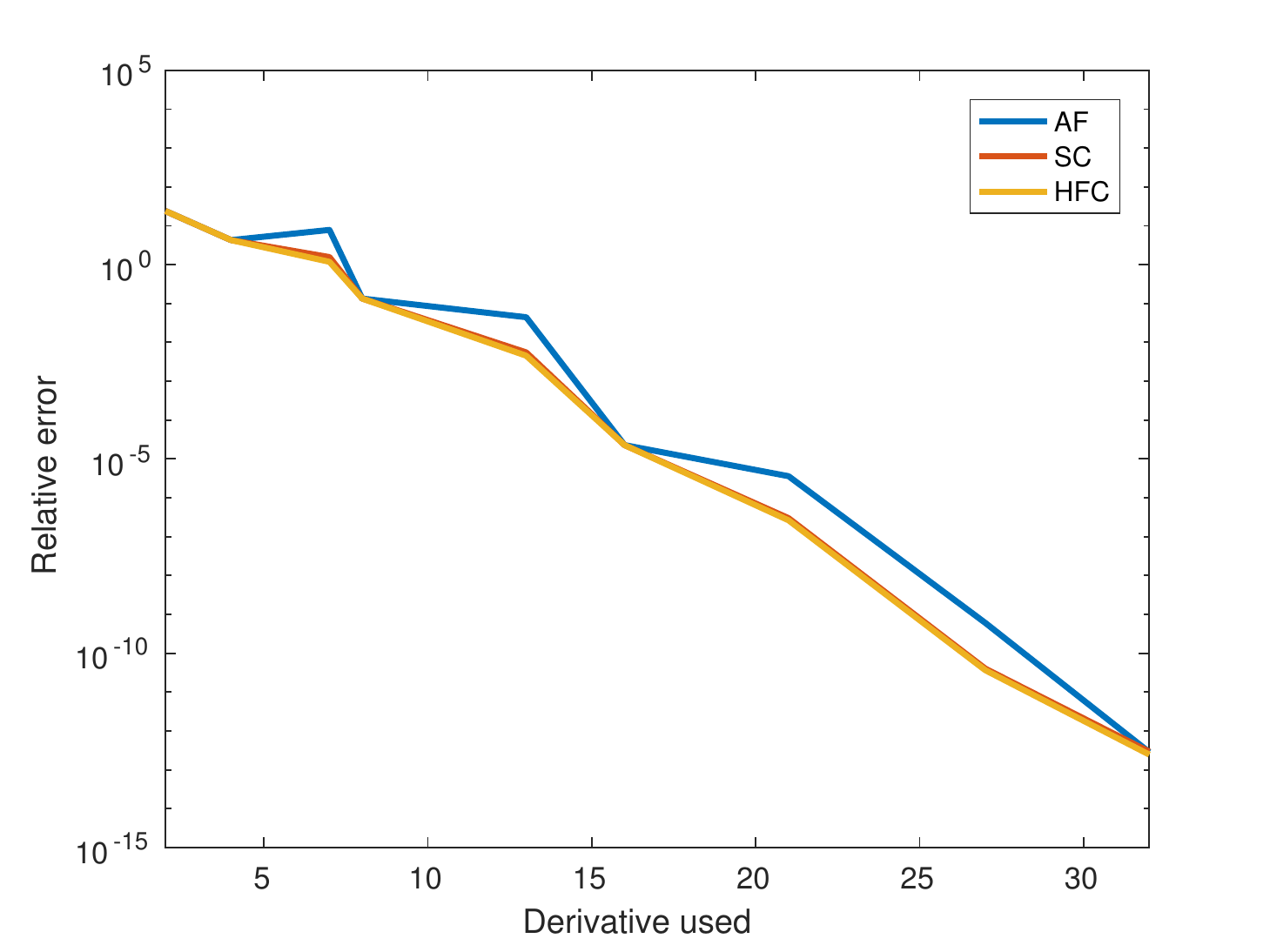}
        \end{tabular}
        \caption{Relative error in the Maxwell equation with respect to the number of generalized gradient derivatives used. Positions $(X_0, Y_0) = (0, 0.01)$ (left) and $(X_0, Y_0) = (0, 0.04)$ (right). Second order harmonic $B_2$ of realistic case (sec. \ref{realCase}).}
        \label{fig:max_err_ND}
    \end{figure}

A final   remark concerns one important consequence of the previously introduced vector potential representation. The vector potential expressions \eqref{eq:vecPotAF}, \eqref{eq:vecPotCouGauge} and \eqref{eq:vecPotHFC} are truncated at a finite number of terms  $ND.$
 As a consequence, many equations that the vector potential should satisfy  only hold up to terms associated to the first generalized gradient derivatives that have been neglected,
 i.e.  $C_{m, s/c}^{[ND + 1]} (z)$~. If $ND$ is an even number, all the vector potential gauges produce exactly the same magnetic field, which satisfies the  Maxwell equation 
 \[
 \grad \times \grad \times \vec{A} = \vec{j}^{SP}~,
 \]
 where $\vec{j}^{SP}$ is a spurious current that depends on the generalized gradients only through the generalized gradient derivatives of order $ND + 1$. This implies that, generally, the largest contributions
 to these currents will be concentrated in the fringe field. For example, using only two derivatives of $C_2^{[0]}(Z)$, the spurious currents are:
 \begin{equation}
 \vec{j}^{SP} = 
 \begin{bmatrix}
 \dfrac{1}{6} \parT{X^3 - 3 X\,Y^2} C_2^{[3]}(Z) \\
 - \dfrac{1}{6} \parT{Y^3 + 3 Y\,X^2} C_2^{[3]}(Z) \\
 0
 \end{bmatrix}~.
 \end{equation}
 The magnitude of this error can be measured using the following quantity:
 \begin{equation}
 \label{eq:rotrot_err}
 \dfrac{\max\limits_{Z}\parT{\norm{2}{\nabla \times \nabla \times \vec{A}\parT{X_0, \, Y_0, \, Z}}}}
 {\max\limits_{Z}\parT{\norm{2}{\vec{A}\parT{X_0, \, Y_0, \, Z}}}}~,
 \end{equation}
 where $(X_0, Y_0)$ refer to a fixed position in the transversal plane.   In figure \ref{fig:max_err_ND}, it is reported the behaviour of the error \eqref{eq:rotrot_err} for two different transversal positions, one closer to the quadrupole axis and another further away. The results refer to vector potentials in different gauges, computed considering a second order harmonic $B_2$ coming from a realistic design of a quadrupole (see section \ref{realCase}). 
 It can be noticed that the error in the different gauges coincides if $ND$ is an even number, coherently with the fact that the magnetic fields are identical.
 It is unclear to which extent this approximation error, which is far from negligible
 at larger distances from the quadrupole axis, entails a limitation for the  accuracy that tracking algorithms
 can achieve.  Mimetic methods like those presented in
 \cite{lipnikov:2011}  guarantee that a discrete
 equivalent of  $\grad \times \grad \times \vec{A} =0 $ is exactly satisfied, so that they could represent
 a potentially interesting alternative to the techniques described in this section.

As a final remark we notice that, in the use of equation \ref{eq:grad}, particular attention  has to be devoted  to the computation of the input harmonics. If harmonics that do not go to zero sufficiently fast
 at the boundaries are employed, significant  errors in the field description may result.
 For example, even if the harmonics vanish at the boundaries, the computed generalized gradients, and thus the vector potential, may not vanish as well, causing a discrepancy between the canonical and the mechanical momenta in a region where they should coincide. In our experience, these issues can be avoided by appropriate
 extension of the harmonics data by sufficiently large regions filled with zero values.

\section{\hspace{-2mm}Numerical experiments with an analytic magnetic vector potential}
\label{analyticCase} \indent
In this section, equations \eqref{eq:ODE6Dparax} will be solved for the case of
a simple vector potential, whose expression is given by a polynomial with just  a few non zero coefficients. It is important to notice that this configuration is not completely realistic, since this vector potential does not exactly satisfy the  Maxwell equation $\Nabla \times \Nabla \times \vec{A}=0.$ As discussed in the previous section, 
this leads to spurious residual currents in the right hand side of this equation, which are however not accounted 
for in equations \eqref{eq:ODE6Dparax}. In spite of this inconsistency,
 this benchmark is useful to assess the impact of the interpolation procedure mentioned in section \ref{ode_rev}, since the simple analytic expression can also be computed cheaply online and comparison between results obtained with and without interpolation can be carried out.

A set of scaled vector potentials in different gauges is built using equations \eqref{eq:vecPotAF}, \eqref{eq:vecPotCouGauge} and \eqref{eq:vecPotHFC} and employing only the
$C_2^{[0]}(Z)$ function defined as follows. We first set
\begin{equation}
\label{eq:stepFun}
\sigma \parT{x} = \left\{
\begin{aligned}
& 0 && \qquad x \in \left(-\infty, -1 \right] \\
& \dfrac{1}{2} \parQ{1 + \erf{\tan{\dfrac{\pi}{2}\, x}}} && \qquad x \in \parT{-1, 1} \\
& 1 && \qquad x \in \left[ 1, +\infty \right)
\end{aligned}
\right.
\end{equation}
where $\erf$ denotes the  error function, defined as 
\[
\erf{x} = \dfrac{2}{\sqrt{\pi}} \int_0^x e^{-t^2} \dd t ~.
\]
 Using \eqref{eq:stepFun}, the function $C_2^{[0]}(Z)$ 
is then defined as
\begin{equation}
\label{eq:genGradSimply}
\hspace{-0.3cm}\resizebox{0.93\linewidth}{!}{%
    $C_2^{[0]}(Z) = \left\{
    \begin{aligned}
    & \alpha \, \parQ{\sigma\parT{-1 + 2\, \dfrac{Z}{L_1}} + \sigma\parT{1 - 2\, \dfrac{Z - Z_2}{L_2}}} - \alpha&& \quad Z \in \parT{0, \, Z_{MAX}} \\
    & 0 && \quad \text{otherwise.}
    \end{aligned}
    \right.$}
\end{equation}
 
\begin{figure}[!htb]
    \centering
    \includegraphics[width=0.7\textwidth]{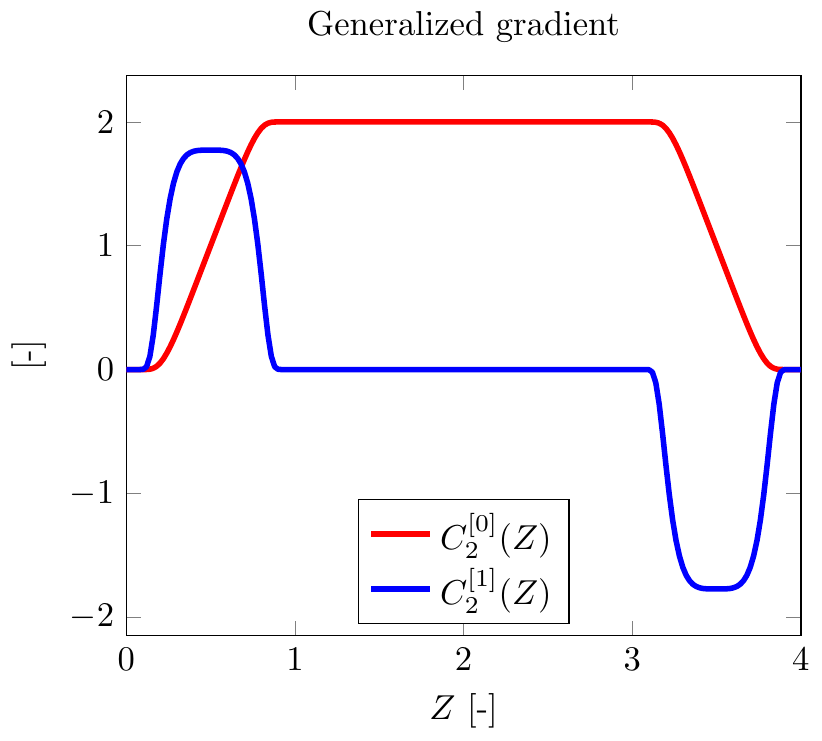}
    \caption{Analytic generalized gradient and its derivative. }
    \label{fig:sipleGenGrad}
\end{figure}

Here, the coefficient $\alpha$ is the plateau value of $C_2^{[0]}(Z)$,
$Z_{MAX}$  denotes the total field length, 
$L_1, L_2$  are the widths of the regions in which the field is not constant and
 $Z_2$ denotes the location of the second of these regions, the first being located at $Z=0 $ (see figure \ref{fig:sipleGenGrad}).
In this work, the value $\alpha$  in \eqref{eq:genGradSimply} has been chosen equal to $6 \cdot 10^{-4}$ with  $L_1 = L_2 = 0.9$ and $Z_2 = 3.1.$  The initial conditions for the transversal positions and momenta of the particle are set to $\ww_0 = \parT{0.02, \, -0.04, \, 0, \, 0}$ and $\delta_0 = 0$.
A reference solution is computed using the exact vector potential and the \matlab ODE solver \texttt{ode45} with a maximum $\Delta Z^{ref} = \Delta Z^{data} / 10$ and a relative error tolerance of $10^{-13}$.

\begin{figure}[!htb]
    \centering
    \newcommand{\relPath}{images/4D/gaugeAF/AnalyticSamp_dz2mm_newVecPot}
    \begin{subfigure}[b]{.45\linewidth}
        \centering
        \includegraphics[width=0.994\textwidth]{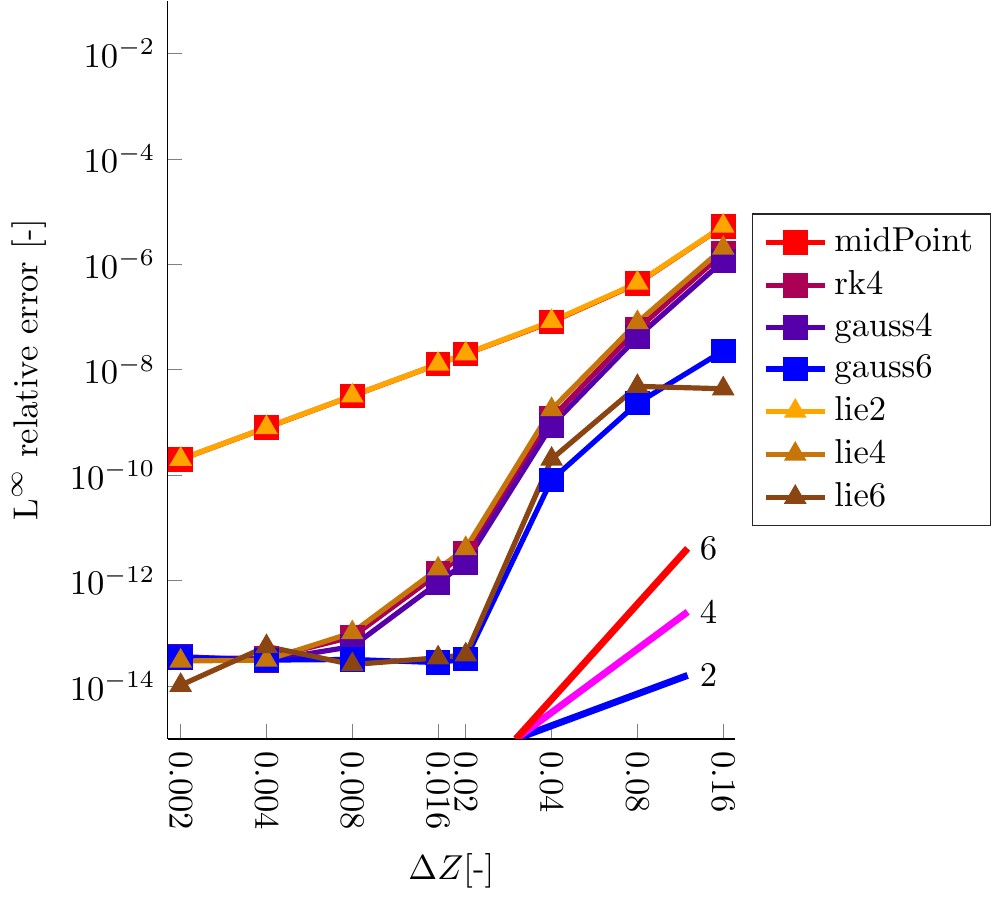}
    \end{subfigure}%
    \begin{subfigure}[b]{.45\linewidth}
        \includegraphics[width=0.994\textwidth]{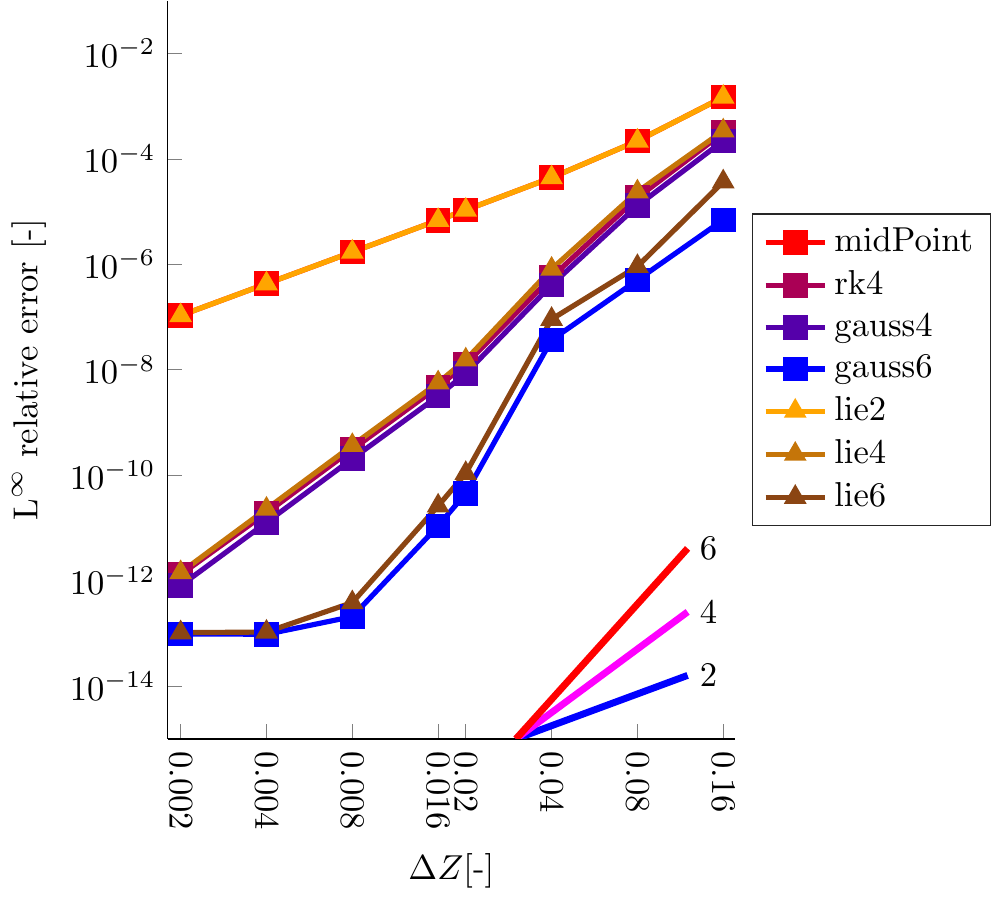}
    \end{subfigure}
    \caption{Test case with analytically defined vector potential. Convergence behaviour 
    in the $l_{\infty} $  norm  of different ODE methods for $X$ (left) and $P_y$ (right) when exact vector potential values are employed at all intermediate steps. The straight lines in the bottom right corner denote   theoretical slopes for different convergence orders.}
    \label{fig:anErrComp2mm}
\end{figure}

In figure \ref{fig:anErrComp2mm},  the errors obtained with the   different ODE methods
are presented, in the case in which no interpolation is used and the
  magnetic vector potential is computed exactly at each required location. 
The  errors on the  $X, Y$ coordinates behave very similarly and the same is true for  
the corresponding moments, but the errors on 
  transversal momenta are larger than those
on positions. As long as the errors are above the tolerance chosen for the reference solution,
both second and fourth order solvers behave in agreement with  theoretical expectations,
while sixth order solvers only seem to display the expected error decay for sufficiently small
values of the interval $\Delta Z.$ Furthermore, very similar errors are obtained 
for second and fourth order methods for the largest 
$\Delta Z $ employed. We attribute these facts to the poor resolution of the  
larger gradient areas at the beginning and at the end of the idealized quadrupole. This effect can also be seen in the numerical results of section \ref{realCase}. It gives a clear indication that there is an upper limit to the 
value $\Delta Z $ that can be employed, independently of the  accuracy of the solver employed.

\begin{figure}[!htb]
    \centering
    \newcommand{\relPath}{images/4D/gaugeAF/AnalyticSamp_dz2mm_newVecPot}
    \begin{subfigure}[b]{.5\linewidth}
        \centering
        \includegraphics[width=0.994\textwidth]{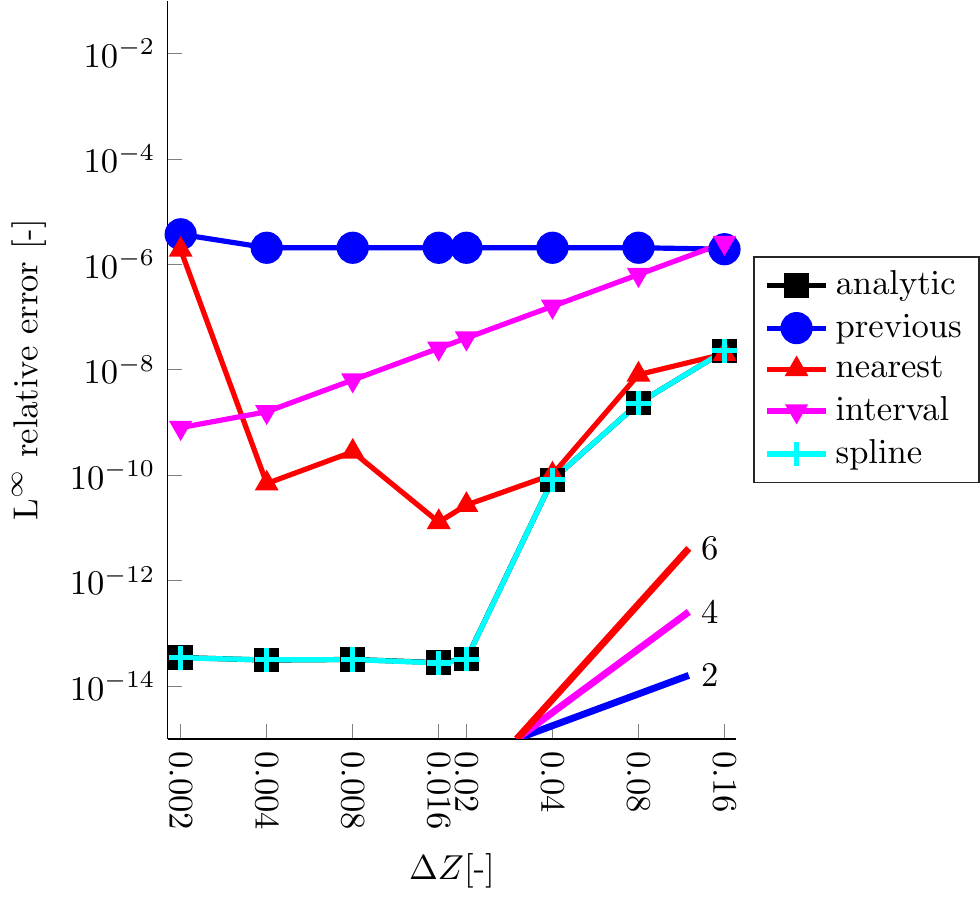}
    \end{subfigure}%
    \begin{subfigure}[b]{.5\linewidth}
        \includegraphics[width=0.994\textwidth]{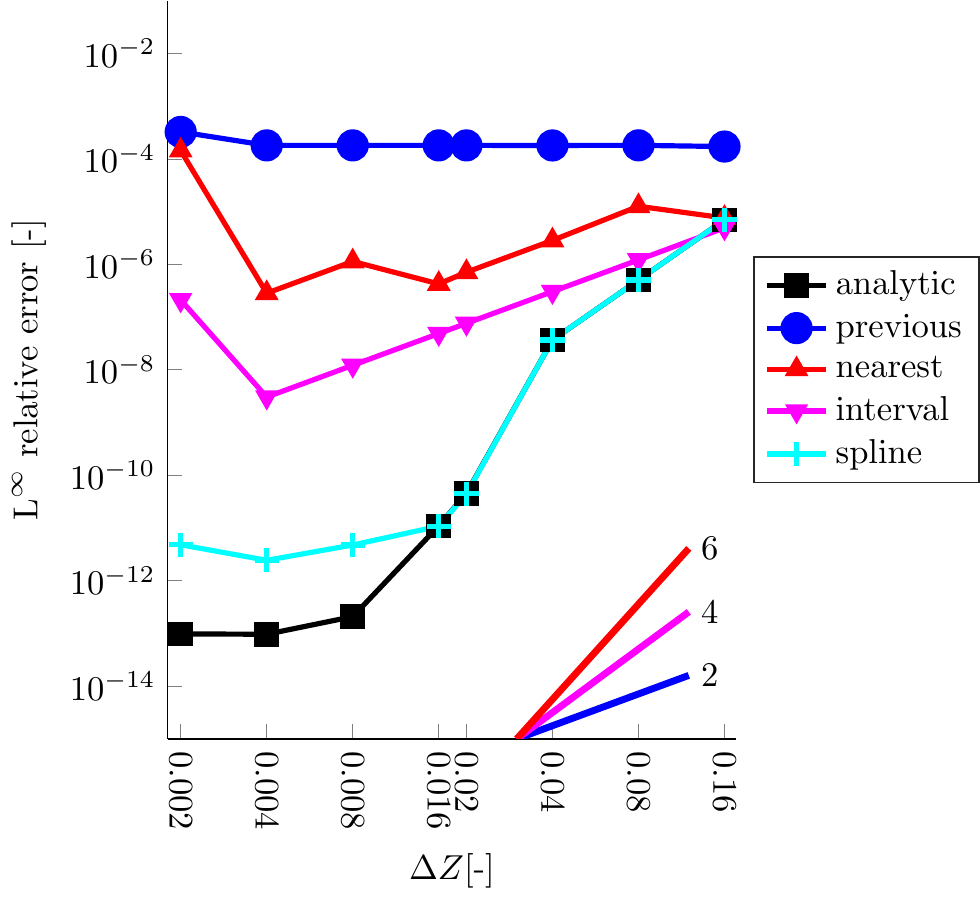}
    \end{subfigure}
    \caption{Test case with analytically defined vector potential. Convergence behaviour 
    in the $l_{\infty} $  norm of different ODE methods for $X$ (left) and $P_y$ (right) when reconstructed vector potential values are employed at all intermediate steps. 
    The straight lines in the bottom right corner denote   theoretical slopes for different convergence orders.}
    \label{fig:anSampGauss6ErrComp2mm}
\end{figure}

In order to study the impact of different interpolation techniques in the more realistic case in which the
vector potential is only available as sampled data, the analytic vector potential
obtained from  \eqref{eq:genGradSimply} are sampled on a fine mesh with $\Delta Z^{data} = 0.002$ over the interval $[0, 4]$. 
For each given position along
the $Z$ axis where the vector potential is not available, we employ    a) 
the value at the last previously available gridded location ({\tt previous}), b) the value at  the nearest gridded location available  ({\tt nearest}), c) the average of the nearest available potential values   ({\tt interval})
d) a cubic spline interpolation  ({\tt spline}). 
The results of this comparison are reported
in figure \ref{fig:anSampGauss6ErrComp2mm}. Only  results obtained with the sixth-order Gauss method
are displayed,   since the highest order methods are those most affected by the accuracy of the field interpolation.
It can be noticed that all interpolation
 methods limit the overall accuracy of the time integrator, to a larger or lesser extent.
Ideally, an interpolator of the same order of the time integration method
should be employed. On the other hand, spline interpolation seems to be sufficient to achieve
errors comparable to those of the exact potential evaluation in most cases.

\begin{figure}[!htb]
    \centering
    \newcommand{\relPath}{images/4D/gaugeAF/AnalyticSamp_dz2mm_newVecPot}
    \begin{subfigure}[b]{.5\linewidth}
        \centering
        \includegraphics[width=0.994\textwidth]{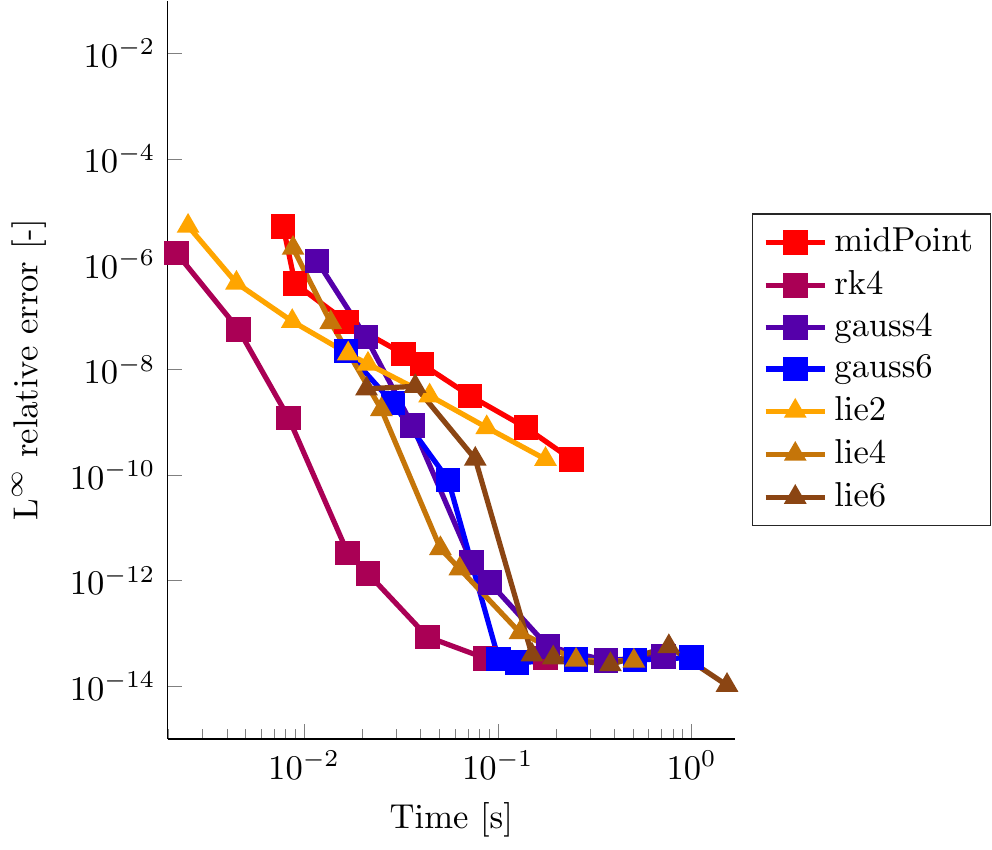}
    \end{subfigure}%
    \begin{subfigure}[b]{.5\linewidth}
        \centering
        \includegraphics[width=0.994\textwidth]{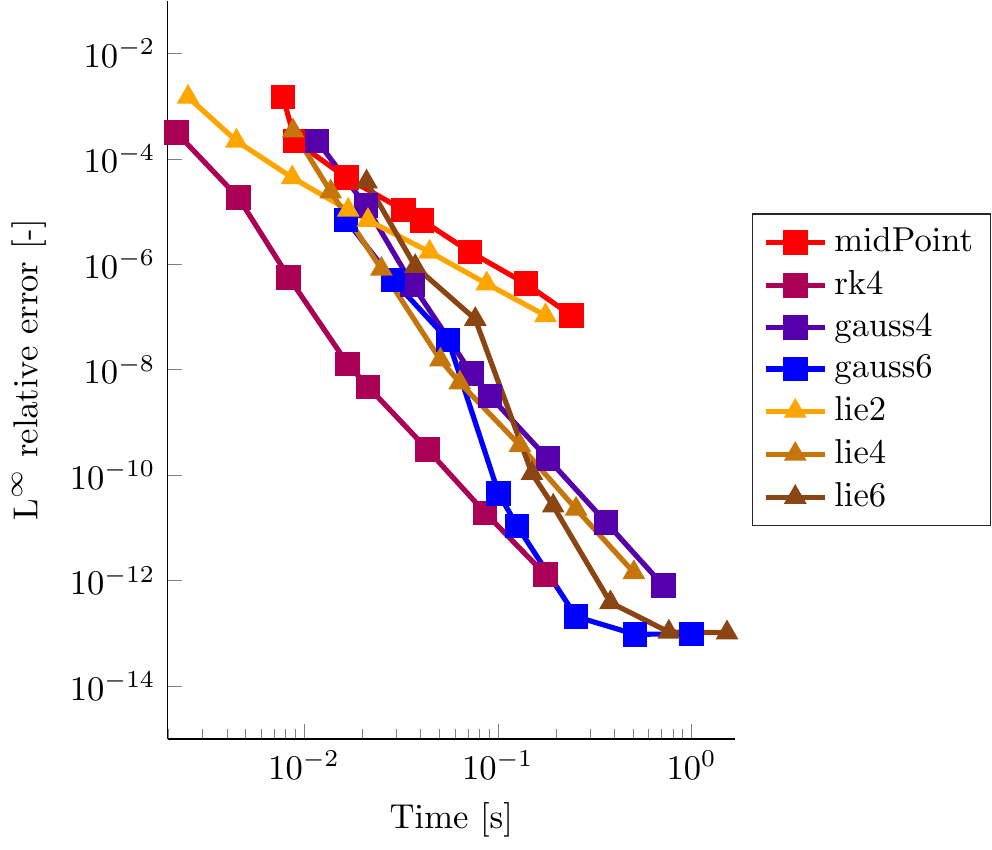}
    \end{subfigure}
    \caption{Analytic case, no interpolation. Efficiency comparison between ODE methods for $X$ (left) and $P_y$ (right).}
    \label{fig:anSpeedComp2mm}
\end{figure}

 In order to compare the efficiencies of the methods employed, we report in figure \ref{fig:anSpeedComp2mm}
 the behavior of the error as a function of the CPU time required by each method for a given resolution.
 Since the interpolation stage is done off-line, the CPU time required does not depend on the interpolation
 method employed. Among the ODE methods, the fourth-order explicit Runge-Kutta method gives the best results, followed by the sixth-order Gauss method and the fourth-order Lie method. Among the symplectic methods, the fastest method is the second-order Lie one, but it has a relatively low accuracy.

\begin{figure}
    \centering
    \newcommand{\relPath}{images/4D/gaugeAF/analytic_phase_space_3000quadPairs_DZ80mm/xm002y004}
    \begin{tabular}{cc}
        \includegraphics[width=0.45\textwidth]{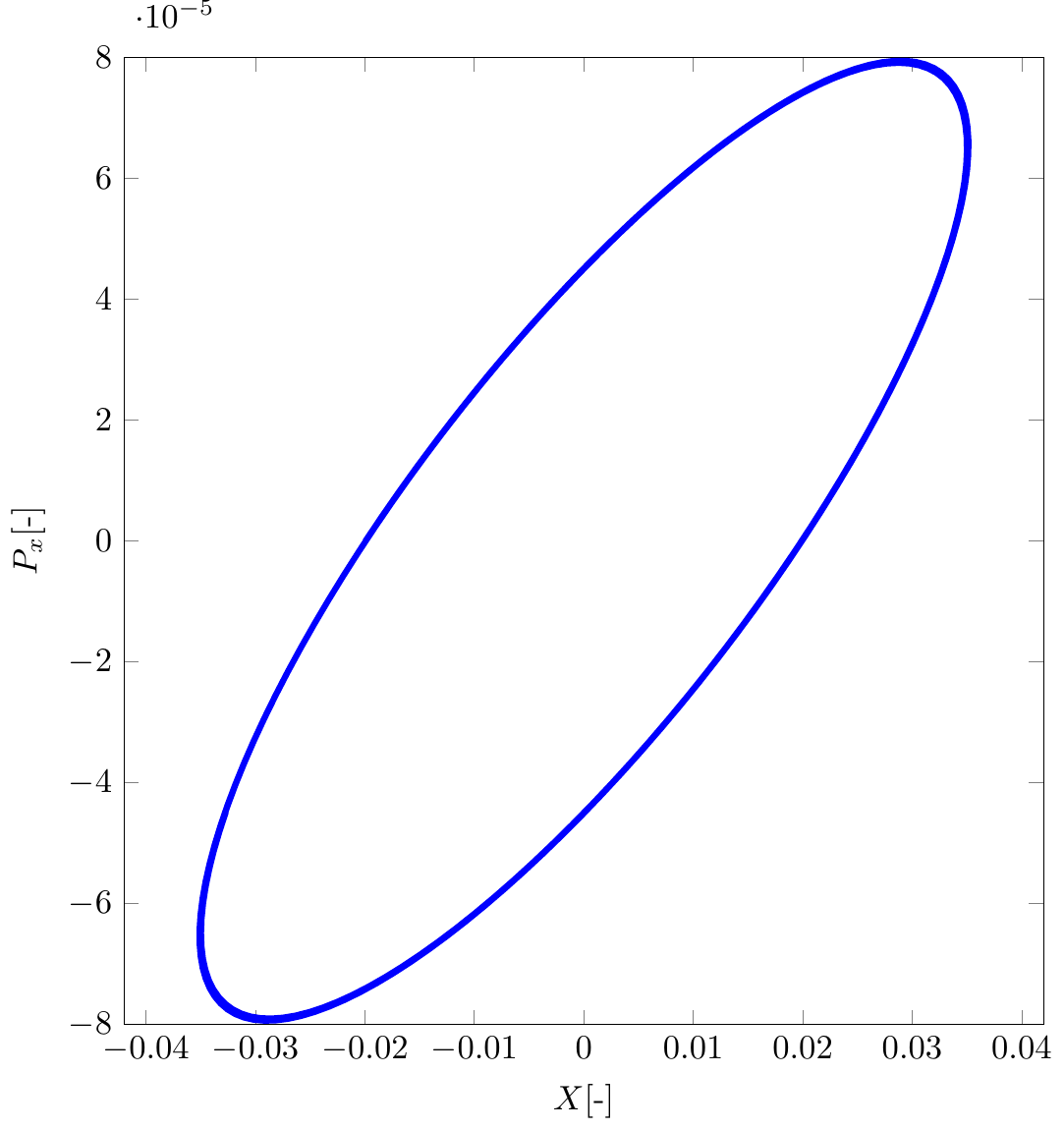}    &
        \includegraphics[width=0.45\textwidth]{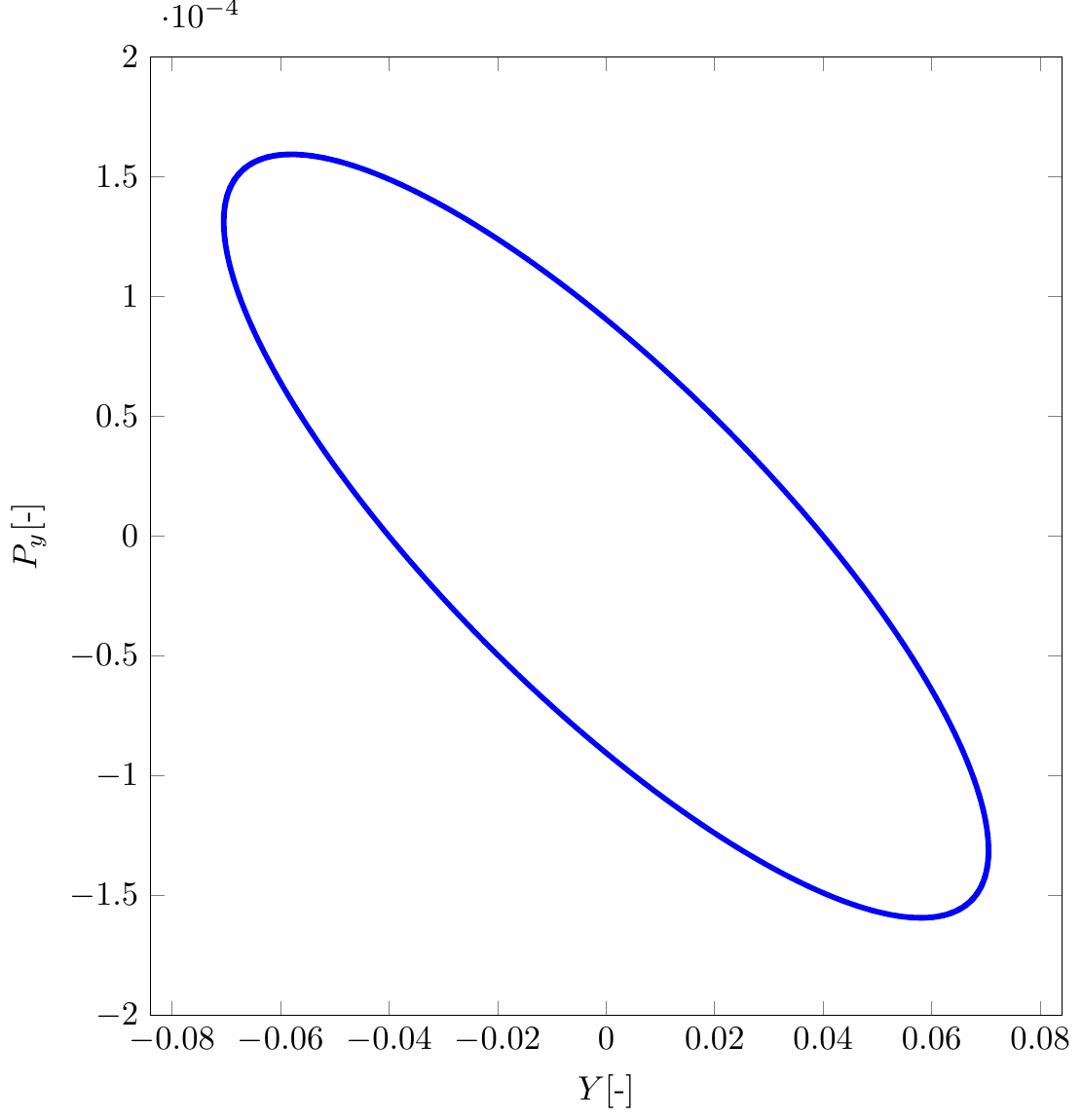}
    \end{tabular}
    \caption{Analytic case, cubic spline interpolation. Phase-space trajectories in the $\parT{X, P_x}$ (left) and $\parT{Y, P_y}$ (right) planes, $X_0=-0.02$, $Y_0=0.04$. Fourth-order Lie method.}
    \label{fig:an_m002004_lie4}
\end{figure}

\begin{figure}
    \centering
    \newcommand{\relPath}{images/4D/gaugeAF/analytic_phase_space_3000quadPairs_DZ80mm/xm002y004}
    \begin{tabular}{cc}
        \includegraphics[width=0.45\textwidth]{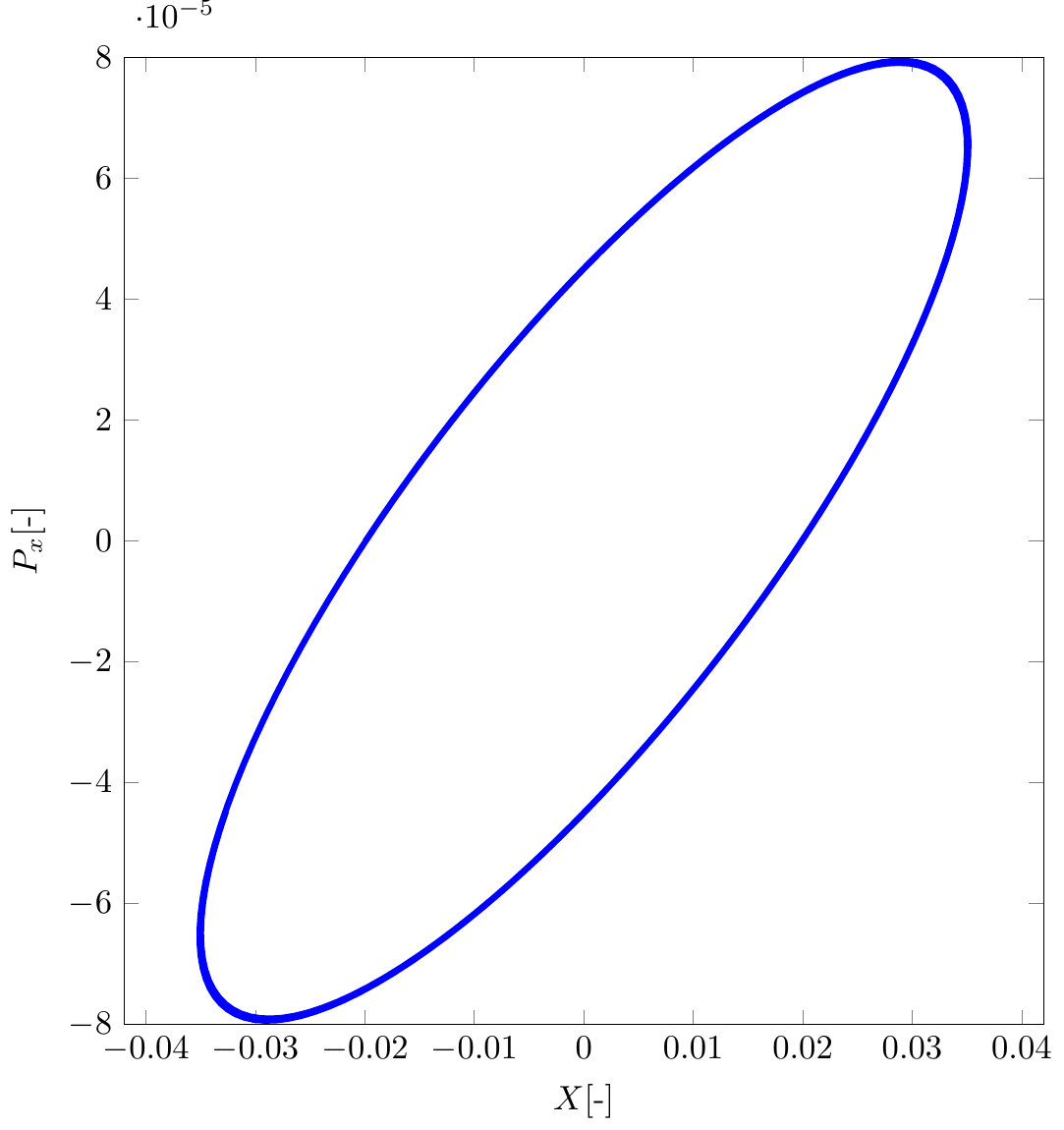}    &
        \includegraphics[width=0.45\textwidth]{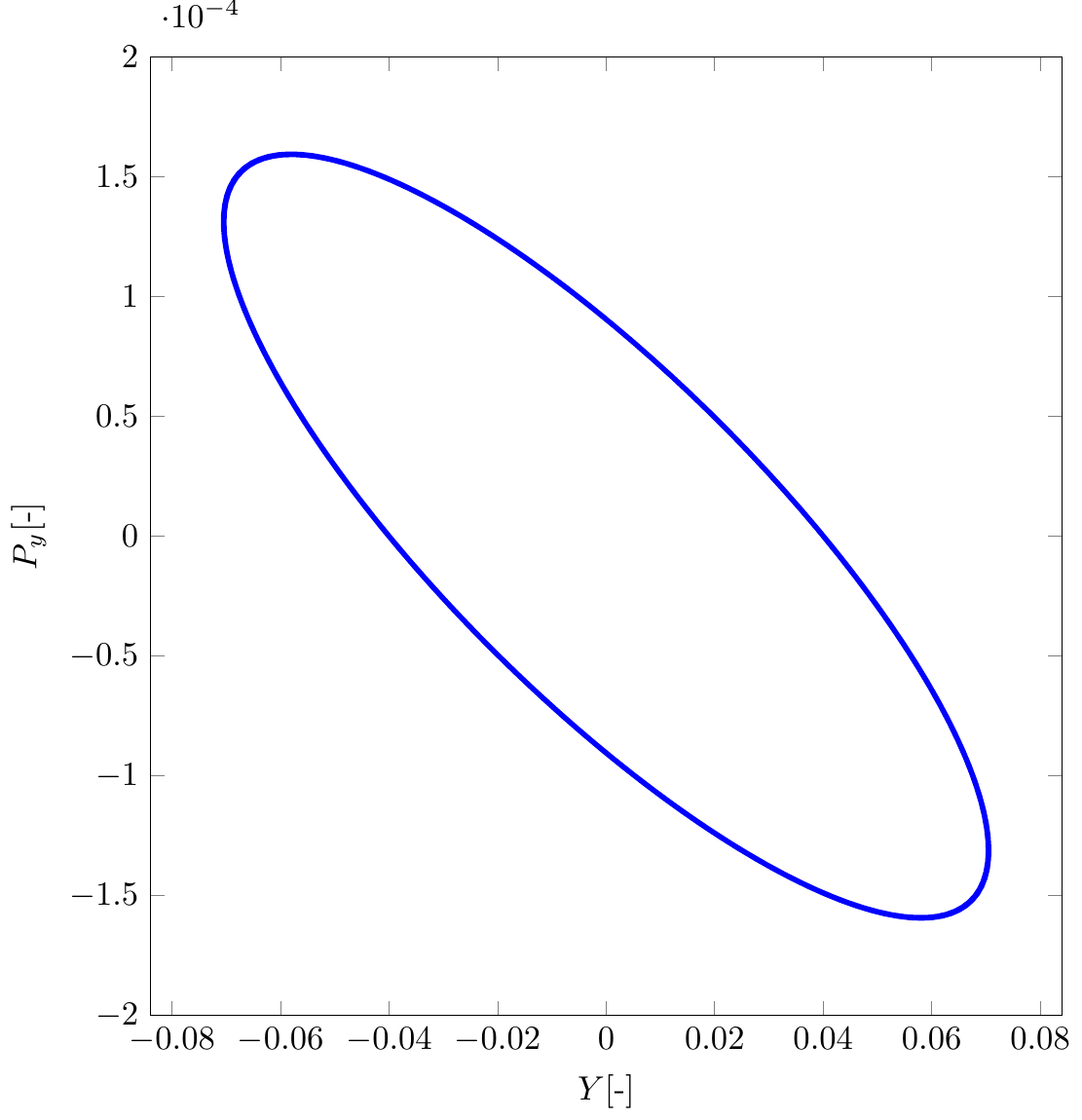}
    \end{tabular}
    \caption{Analytic case, cubic spline interpolation. Phase-space trajectories in the $\parT{X, P_x}$ (left) and $\parT{Y, P_y}$ (right) planes, $X_0=-0.02$, $Y_0=0.04$. Fourth-order explicit Runge-Kutta method.}
    \label{fig:an_m002004_rk4}
\end{figure}

As mentioned in the introduction, the   beam stability assessment requires long-term simulations. 
In order to verify the solution quality in this framework, the phase-space orbits of a particle which travels through a sequence of focusing-defocusing quadrupole couples are measured. This test is carried out considering a sequence of $3000$ consecutive quadrupole couples and an integration step of $\Delta Z = 0.08$. All the numerical methods produced stable orbits in this test. As an example, in figure \ref{fig:an_m002004_lie4} the results for the fourth-order Lie method are reported.  Remarkably, the non-symplectic fourth-order Runge-Kutta method gives results entirely analogous to those  of the other (symplectic) methods considered, see figure \ref{fig:an_m002004_rk4}.
\section{Numerical experiments with a realistic vector potential}
\label{realCase}
\begin{figure}[!htb]
    \centering
    \includegraphics[width=0.8\textwidth]{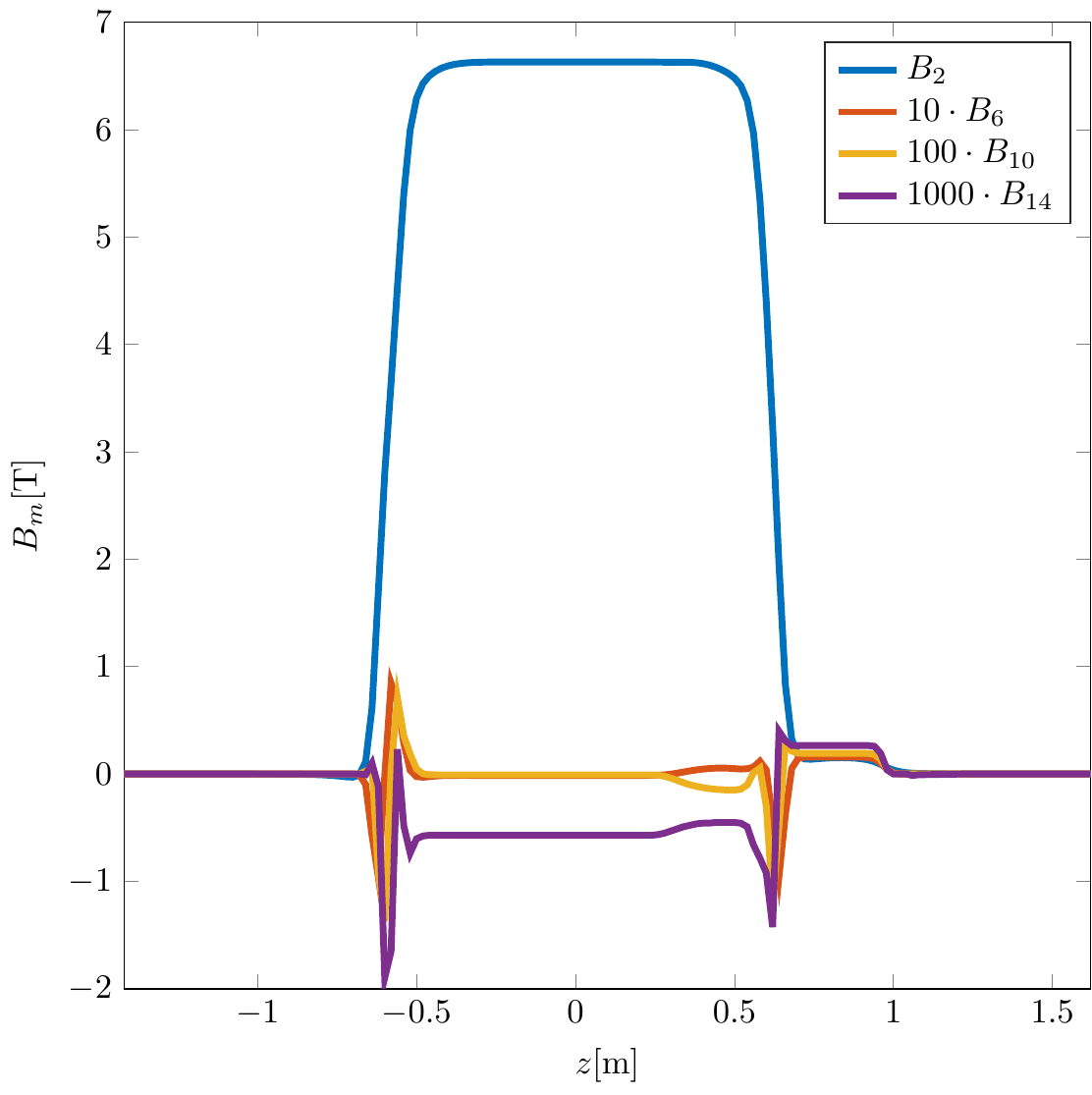}
    \caption{Realistic case. Normal harmonics sampled at $\Delta z^{data} = 0.02\, \text{m}$.}
    \label{fig:realHarmo}
\end{figure}

In this section, equations \eqref{eq:ODE6Dparax} will be solved for the case of
a   magnetic vector potential  that   corresponds to the design of a realistic quadrupole.
The harmonics $m=2, \, 6, \, 10, \, 14$, at a radius of analysis of $0.05$ and sampled at $\Delta Z^{data} = 0.02$ (figure \ref{fig:realHarmo}) are provided by numerical simulations performed by the FEM/BEM software  ROXIE
\cite{roxie:web},
 used at CERN to design the accelerator magnets.
The presence of connectors on one side of the quadrupole causes  asymmetries in the harmonics along the $Z$ axis and the presence of skew harmonics.
The generalized gradients are computed using up to $16$ derivatives. We will consider the case of a particle quite far from the quadrupole axis: indeed, the initial conditions for the transversal positions and momenta are set to $\ww_0 = \parT{0.02, \, -0.04, \, 0, \, 0}$ and $\delta_0 = 0$.
Using the horizontal-free Coulomb gauge, we obtain a gain in efficiency quite in agreement with the estimates given in section \ref{vecpot},  see tables \ref{tab:gain_speed_rk4} and \ref{tab:gain_speed_lie4}.
\begin{table}
    \centering
    \begin{tabular}{|c|cc|c|}
        \hline
        & \multicolumn{2}{c|}{Time [s]} & Ratio [-] \\
        \hline
        $\Delta Z$ [-] & AF             & HFC           & HFC/AF \\
        \hline
        $0.02$   & $0.0599$  & $0.0341$  & $0.5704$ \\
        $0.04$   & $0.0271$  & $0.0166$  & $0.6131$ \\
        $0.08$   & $0.0133$  & $0.0084$  & $0.6348$ \\
        $0.16$   & $0.0071$  & $0.0043$  & $0.6041$ \\
        \hline
        \multicolumn{3}{|r|}{Average}        &
        $\mathbf{0.606}$\\
        \hline
    \end{tabular}
    \caption{Realistic case. CPU time and speed-up obtained using different vector potential gauges and the fourth-order explicit Runge-Kutta method.}
    \label{tab:gain_speed_rk4}
\end{table}
\begin{table}
    \centering
    \begin{tabular}{|c|cc|c|}
        \hline
        & \multicolumn{2}{c|}{Time [s]} & Ratio [-] \\
        \hline
        $\Delta Z$ [-] & AF             & HFC           & HFC/AF \\
        \hline
        $0.02$   & $0.0966$  & $0.0548$  & $0.5680$ \\
        $0.04$   & $0.0474$  & $0.0256$  & $0.5401$ \\
        $0.08$   & $0.0241$  & $0.0132$  & $0.5477$ \\
        $0.16$   & $0.0118$  & $0.0068$  & $0.5726$ \\
        \hline
        \multicolumn{3}{|r|}{Average}        &
        $\mathbf{0.557}$\\
        \hline
    \end{tabular}
    \caption{Realistic case. CPU time and speed-up obtained using different vector potential gauges and the fourth-order Lie method.}
    \label{tab:gain_speed_lie4}
\end{table}
In the following, only the results with the horizontal-free Coulomb  gauge and using cubic spline interpolation 
of gridded data are reported.

\begin{figure}[!htb]
    \centering
    \newcommand{\relPath}{images/Nb3Sn/ode_test/HFC_err_comp/AbsLast}
    \begin{subfigure}[b]{.5\linewidth}
        \centering
        \includegraphics[width=0.994\textwidth]{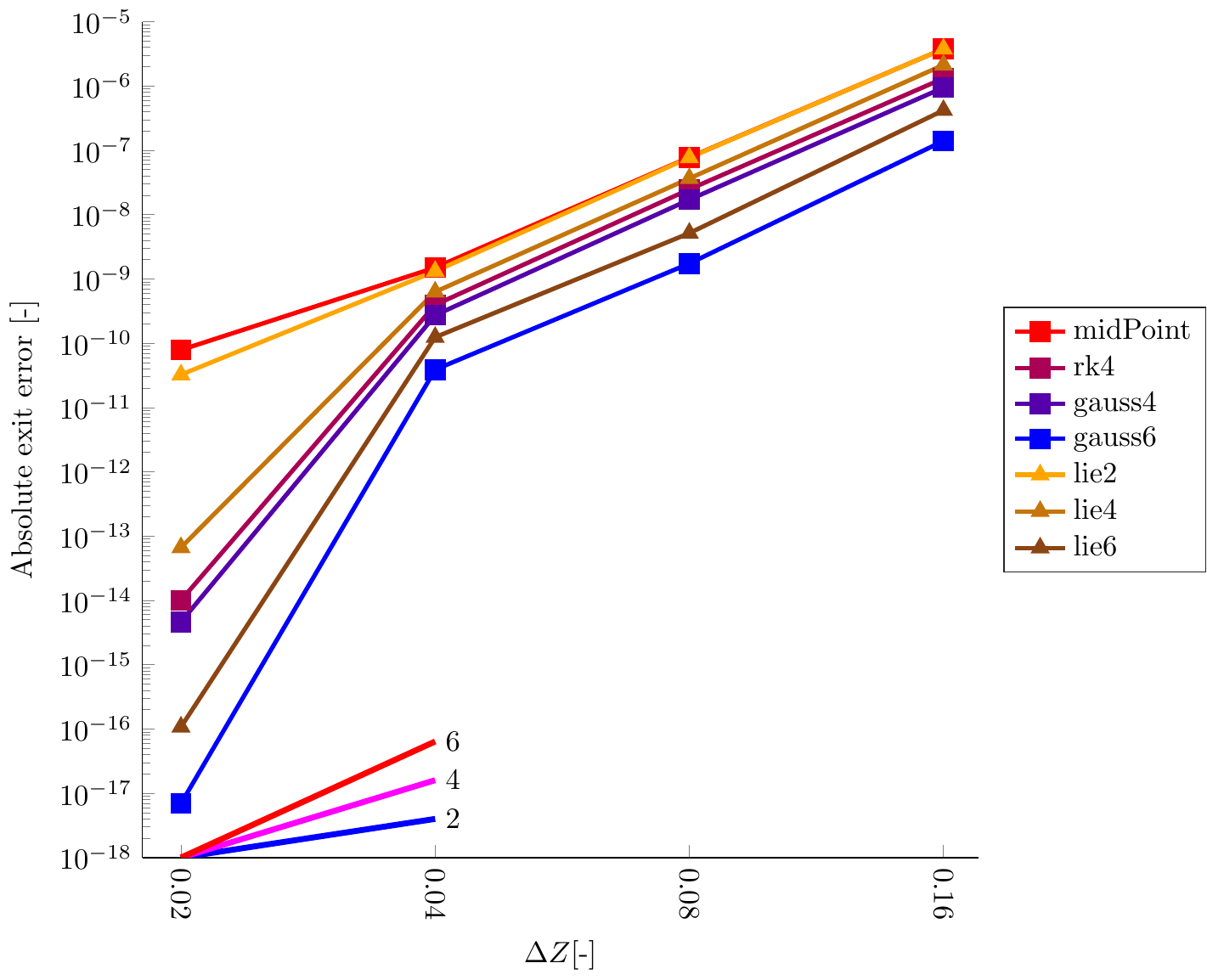}
    \end{subfigure}%
    \begin{subfigure}[b]{.5\linewidth}
        \includegraphics[width=0.994\textwidth]{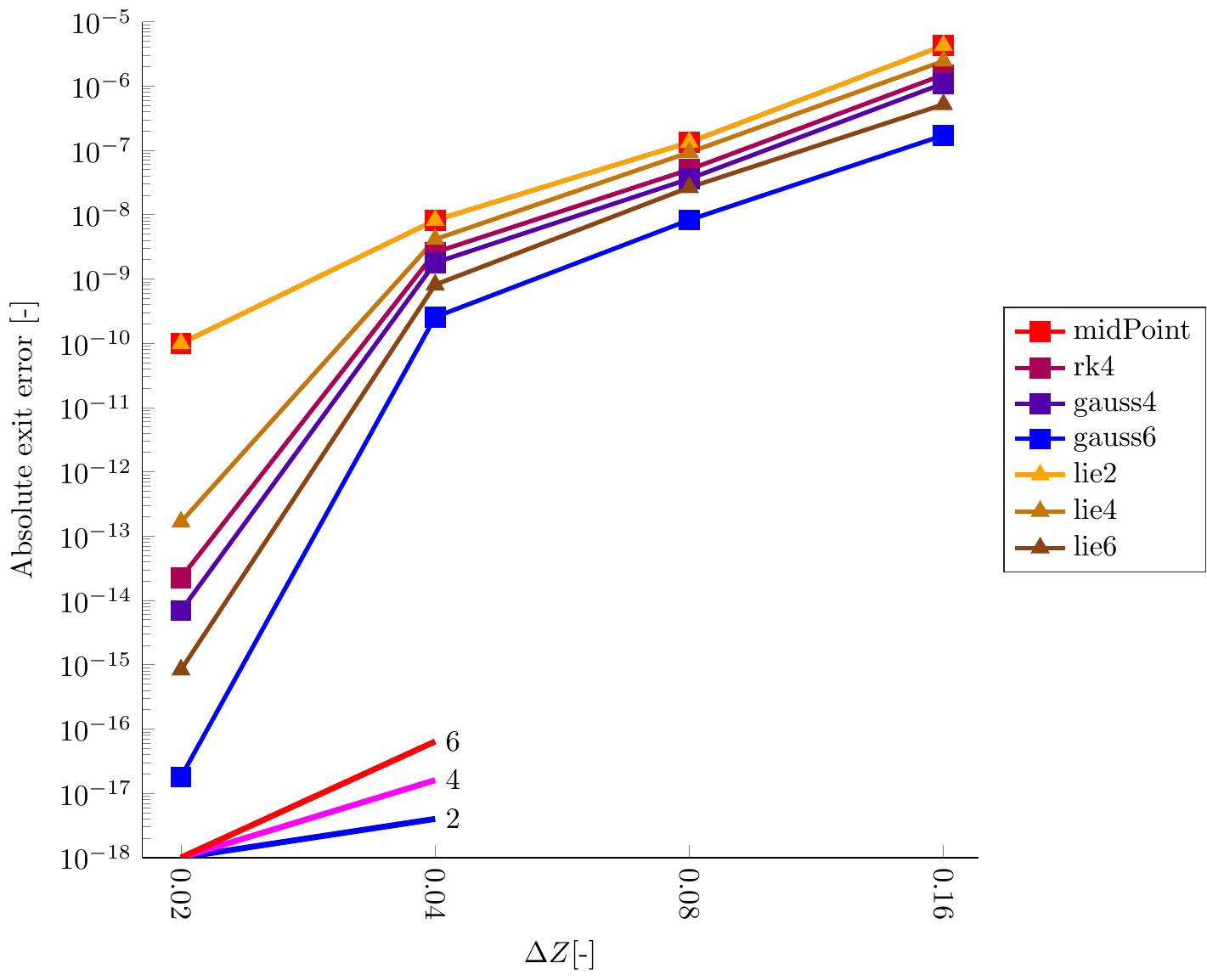}
    \end{subfigure}
    \caption{Realistic case, cubic spline interpolation. Error comparison between ODE methods for $X$ (left) and $P_y$ (right).
        The straight lines in the bottom right corner correspond to the theoretical slopes of the error curves for different convergence orders.}
    \label{fig:realErrComp2mmSpline}
\end{figure}

\begin{figure}[!htb]
    \centering
    \newcommand{\relPath}{images/Nb3Sn/ode_test/HFC_err_comp/AbsLast}
    \begin{subfigure}[b]{.5\linewidth}
        \centering
        \includegraphics[width=0.994\textwidth]{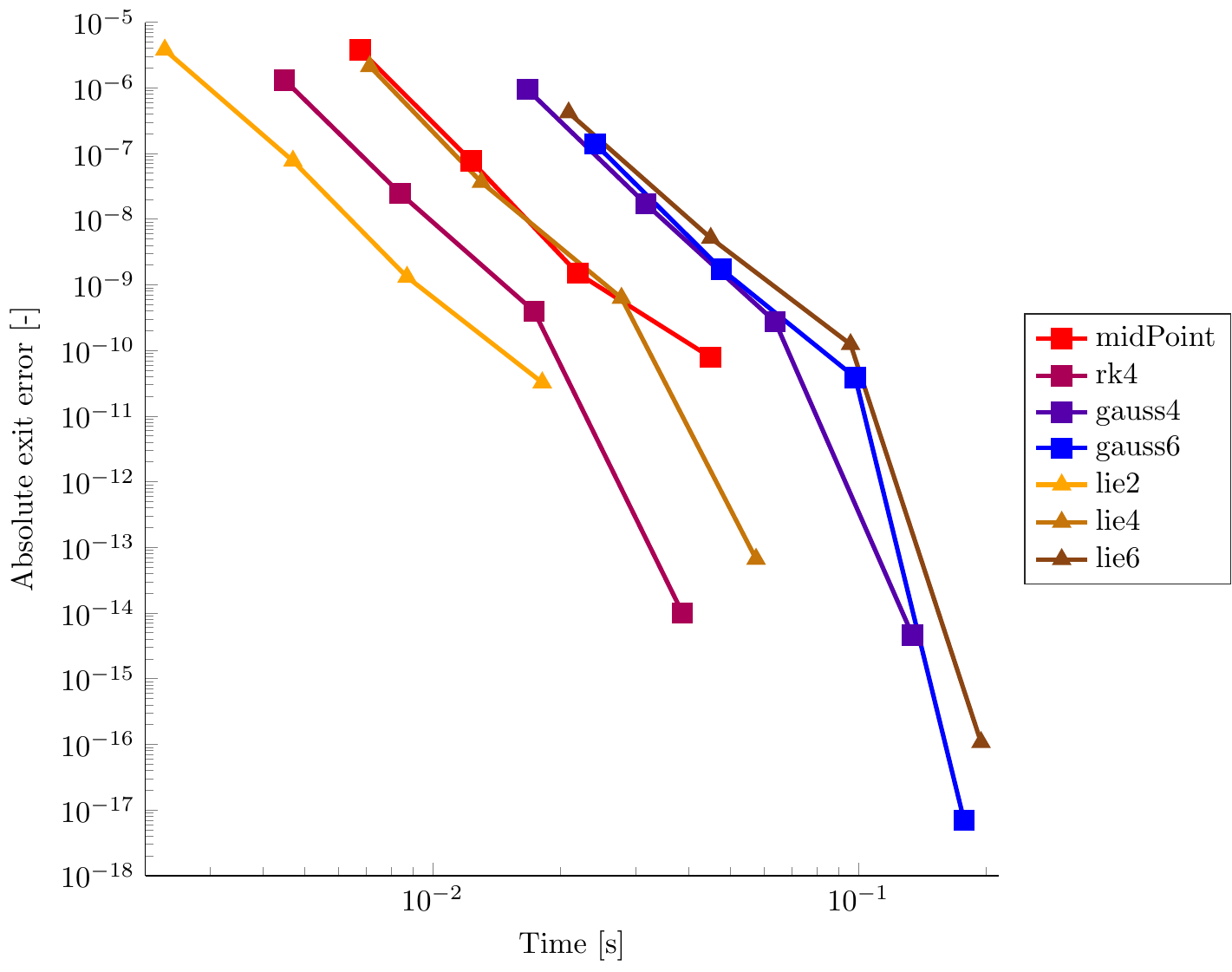}
    \end{subfigure}%
    \begin{subfigure}[b]{.5\linewidth}
        \includegraphics[width=0.994\textwidth]{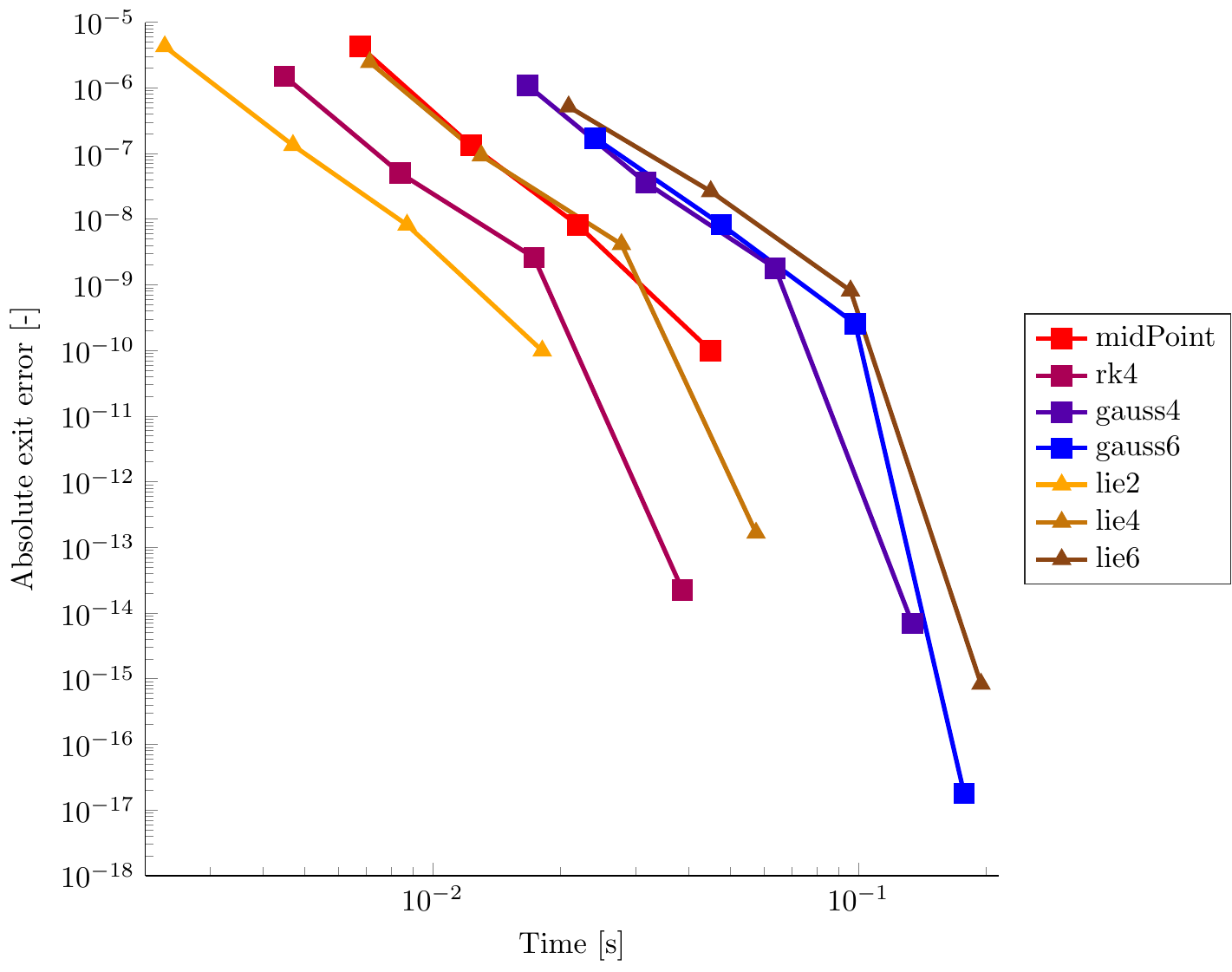}
    \end{subfigure}
    \caption{Realistic case, cubic spline interpolation. Efficiency comparison between ODE methods for $X$ (left) and $P_y$ (right).
    }
    \label{fig:Nb3SnSampSplineSpeedComp20mmHFC}
\end{figure}

The error comparisons are shown in figure \ref{fig:realErrComp2mmSpline}.	
In order to measure the methods' accuracy, the absolute error on the positions and the momenta at the quadrupole exit are used, because these values 
are useful to to understand if the non-linear effects, mentioned in the introduction, can be described correctly.
 The trend is similar to those seen in the previous sections, but with a general worsening in accuracy.
Longer time integration steps prevent the correct description of the field and consequently the achievement of the theoretical convergence orders for the numerical methods. The  final error drops in the high-order methods can be due to the fact that the reference solution has been computed using the same starting vector potential data and only a finer mesh of interpolated values.
 In order to compare the efficiencies of the methods employed, we plot in figure \ref{fig:Nb3SnSampSplineSpeedComp20mmHFC} 
 the behaviour of the error as a function of the CPU time required by each method for a given resolution.

In this case, the large $\Delta Z^{data}$ of the input data limits the possibility of achieving high accuracies, therefore there is no clear advantage in the use of high order methods and the second-order Lie method is the most efficient.

\begin{figure}
    \centering
    \newcommand{\relPath}{images/Nb3Sn/ode_test/HFC_en_comp/DZ40mm_24000qp/x002ym004}
    \begin{tabular}{cc}
        \includegraphics[width=0.45\textwidth]{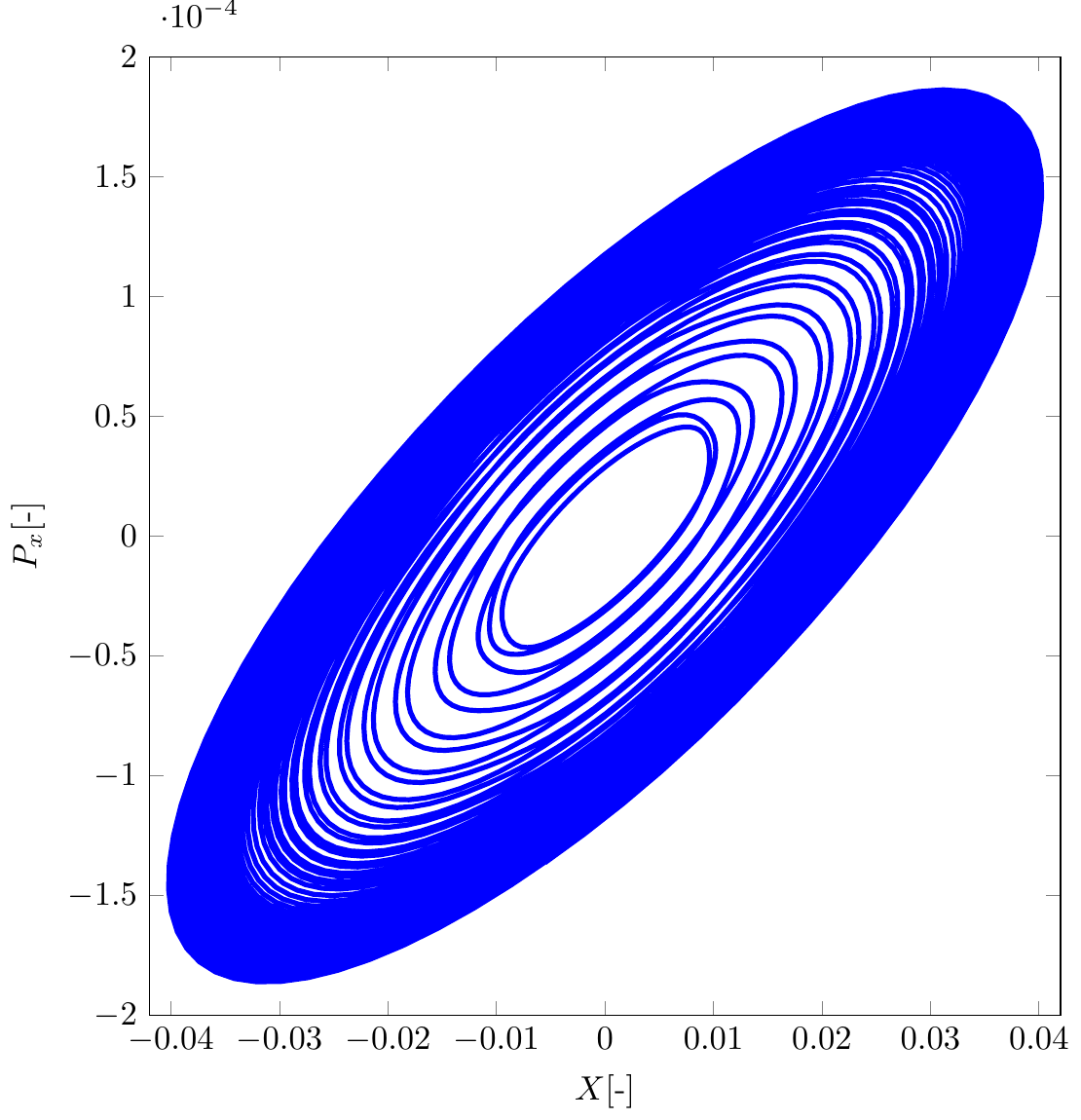}    &
        \includegraphics[width=0.45\textwidth]{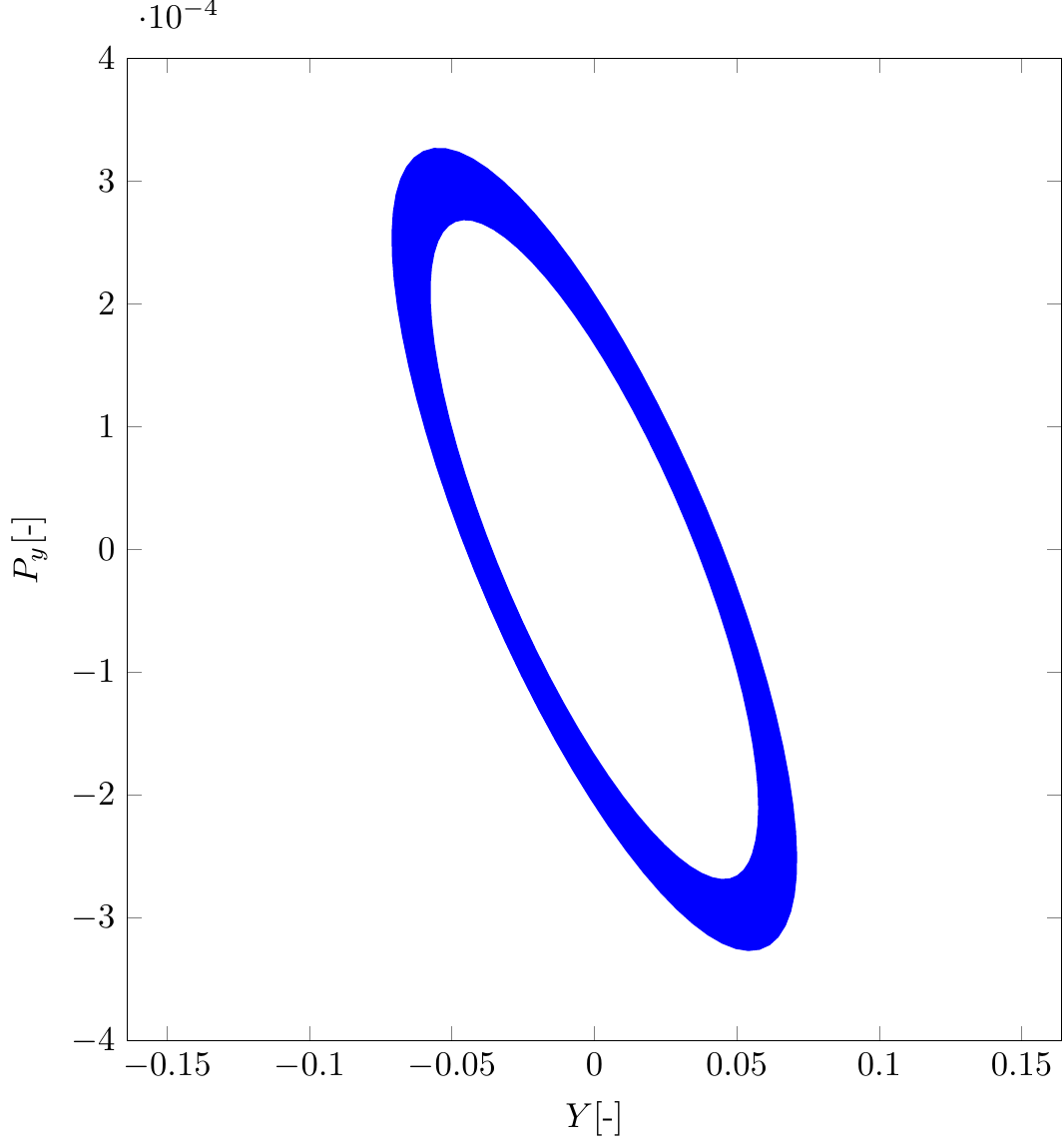}
    \end{tabular}
    \caption{Realistic case, cubic spline interpolation. Phase-space trajectories in the $\parT{X, P_x}$ (left) and $\parT{Y, P_y}$ (right) planes, $X_0=0.02$, $Y_0=-0.04$. Fourth-order Lie method.}
    \label{fig:ps_002m004_lie4}
\end{figure}

\begin{figure}
    \centering
    \newcommand{\relPath}{images/Nb3Sn/ode_test/HFC_en_comp/DZ40mm_24000qp/x002ym004}
    \begin{tabular}{cc}
        \includegraphics[width=0.45\textwidth]{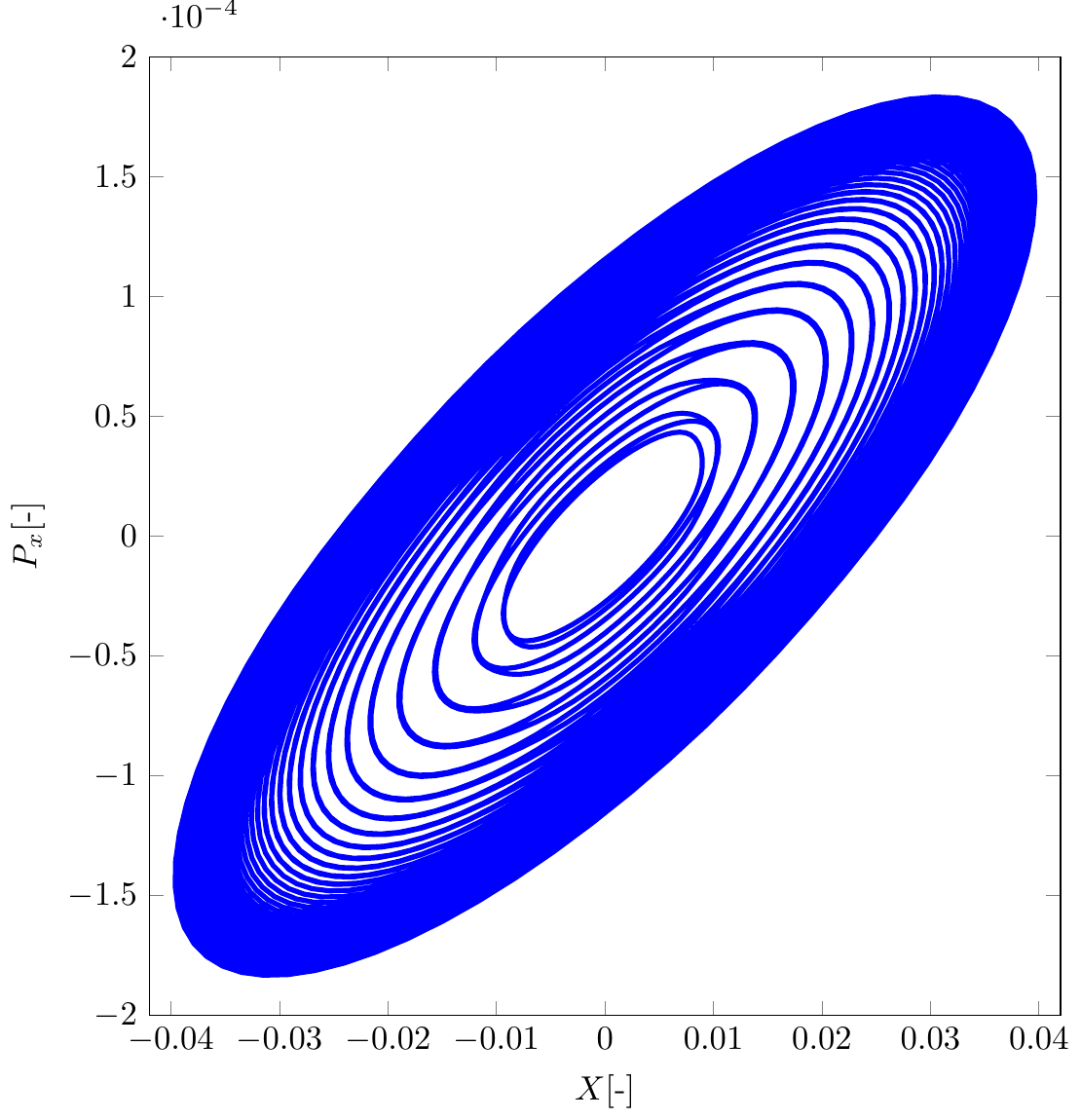}    &
        \includegraphics[width=0.45\textwidth]{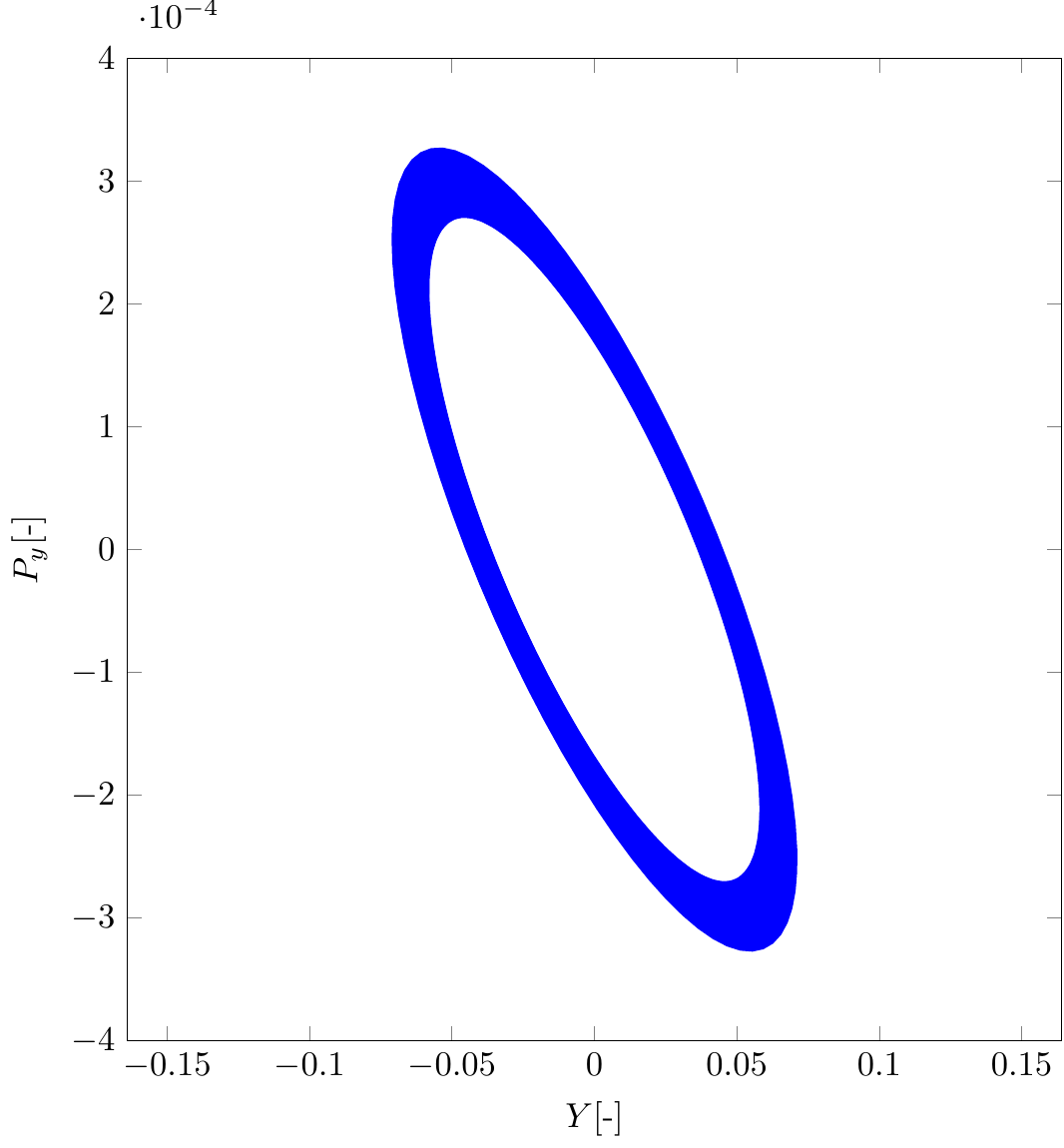}
    \end{tabular}
    \caption{Realistic case, cubic spline interpolation. Phase-space trajectories in the $\parT{X, P_x}$ (left) and $\parT{Y, P_y}$ (right) planes, $X_0=0.02$, $Y_0=-0.04$. Fourth-order explicit Runge-Kutta method.}
    \label{fig:ps_002m004_rk4}
\end{figure}

Finally, we check the beam stability looking at the phase-space orbits of a particle which travels through a sequence of focusing-defocusing quadrupole couples. This test is carried out considering a sequence of $24000$ consecutive quadrupole couples and an integration step of $\Delta Z = 0.04$.
 In this case it is possible to notice (see figures \ref{fig:ps_002m004_lie4} and \ref{fig:ps_002m004_rk4}) that the trend in the $\parT{X, P_x}$ phase space appears unstable. 

\begin{figure}
    \centering
    \newcommand{\relPath}{images/4D/gaugeHFC/Nb3Sn_phase_space_3000quadPairs_DZ80mm_H2ND2/xm002y004}
    \begin{tabular}{cc}
        \includegraphics[width=0.45\textwidth]{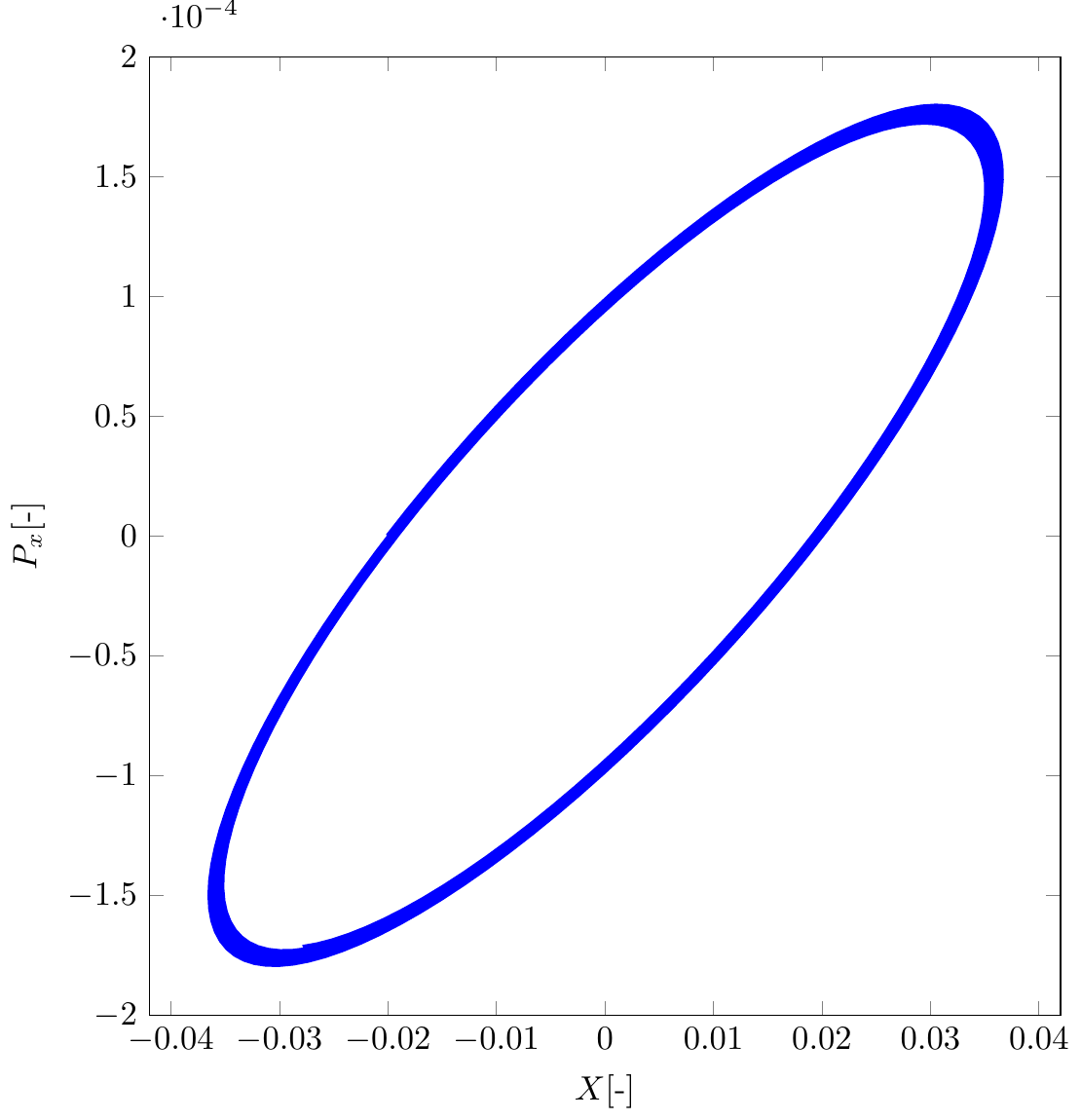}    &
        \includegraphics[width=0.45\textwidth]{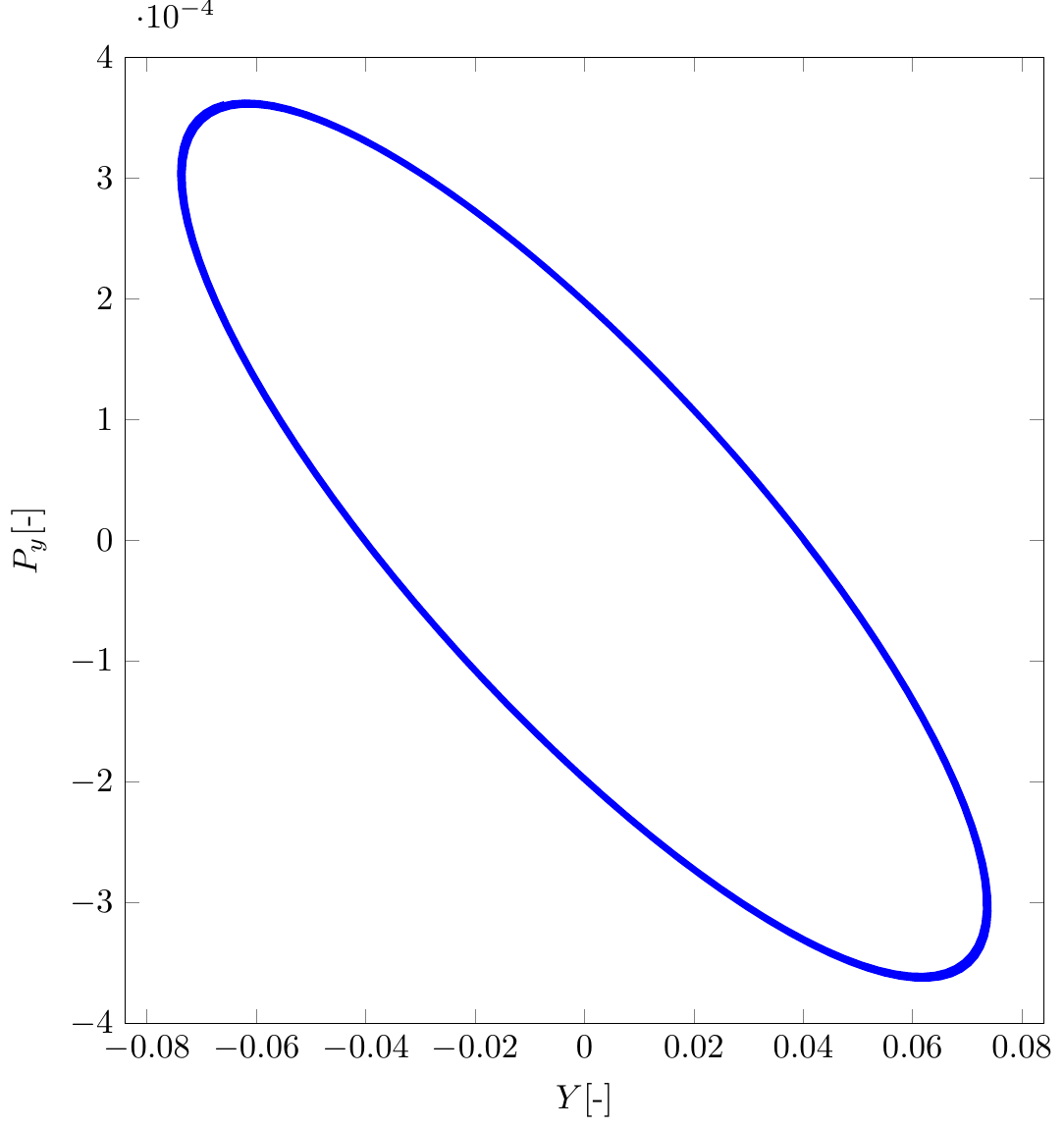}
    \end{tabular}
    \caption{Realistic case II, second order harmonic, two generalized gradient derivatives, cubic spline interpolation. Phase-space trajectories in the $\parT{X, P_x}$ (left) and $\parT{Y, P_y}$ (right) planes, $X_0=-0.02$, $Y_0=0.04$. Fourth-order Lie method.}
    \label{fig:simple_m002004_lie4}
\end{figure}

\begin{figure}
    \centering
    \newcommand{\relPath}{images/4D/gaugeHFC/Nb3Sn_phase_space_3000quadPairs_DZ80mm/xm002y001}
    \begin{tabular}{cc}
        \includegraphics[width=0.45\textwidth]{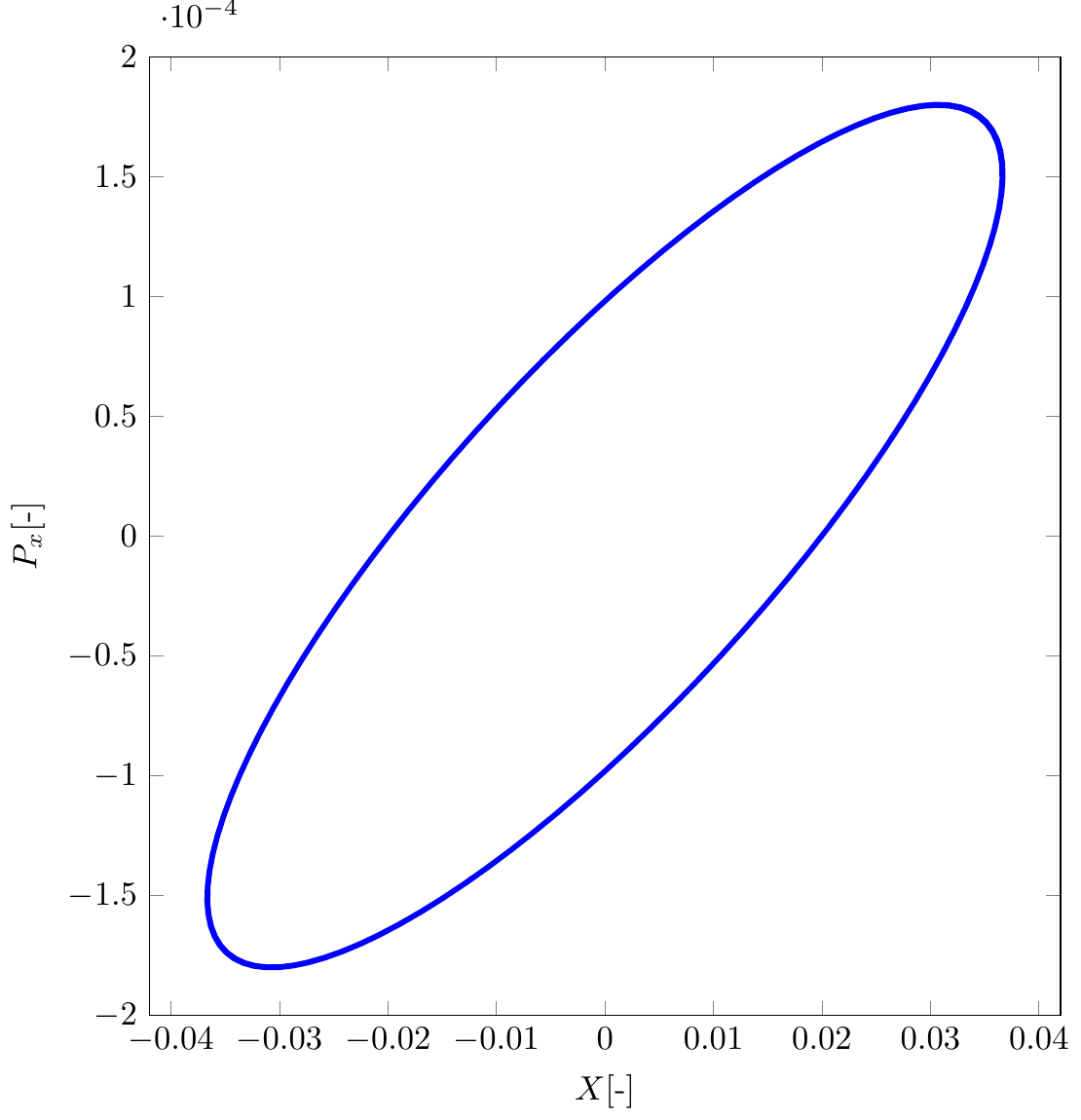}    &
        \includegraphics[width=0.45\textwidth]{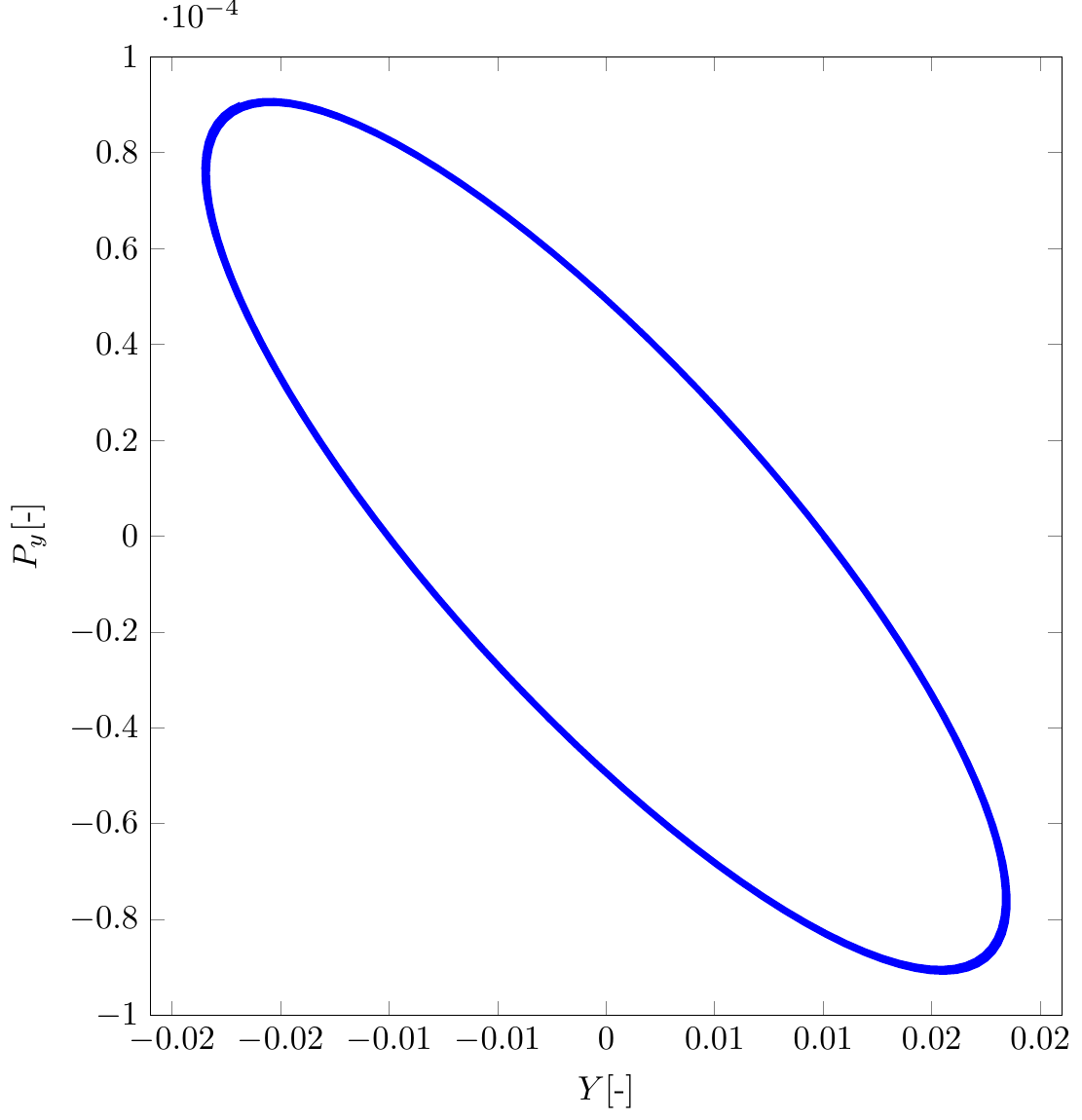}
    \end{tabular}
    \caption{Realistic case II, cubic spline interpolation. Phase-space trajectories in the $\parT{X, P_x}$ (left) and $\parT{Y, P_y}$ (right) planes, $X_0=-0.02$, $Y_0=0.01$. Fourth-order Lie method.}
    \label{fig:ps_m002001_lie4}
\end{figure}

\begin{figure}
    \centering
    \newcommand{\relPath}{images/4D/gaugeHFC/Nb3Sn_phase_space_3000quadPairs_DZ80mm_H2ND2/xm002y004}
    \begin{tabular}{cc}
        \includegraphics[width=0.45\textwidth]{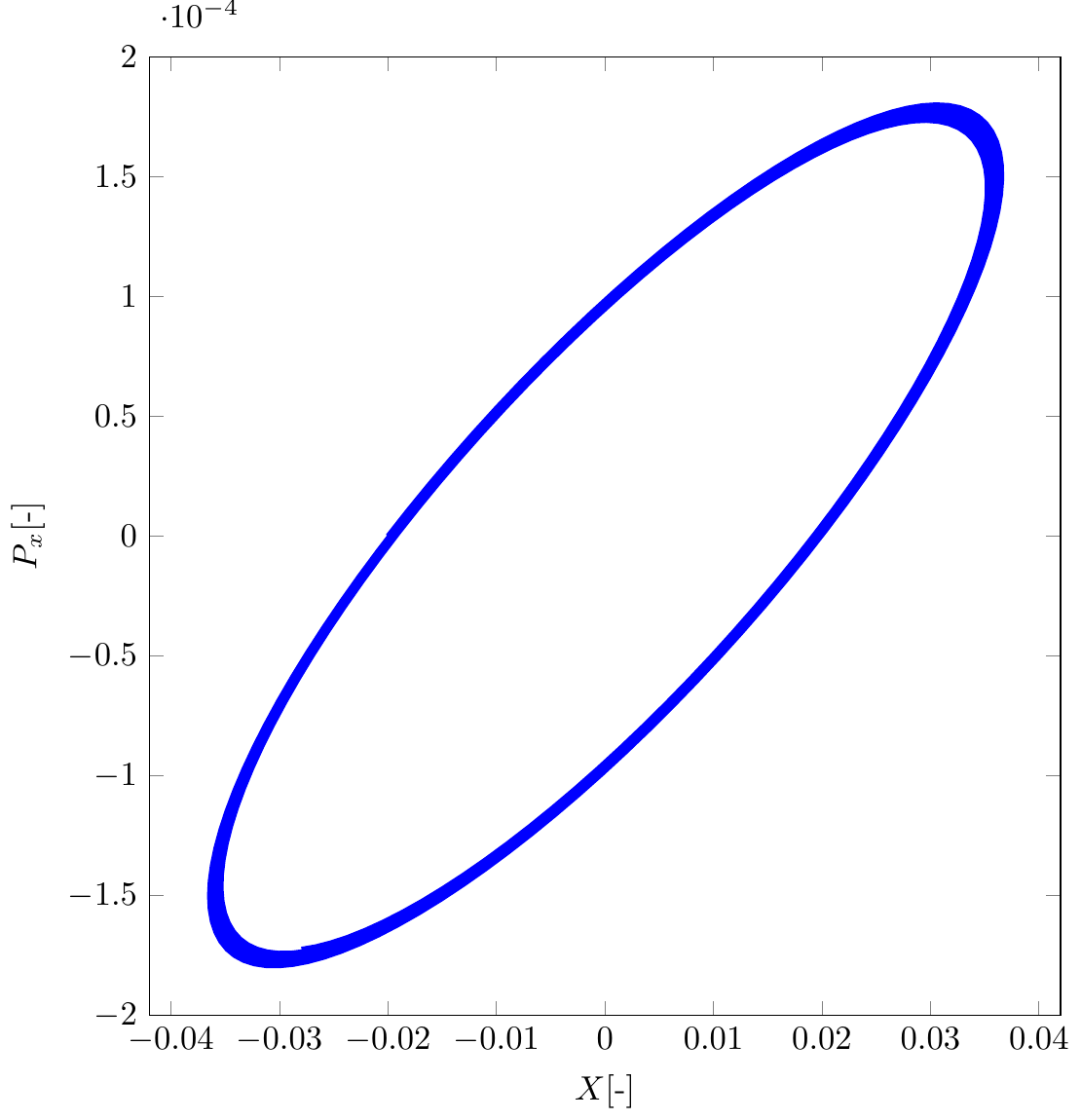}    &
        \includegraphics[width=0.45\textwidth]{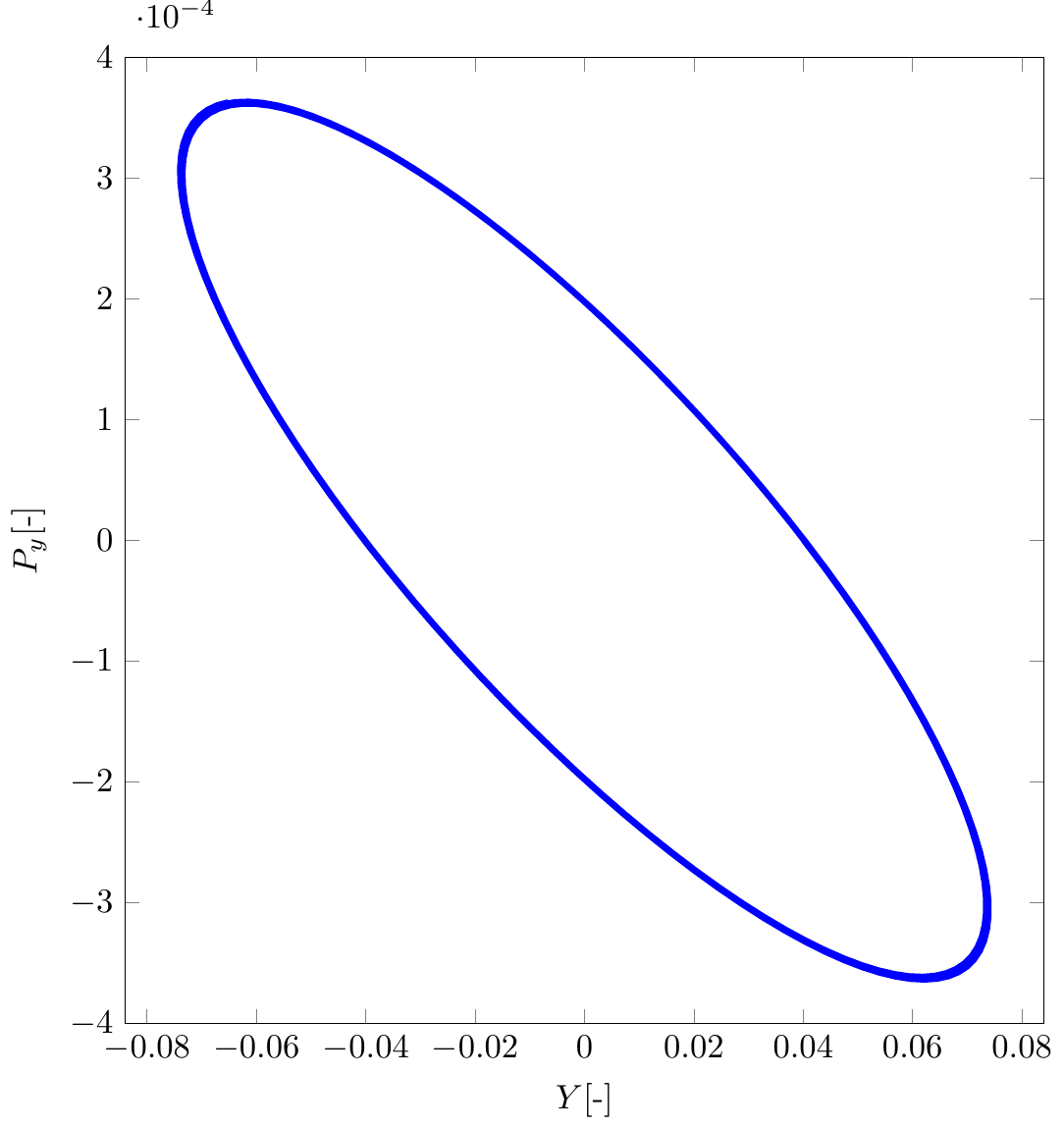}
    \end{tabular}
    \caption{Realistic case II, second order harmonic, two generalized gradient derivatives, cubic spline interpolation. Phase-space trajectories in the $\parT{X, P_x}$ (left) and $\parT{Y, P_y}$ (right) planes, $X_0=-0.02$, $Y_0=0.04$. Fourth-order explicit Runge-Kutta method.}
    \label{fig:simple_m002004_rk4}
\end{figure}

\begin{figure}
    \centering
    \newcommand{\relPath}{images/4D/gaugeHFC/Nb3Sn_phase_space_3000quadPairs_DZ80mm/xm002y001}
    \begin{tabular}{cc}
        \includegraphics[width=0.45\textwidth]{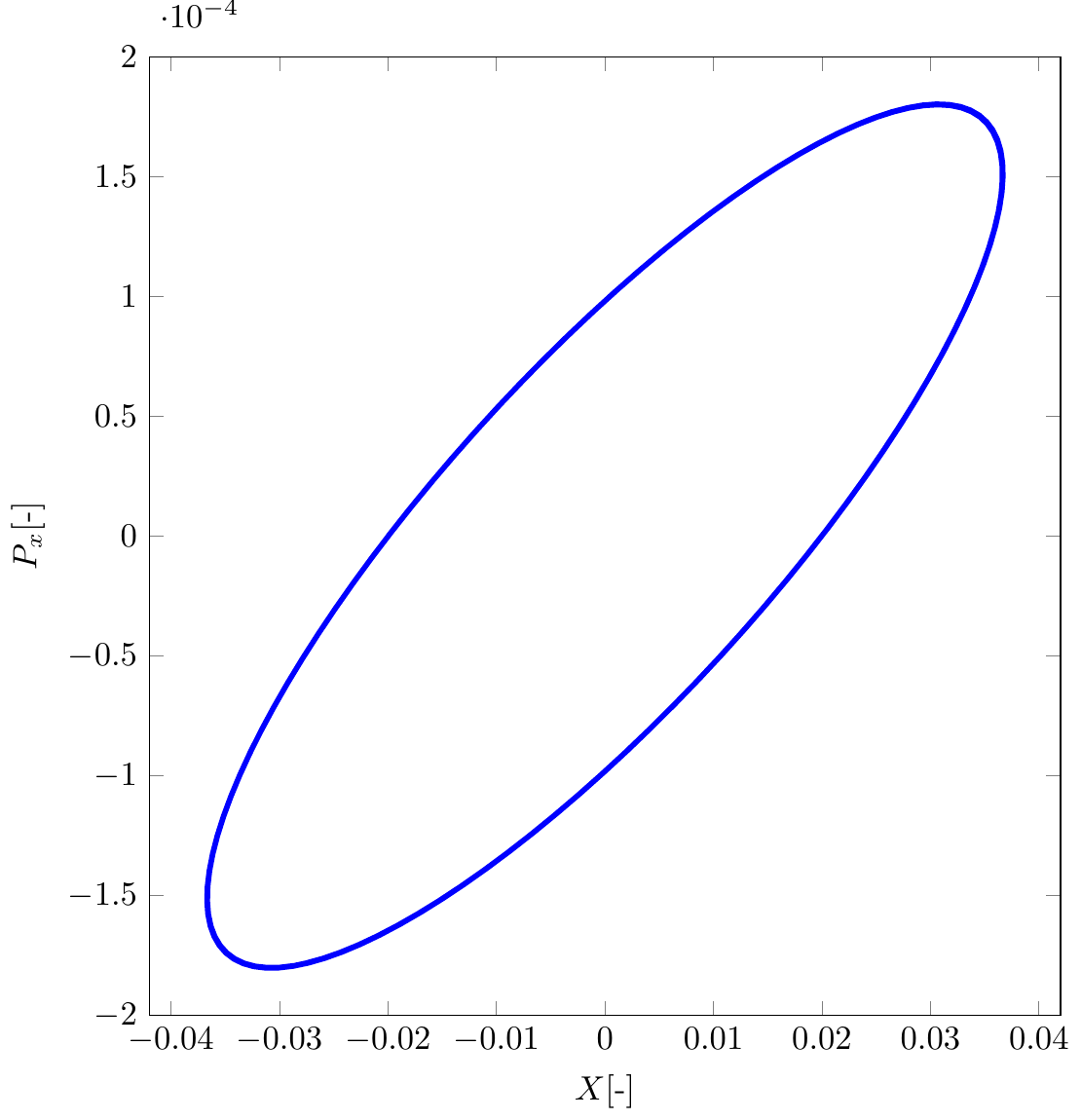}    &
        \includegraphics[width=0.45\textwidth]{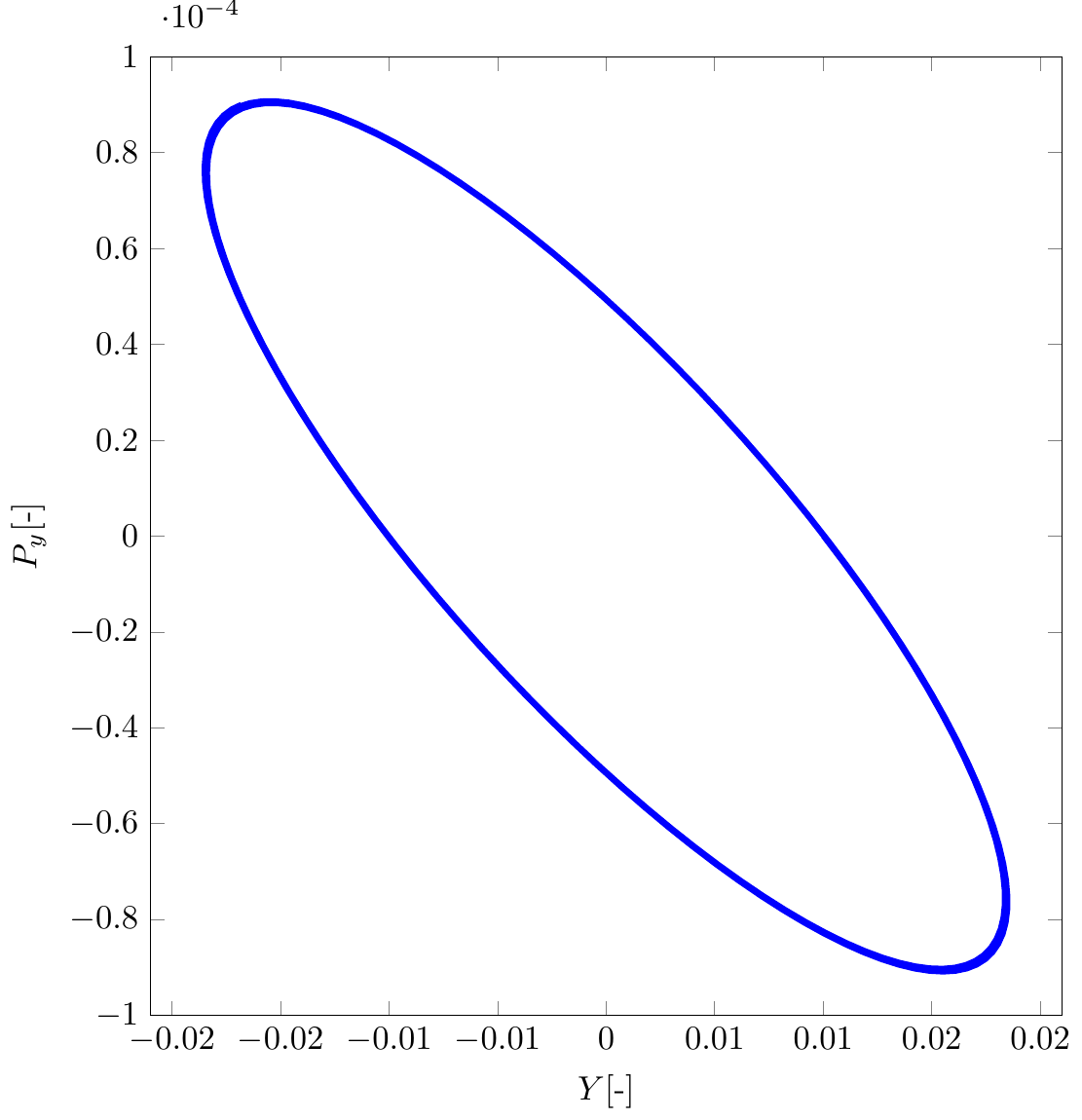}
    \end{tabular}
    \caption{Realistic case II, cubic spline interpolation. Phase-space trajectories in the $\parT{X, P_x}$ (left) and $\parT{Y, P_y}$ (right) planes, $X_0=-0.02$, $Y_0=0.01$. Fourth-order explicit Runge-Kutta method.}
    \label{fig:ps_m002001_rk4}
\end{figure}

The instability is probably due to the high degree of the vector potential polynomial, obtained using various harmonics and many generalized gradient derivatives. In fact, using only the second order harmonic and two generalized gradient derivatives (figure \ref{fig:simple_m002004_lie4}), or considering a motion closer to the center (figure \ref{fig:ps_m002001_lie4}), it is possible to obtain stable orbits  similar to the ones obtained in the analytical case presented in section \ref{analyticCase}.
It is especially noteworthy that, in spite of its lack of symplectic properties, the classical fourth order RK method does not appear to behave differently with respect to the symplectic ones (figures \ref{fig:simple_m002004_rk4} and \ref{fig:ps_m002001_rk4}).

\begin{figure}[!htb]
    \centering
    \newcommand{\relPath}{images/4D/gaugeHFC/Nb3Sn_dz20mm_enCons8000quadPair_DZ80mm/}
    \begin{subfigure}[b]{.5\linewidth}
        \centering
        \includegraphics[width=0.994\textwidth]{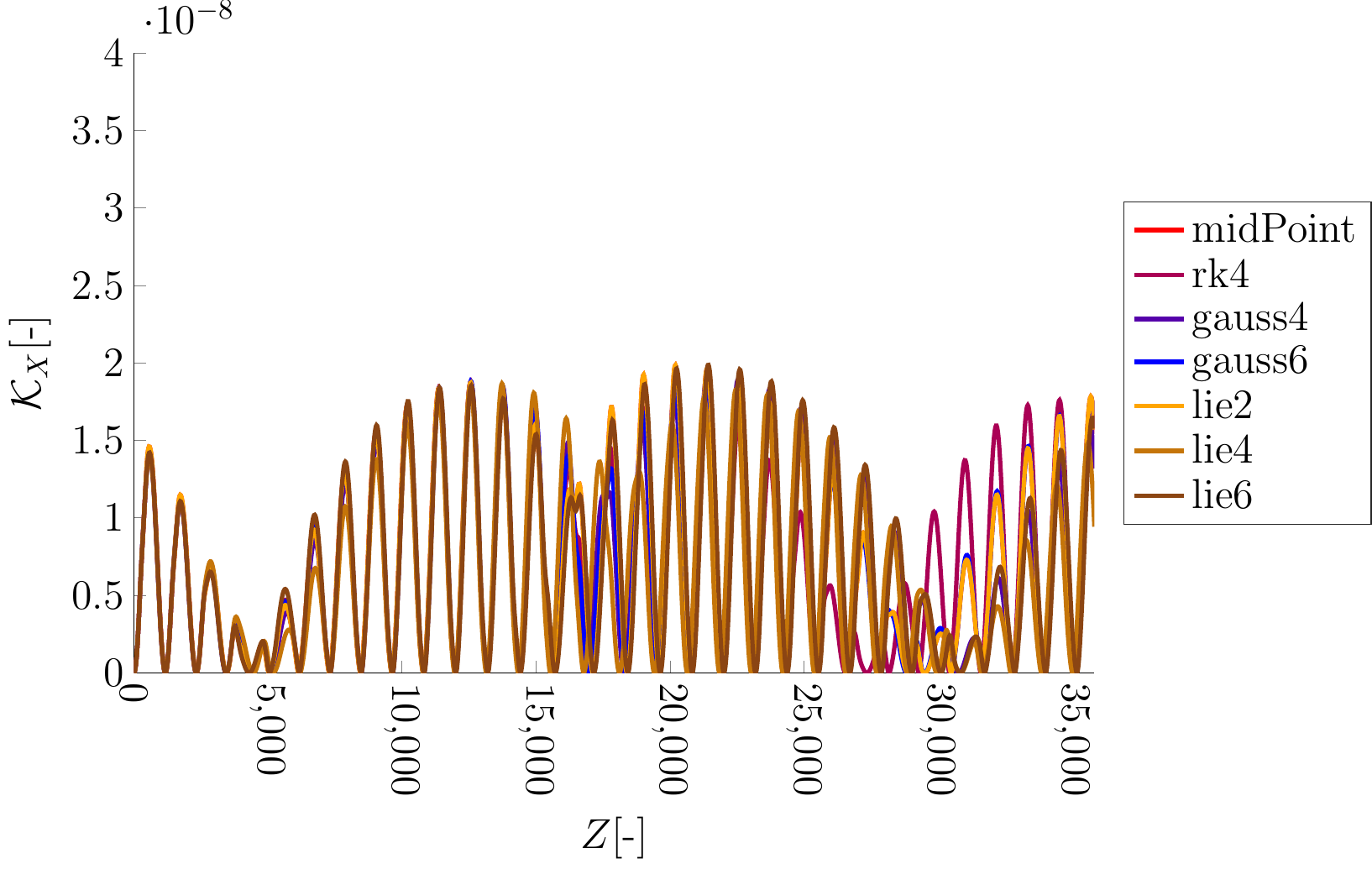}
    \end{subfigure}%
    \begin{subfigure}[b]{.5\linewidth}
        \centering
        \includegraphics[width=0.994\textwidth]{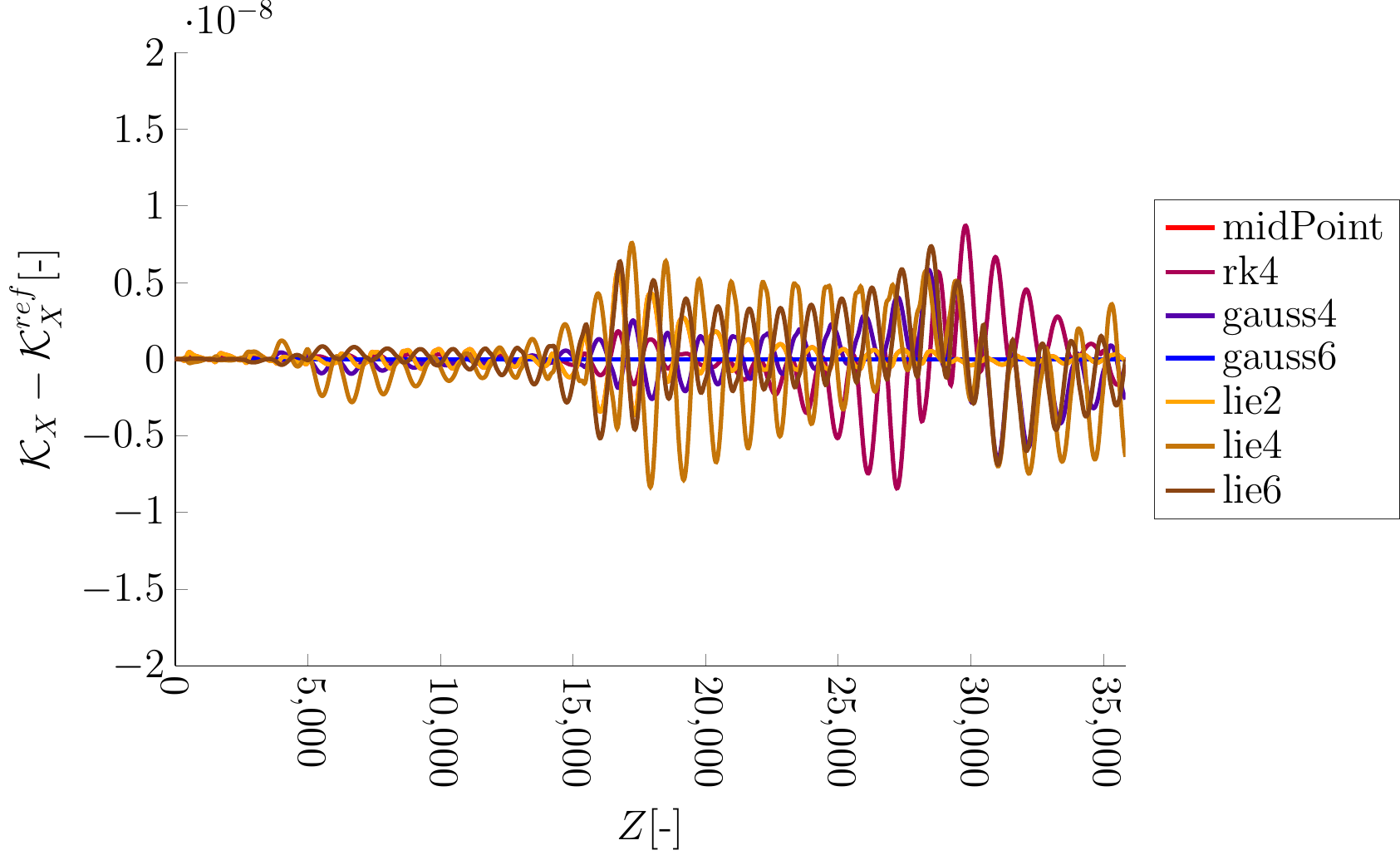}
    \end{subfigure}
    \caption{Realistic case II, cubic spline interpolation. Trend of the $\KK_X$ (left)  energy component along $8000$ focusing-defocusing quadrupole couples for different ODE methods and deviation of $\KK_X$ associated to the different methods with respect to the sixth-order Gauss method (right).}
    \label{fig:Nb3SnSampEnergyComp20mmSpline}
\end{figure}

Finally, we check the energy conservation for a particle which travels through $8000$ focusing-defocusing quadrupole pairs using an integration step $\Delta Z = 0.08$. Results  are shown in figure \ref{fig:Nb3SnSampEnergyComp20mmSpline}.
In this realistic case the magnetic vector potential has a more complex shape and the differences  in the $\KK_X$ trend among the numerical methods appear sooner. Nevertheless, the $\KK_X$ oscillations appear to be stable and no significant differences between symplectic and non symplectic methods concerning the energy conservation
are observed.

\section{Conclusions and future developments}
\label{conclu} \indent
We have discussed and analyzed several issues that arise in the numerical approximation
of  charged particles  trajectories  in magnetic quadrupoles.
We have shown that a specific gauge transformation that allows to reduce by approximately 50\%  the computational cost of each vector potential evaluation, thus significantly enhancing the efficiency of any numerical approximation method employed for the particle trajectory simulation.

 The impact of the interpolation technique employed to compute magnetic vector potential values
at arbitrary locations from gridded data has also been assessed, highlighting potential limitations in accuracy induced by insufficiently accurate interpolation methods, if high order time integration
techniques are to be applied. However, cubic spline interpolation was found to be sufficient
for preserving the accuracy of all the methods considered in this work over a wide range of values for the
integration step.
		  
We  have then compared several high order integration techniques, which  allow to maintain   high accuracy
even with relatively large integration step values, 
in order to assess their accuracy and efficiency for   long-term simulations. 
Explicit high order Lie methods have been considered, along with implicit high order symplectic integrators  and a more conventional, non symplectic explicit Runge-Kutta methods. 

 In the case of  realistic  vector potentials, the errors induced by the vector potential representation and interpolation become significant and reduce the highest possible accuracy that can be attained. Furthermore, since in realistic cases the magnetic vector potential evaluation   costlier, numerical methods which require less evaluations, such as the second-order Lie method, appear to be more competitive in terms of efficiency. 
 On the other hand, if these errors could be reduced by different approaches for the vector field representation,
  higher order methods could be advantageous if a more precise approximation is required. The speed gain obtained by the horizontal-free Coulomb gauge would also allow to use more expensive methods. In particular, the explicit fourth-order Runge-Kutta appears to be the most efficient method
  and the fourth-order Lie the most efficient among  symplectic methods.  

All the computations carried out in this work employed vector potential values sampled on a uniform 
grid along the longitudinal axis of the quadrupole. 
 In order to increase the efficiency of the methods employed, using  vector potentials sampled over a non-uniform grid   appears to be a straightforward and necessary improvement. In such way, it could be possible to use a smaller step in the fringe field regions, where the vector potential has more complex shape, and a larger step where the field is uniform and has  simpler structure.  
 
\begin{figure}[!htb]
    \centering
        \includegraphics[width=0.75\textwidth]{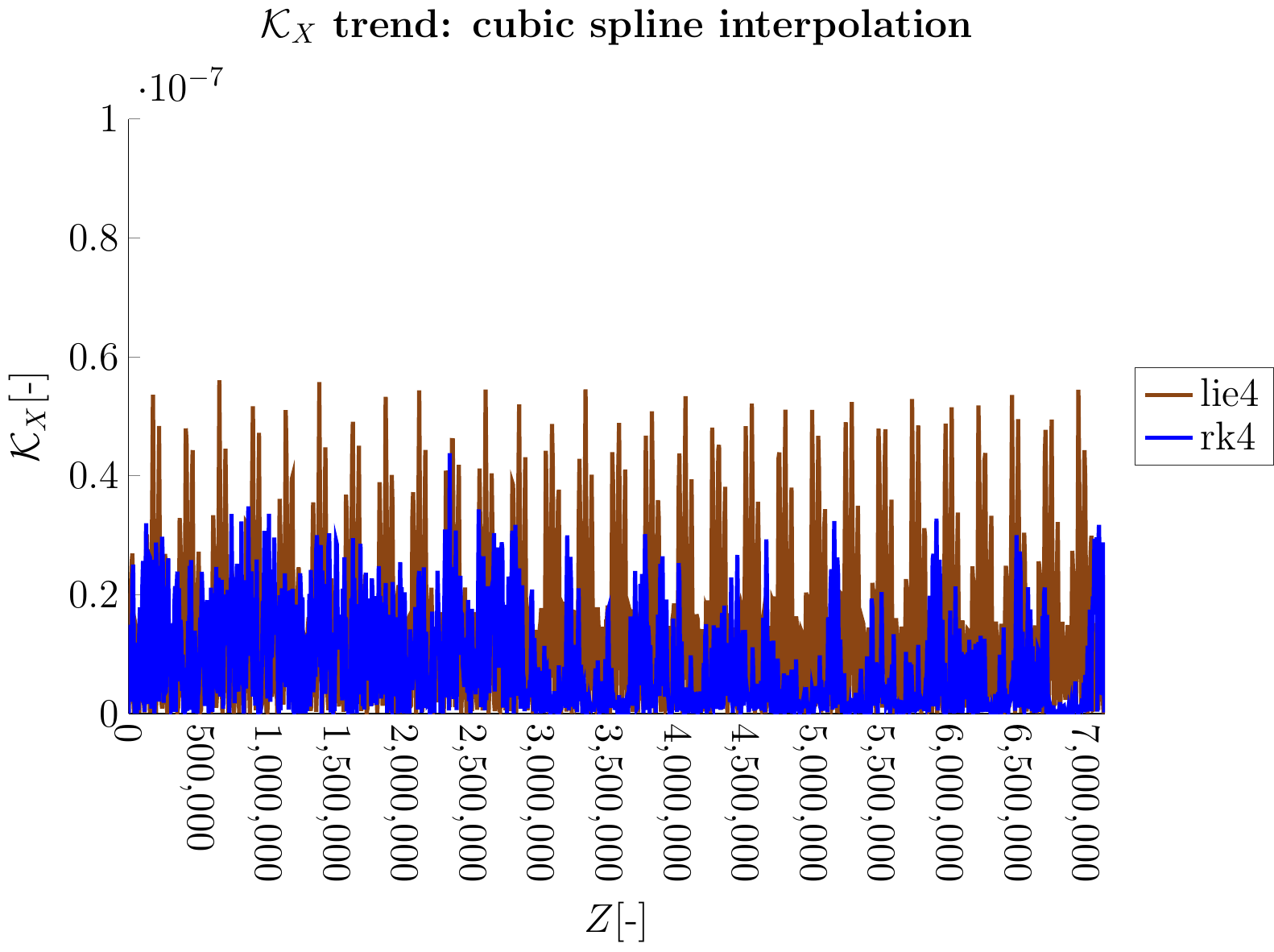}
    \caption{Energy conservation of a particle through $1600000$ focusing-defocusing quadrupole couples using the fourth-order explicit Runge-Kutta and the fourth-order Lie method.}
    \label{fig:1600000rk4}
\end{figure}

 A particularly interesting aspect of the results obtained was the fact that non symplectic methods
appeared to be competitive with symplectic ones, even on relatively long integrations. An even more extreme
example of this behaviour was seen in the simulation of a particle crossing $1600000$ quadrupole couples of realistic design ($32$ quadrupoles in $100000$ revolutions). An integration step $\Delta Z = 0.16 $
was employed. As shown in  figure \ref{fig:1600000rk4}, even in this extremely long simulation
the behaviour of a symplectic  Lie method and of a non symplectic Runge Kutta method was
entirely analogous in terms of energy conservation, in contrast to the theoretical expectations.
It would be interesting to understand whether more realistic test cases could actually highlight  negative effects
induced by  non-symplectic methods and whether more detailed monitoring of the particle motion
could be employed for this purpose. For example, 
in this work only the energy associated to the transverse variable was used to measure the good behaviour in long-term simulations. Other conserved quantities could also be used to compare the performance of  the
different  methods.

\section*{Acknowledgements} 
This paper contains an extension of results initially presented in the Master Thesis of A.S.,
who was partially supported  by CEA Saclay, where part of this work was carried out.

\bibliographystyle{plain}
\bibliography{quadrupole}

\begin{thebibliography}{15}
\providecommand{\natexlab}[1]{#1}
\providecommand{\url}[1]{\texttt{#1}}
\expandafter\ifx\csname urlstyle\endcsname\relax
  \providecommand{\doi}[1]{doi: #1}\else
  \providecommand{\doi}{doi: \begingroup \urlstyle{rm}\Url}\fi

\bibitem[rox(2018)]{roxie:web}
{ROXIE} - software for electromagnetic simulation and optimization of
  accelerator magnets.
\newblock \url{https://espace.cern.ch/roxie/default.aspx}, 2018.

\bibitem[Carey(1988)]{carey:1988}
D.~C. Carey.
\newblock The optics of charged particle beams.
\newblock \emph{Applied Optics}, 27:\penalty0 1002, 1988.

\bibitem[Corno et~al.(2016)Corno, de~Falco, Gersem, and Sch{\"o}ps]{corno:2016}
J.~Corno, C.~de~Falco, H.~De Gersem, and S.~Sch{\"o}ps.
\newblock Isogeometric simulation of {L}orentz detuning in superconducting
  accelerator cavities.
\newblock \emph{Computer Physics Communications}, 201:\penalty0 1--7, 2016.

\bibitem[Dalena and Pugnat(2015)]{pugnat:2015}
B.~Dalena and T.~Pugnat.
\newblock Calcul d'une carte de transport r\'ealiste pour particules
  charg\'ees.
\newblock Technical report, Centre Energie Atomique, 2015.

\bibitem[Dalena et~al.(2014)Dalena, Gabouev, Giovannozzi, Maria, Appleby,
  Chanc{\'e}, Payet, and Brett]{dalena:2014}
B.~Dalena, O.~Gabouev, M.~Giovannozzi, R.~De Maria, R.B. Appleby,
  A.~Chanc{\'e}, J.~Payet, and D.R. Brett.
\newblock Fringe fields modeling for the high luminosity {LHC} large aperture
  quadrupoles.
\newblock Technical Report CERN-ACC-2014-0175, CERN, 2014.

\bibitem[Dragt(1997)]{dragt:1997}
A.J. Dragt.
\newblock \emph{Lie methods for nonlinear dynamics with applications to
  accelerator physics}.
\newblock University of Maryland, Center for Theoretical Physics, Department of
  Physics, 1997.
\newblock URL \url{http://www.physics.umd.edu/dsat/}.

\bibitem[Forest(1998)]{forest:1998}
E.~Forest.
\newblock \emph{Beam {D}ynamics}.
\newblock CRC Press, 1998.

\bibitem[Hairer et~al.(2006)Hairer, Lubich, and Wanner]{hairer:2006}
E.~Hairer, C.~Lubich, and G.~Wanner.
\newblock \emph{Geometric numerical integration: structure-preserving
  algorithms for ordinary differential equations}.
\newblock Springer Science \& Business Media, 2006.

\bibitem[Kotkin and Serbo(2013)]{kotkin:2013}
G.L. Kotkin and V.~G. Serbo.
\newblock \emph{Collection of Problems in Classical Mechanics: International
  Series of Monographs in Natural Philosophy}, volume~31.
\newblock Elsevier, 2013.

\bibitem[Lipnikov et~al.(2011)Lipnikov, Manzini, Brezzi, and
  Buffa]{lipnikov:2011}
K.~Lipnikov, G.~Manzini, F.~Brezzi, and A.~Buffa.
\newblock The mimetic finite difference method for the 3{D} magnetostatic field
  problems on polyhedral meshes.
\newblock \emph{Journal of Computational Physics}, 230\penalty0 (2):\penalty0
  305--328, 2011.

\bibitem[Panofsky and Phillips(1962)]{panofsky:1962}
W.K.~H. Panofsky and M.~Phillips.
\newblock \emph{Classical electricity and magnetism}.
\newblock Addison Wesley, 1962.

\bibitem[Rossi(2011)]{rossi:2011}
L.~Rossi.
\newblock {LHC} upgrade plans: Options and strategy.
\newblock In \emph{Second International Particle Accelerator Conference, San
  Sebastian}, 2011.

\bibitem[Venturini and Dragt(1999)]{venturini:1999}
M.~Venturini and A.J. Dragt.
\newblock Accurate computation of transfer maps from magnetic field data.
\newblock \emph{Nuclear Instruments and Methods in Physics Research Section A:
  Accelerators, Spectrometers, Detectors and Associated Equipment},
  427\penalty0 (1):\penalty0 387--392, 1999.

\bibitem[Wu et~al.(2003)Wu, Forest, and Robin]{wu:2003}
Y.K. Wu, E.~Forest, and D.S. Robin.
\newblock Explicit symplectic integrator for $s-$dependent static magnetic
  field.
\newblock \emph{Physical Review E}, 68\penalty0 (4):\penalty0 046502, 2003.

\bibitem[Yoshida(1990)]{yoshida:1990}
H.~Yoshida.
\newblock Construction of higher order symplectic integrators.
\newblock \emph{Physics Letters A}, 150\penalty0 (5):\penalty0 262--268, 1990.

\end{thebibliography}

\end{document}